# Computational and Molecular Dissection of an X-box cis-Regulatory Module

by

**Timothy Burton Warrington**

M.Sc., University of Alberta, 2010
B.Sc., University of Alberta, 2005

Thesis Submitted in Partial Fulfillment of the
Requirements for the Degree of
Doctor of Philosophy

in the
Department of Molecular Biology and Biochemistry
Faculty of Science

**SIMON FRASER UNIVERSITY**

**Fall 2015**

## Approval

**Name:** **Timothy Burton Warrington**

**Degree:** **Doctor of Philosophy**

**Title:** ***Computational and Molecular Dissection of an X-box cis-Regulatory Module***

**Examining Committee:** **Chair:** Rosemary B. Cornell
Professor

**Jack N. Chen**
Senior Supervisor
Professor

**David L. Baillie**
Supervisor
Professor

**Bruce P. Brandhorst**
Supervisor
Professor Emeritus

**Michel Leroux**
Internal Examiner
Professor
Department of Molecular Biology and Biochemistry

**Catharine Rankin**
External Examiner
Professor
Department of Psychology
University of British Columbia

**Date Defended/Approved:** October 06, 2015

## Abstract

Ciliopathies are a class of human diseases marked by dysfunction of the cellular organelle, cilia. While many of the molecular components that make up cilia have been identified and studied, comparatively little is understood about the transcriptional regulation of genes encoding these components. The conserved transcription factor Regulatory Factor X (*RFX)/DAF-19*, which acts through binding to the cis-regulatory motif known as X-box, has been shown to regulate ciliary genes in many animals from *Caenorhabditis elegans* to humans. However, accumulating evidence suggests that *RFX* is unable to initiate transcription on its own. Therefore, other factors and cis-regulatory elements are likely required. One such element, a DNA motif called the C-box, has recently been identified in *C. elegans*. It is still unclear if the X-box and C-boxes are the only regulatory elements involved and how they interact. To this end, I analyzed the transcriptional regulation of *dyf-5,* the *C. elegans* ortholog of the human ciliopathy gene Male-Associated Kinase (*MAK*). Using computational methods, I was able to confirm the presence of the previously reported X-box and C-boxes as well as identifying an additional C-box. By sequentially mutating each of the identified motifs, I identified the role each potential motif plays in transcriptional regulation of *dyf-5*. My results showed that only the X-box and the three C-boxes are necessary and are sufficient to drive transcription, with the X-box and the centre C-box being the major contributors and the other two C-boxes enhancing expression. This study advances the knowledge of gene regulation in general and will further our understanding of ciliopathies and the mutations that cause them.

*For Teresa,*

*You made me believe dreams can come true*

# Acknowledgements

A project like this cannot be accomplished by one person. Many people deserve thanks for their contributions. First I would like to thank Christian Frech and Tammy Wong for their help with the bioinformatics, they helped me get going when I didn't know where to start, and Chris even re-wrote his gene model script to display upstream promoter elements for me. Next, I would like to thank my compatriots in the wet lab, Ting Zhang, Zhaozhao Qin, and Jun Wang. Technical discussions with them were instrumental to this thesis. My committee members, Lynne Quarmby, Bruce Brandhorst, and David Baillie, all helped keep this thesis focused and on track and helped me avoid getting pulled onto tangents. I would also like to thank my senior supervisor, Jack Chen. Finally, I'd like to thank Teresa Pearen. She grudgingly came with me on this adventure but she kept me sane and kept me going when I was ready to give up. Thank-you all!

# Table of Contents

## List of Tables

## List of Figures

## List of Acronyms

| | |
|---|---|
| BBS | Bardet-Biedl syndrome |
| BRE | B recognition element |
| ChIP | Chromatin immunoprecipiation |
| CRM | *cis*-regulatory element |
| DPE | Downstream promoter element |
| EMSA | Electrophoretic mobility shift assay |
| GFP | Green flourescent protein |
| GRN | Gene regulatory network |
| IFT | Intraflagellar Transport |
| MosSCI | Mos transposase mediated Single Copy Insertion |
| PCR | Polymerase Chain Reaction |
| PKD | Polycistic kidney disease |
| RFX | Regulatory Factor X |
| TF | Transcription factor |
| TFBS | Transcription factor binding site |
| TSS | Transcription Start Site |

## Glossary

| | |
|---|---|
| Amphid | Chemosensory organ found in head of nematodes, enervated by ciliated amphid neurons |
| C-box | A DNA motif found in pan-ciliary expressed genes that is necessary for gene expression |
| CTCF | CCCTC-binding factor, a protein that facilitates DNA looping |
| Dynein | A molecular motor that moves along microtubules, powers retrograde intraflagellar transport |
| GFP | Green flourescent protein, a protein that fluoresces green |
| HMMER | Computer software for detecting DNA motifs, uses hidden Markov models |
| Intergenic | The genomic region between two genes |
| Kinesin | A molecular motor that moves along microtubules, powers anterograde intraflagellar transport |
| mCherry | A protein that fluoresces red |
| MosSCI | Mos-mediated single copy insertion, a technique to introduce a single copy of a transgene into the genome of a worm |
| Phasmid | Sensilia found at tail end of nematodes, enervated by the ciliated phasmid neurons |
| tdTomato | A protein that fluoresces red |
| X-box | A DNA motif that is necessary for the expression of ciliary genes, bound by *RFX/Daf-19* |
| XXmotif | Web-server based software for identifying shared DNA motifs |

# Chapter 1. Introduction

## 1.1. Cilia

Cilia are small microtubule-based organelles that project outwards from cells. They are very well conserved throughout many eukaryotic lineages and are present on most cell types in vertebrates (Wheatley et al. 1996). There are two main types of cilia: motile and non-motile. Motile cilia, as their name suggests, are capable of movement and are responsible for cell motility (e.g. sperm) and fluid flow (e.g. lung epithelium) (Ainsworth 2007; Drummond 2012). Non-motile cilia do not move and perform sensory functions (e.g. olfaction, vision). Cilia are found on almost all vertebrate cell types and can have many diverse roles, including olfactory sensing, fluid movement, and even play roles in patterning the embryo such as establishing left-right asymmetry via the node cilia and hedgehog receptors localised to cilia (Drummond 2012) (Figure 1-1).

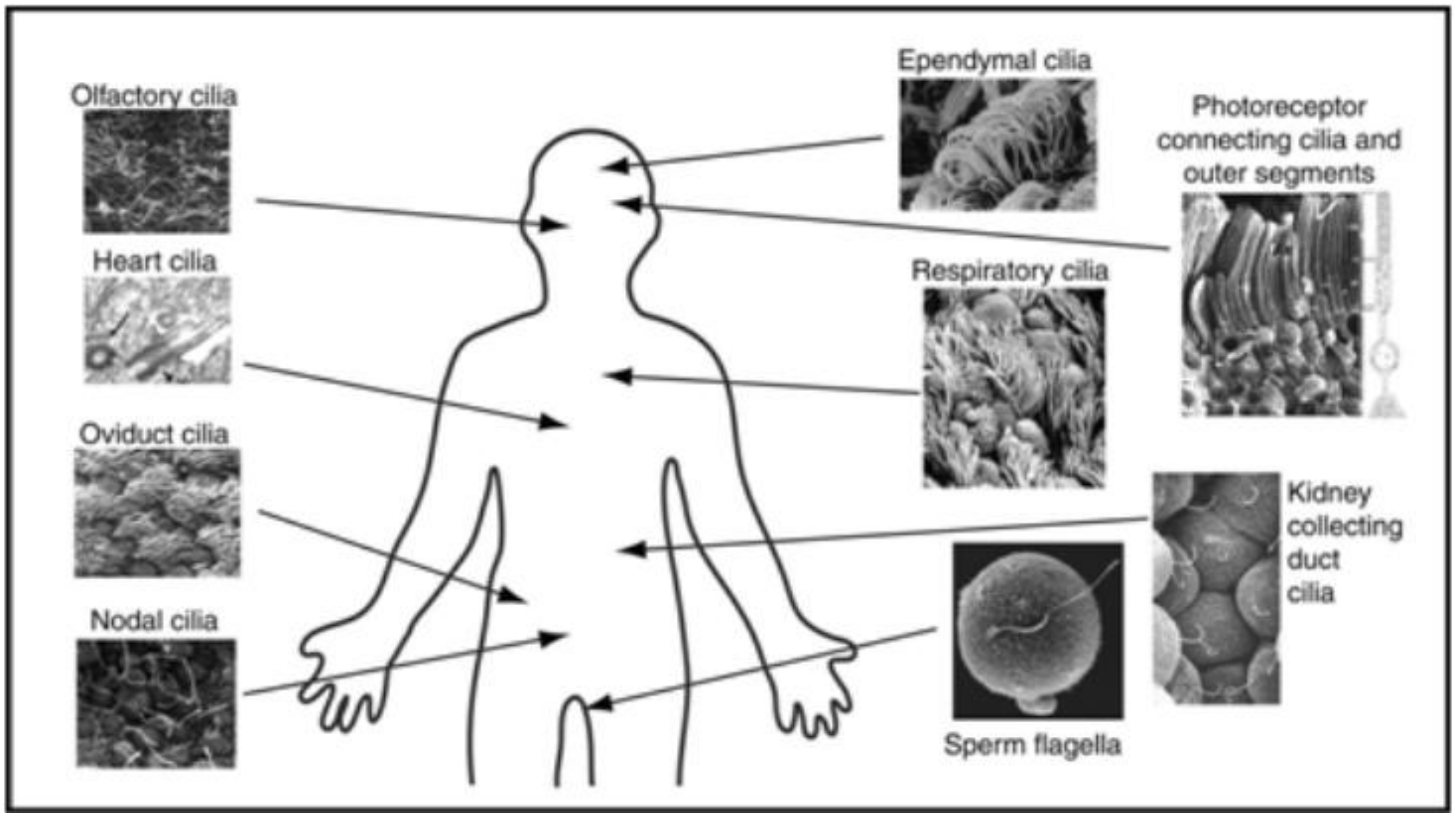

**Figure 1-1. Human cilia types.**
Examples of cilia types found in the human body. Original sources for images: Olfactory, oviduct, photoreceptor, and kidney cilia (Kessel and Kardon 1979); heart cilia (Willaredt et al. 2012); nodal cilia (Follit et al. 2014); ependymal cilia (O'Callaghan et al. 1999); respiratory cilia (Rosenbaum and Witman 2002); sperm on oocyte (Brown and Witman 2014). Figure reprinted with permission from Brown and Witman (Brown and Witman 2014)

### 1.1.1. Cilia structure

The cilium is composed of an axoneme extending from a basal body. The basal body is formed from a centriole that docks with the cell membrane. Nine pairs of microtubules extend from the basal body to form the proximal segment of the axoneme. This is reduced to nine microtubule singlets in the distal segment (Silverman and Leroux 2009). In motile cilia, an additional pair of microtubules is present in the centre of the axoneme (Figure 1-2). An additional region known as the transition zone is found between the basal body and the axoneme. This region often has Y-links that anchor it to the surrounding membrane (Shiba and Yokoyama 2012). Finally, the entire cilium is surrounded with membrane which is continuous with the plasma membrane. Receptor molecules are often localised to this region (Goetz and Anderson 2010).

Intraflagellar transport (IFT) is responsible for transporting molecules throughout the cilium. IFT particles are assembled at the basal body transition fibers for transport along the cilium (Williams et al. 2011). The IFT complex contains three distinct modules consisting of two separate cargos termed, IFT-A and IFT-B, and the BBSome, which stabilises the complex of the two (Ou et al. 2007; Burghoorn et al. 2007). Two kinesin-2 motors, kinesin-II and OSM-3, transport molecules to the tip of the ciloum during anterograde transport. Kinesin-II, made up of the products of *klp-11, klp-20,* and *kap-1*, and OSM-3 function together in the middle segment while OSM-3 alone is responsible for anterograde transport in the distal segment. (Hao and Scholey 2009; Burghoorn et al. 2007). A single type of dynein motors is responsible for retrograde transport back to the cilium base (Hao and Scholey 2009).

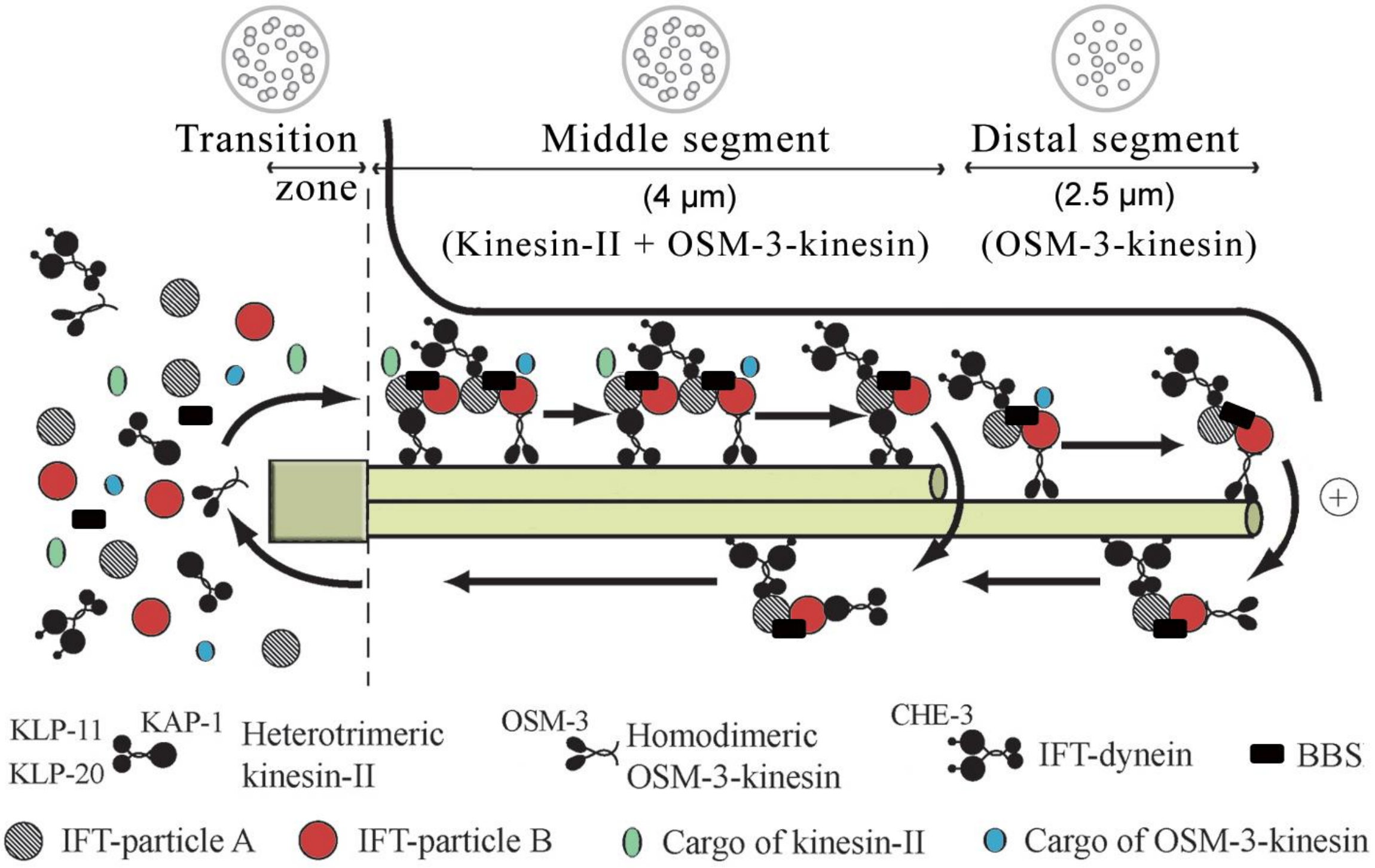

**Figure 1-2. Cilia structure.**
Top: Lateral cross-sections of cilia. Transition zone has doublet microtubules and Y-links, proximal and distal segments have microtuble doublets and singlets respectively. Bottom: Longitudinal cross-section of cilium showing IFT. Anterograde IFT involves both Kinesin-II and OSM-3 in the proximal segment. Only OSM-3 is used in the distal segment. Retrograde IFT mediated by CHE-3 dynein. Image source: (Inglis et al. 2007). Used under creative commons licence (http://creativecommons.org/licenses/by/2.5/).

## 1.1.2. Ciliopathies

Dysfunction of cilia causes a variety of human disorders collectively known as ciliopathies. More than fifteen distinct syndromes are recognised and altogether affect upwards of 1/1000 births (Davis and Katsanis 2012; Badano et al. 2006). These disorders can affect virtually any organ and can have a broad range of symptoms including loss of smell (anosmia), blindness, kidney problems, and even developmental defects such as polydactyly and *situs inversus* (Badano et al. 2006; Hildebrandt et al. 2011; Lee and Gleeson 2011).

A number of diseases and syndromes have been directly linked to mutations in cilia genes. Primary ciliary dyskinesia is caused by defects in motile cilia. Symptoms include bronchitis and sinusitis, as well as infertility and *situs inversus*. Twenty-one associated genes have been identified (Knowles et al. 2013). Polycystic kidney disease, another ciliopathy, is marked by kidney failure caused by the over-proliferation of kidney epithelial cells, which ultimately block kidney ducts. The genetic cause of this seems to be disruption of the formation of calcium channels localised to cilia present on kidney cells (Badano et al. 2006). Cilia dysfunction can also cause blindness. For example, retinitis pigmentosa causes progressive vision loss throughout life. More than seventy-five genes have been identified to play a role in this disease (Rivolta et al. 2002). Finally, ciliopathy syndromes are associated with multiple organs and include symptoms mentioned previously in addition to other unique features. For example, Bardet-Beidl syndrome (BBS) symptoms include polydactyly, brachydactyly obesity, blindness, and cystic kidneys. Most causative genes are involved in the BBSome, which is involved in loading IFT particles (Avasthi and Marshall 2012). Obviously, these syndromic ciliopathies are more severe and in many cases fatal. Although many causative genes have been identified, how these genes are regulated is still largely unknown.

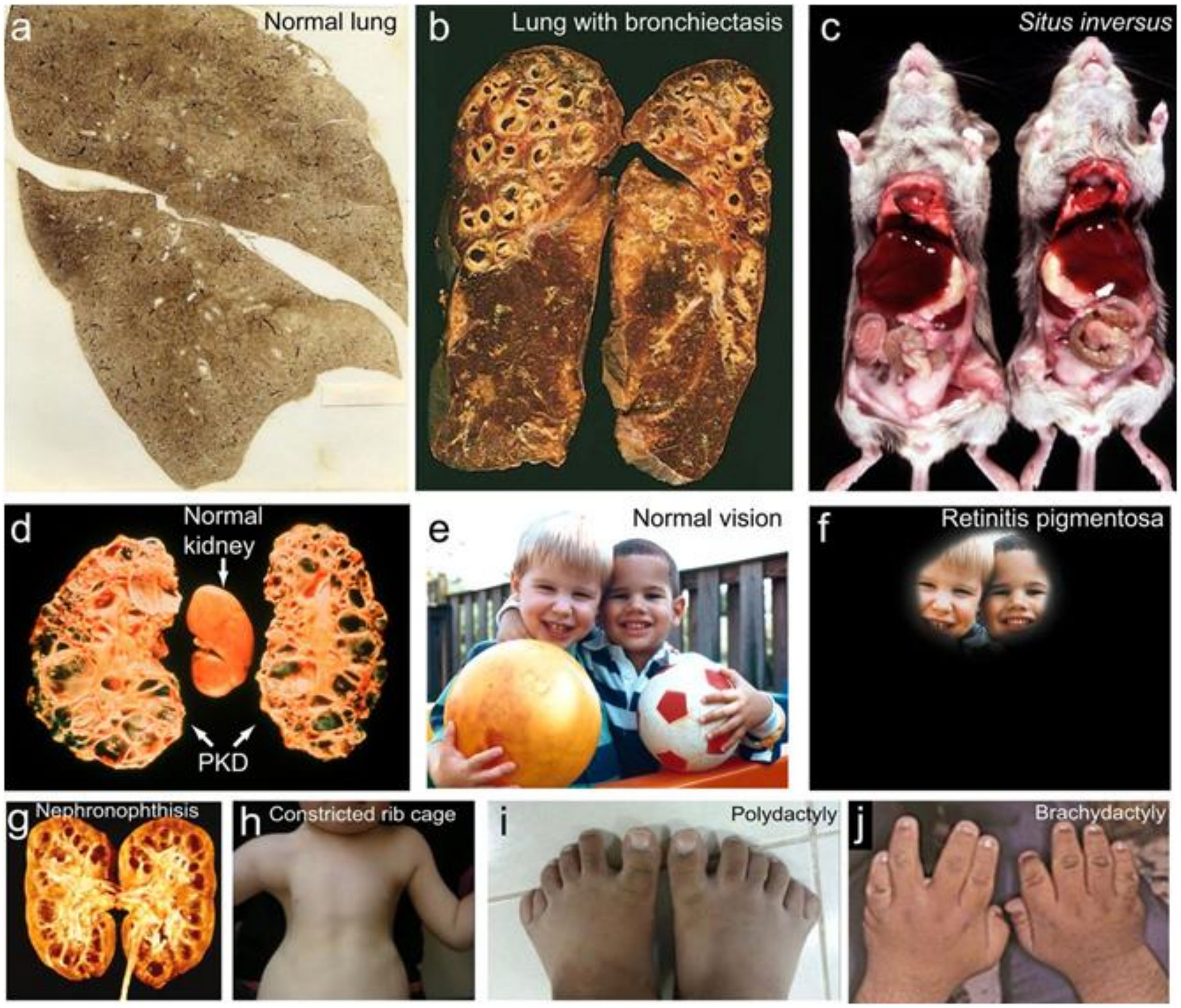

**Figure 1-3. Symptoms of ciliopathies.**
Various ciliopathy phenotypes. (a) Normal human lungs. (b) Lungs with bronchiectasis. (c) Left: a normal mouse, Right: a mouse with situs inversus (d) polycystic kidney disease (PKD). (e) Normal vision (f) how someone suffering from retinitis pigmentosa ight view (e).(g) Nephronophthisis. (h) Constricted rib cage of someone suffering from asphyxiating thoracic dystrophies. (i) Polydactyly. (j) Brachydactyly. Original sources of images: (a) National Institute for Occupational Safety, Centers for Disease Control; (b) Matthew M. Fitz, Loyola University Chicago, Stritch School of Medicine; (c) Noah's Arkive Database, Department of Pathology, University of Georgia College of Veterinary Medicine (http://dlab.vet.uga.edu/NA); (d) Vicente E. Torres, Mayo Clinic; (e, f) National Eye Institute, National Institutes of Health, reference nos: EDS01 and EDS07; (g) (Hildebrandt and Zhou 2007); (h) (Huber and Cormier-Daire 2012); (i) (Aldahmesh et al. 2014); (j) (Forsythe and Beales 2013) Figure reprinted with permission from Brown and Witman (Brown and Witman 2014)

## 1.2. Gene Regulation

### 1.2.1. Transcriptional Regulation

Transcriptional regulation refers to the complex regulation of transcription initiation allowing for correct temporospatial expression of a gene. In eukaryotes, transcription of DNA into RNA is accomplished by three enzymes RNA polymerase I (Pol I) which transcribes ribosomal RNA (rRNA), RNA polymerase II (Pol II) which primarily transcribes messenger RNA (mRNA), and RNA polymerase III (Pol III) which primarily transcribes transfer RNA (tRNA) (Vannini and Cramer 2012). For the purposes of this thesis, we will only consider regulation of Pol II transcribed genes as all known cilia genes are protein coding and therefore fall into this category.

In general, transcription of a gene occurs when Pol II is recruited to the promoter of that gene. The classical promoter consists of two components: the core promoter and the promoter proximal region. The core promoter is typically considered sufficient for initiation of transcription. It is located between forty base pairs upstream to forty base pairs downstream of the transcription start site (TSS). As the name suggests, the TSS is the location where transcription is initiated. The core promoter may contain several binding sites for core transcription machinery including TATA box, B recognition element (BRE), which is bound by Transcription Factor IIB, Initiator (Inr), and downstream promoter element (DPE). These elements are seldom all present together (Blackwood and Kadonaga 1998; Chen et al. 2013). The promoter proximal region is located immediately upstream of the core promoter, usually between fifty and two-hundred base pairs upstream of the TSS (Blackwood and Kadonaga 1998). This region frequently contains multiple transcription factor binding sites including GC-box and CAAT box as well as binding sites for cell type specific transcription factors (Blackwood and Kadonaga 1998) (Figure 1-4).

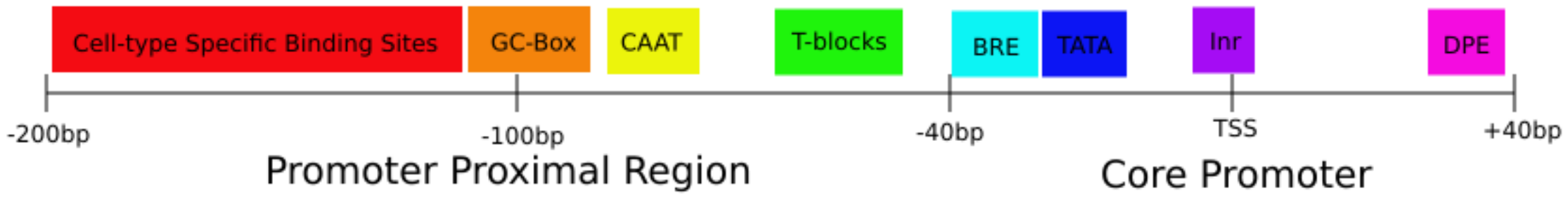

**Figure 1-4. Relative arrangement of promoter elements.**
Core promoter found between -40bp and +40bp relative to the transcription start site (TSS). Promoter proximal region is found upstream of the core promoter, usually within 200bp. Not all elements are present in all genes/species. T-blocks are unique to *C. elegans*.

The transcription of genes can also be controlled by distant enhancers. These enhancer regions function in much the same way as the promoter proximal transcription factor binding sites. They can, however, be quite distant from the genes they control and may even be located downstream or even within introns (Blackwood and Kadonaga 1998; Jeziorska et al. 2009). Activity of enhancers is restricted to particular domains by insulators. Insulators are thought to function by controlling the three dimensional organisation of DNA within the nucleus by interacting with a protein called CTCF. CTCF is a zinc-finger transcription factor which can interact with both DNA and other proteins. It is able to form DNA loops by interacting with other CTCF molecules or associate with the nuclear lamina via cohesin (Guelen et al. 2008; Lee and Iyer 2012).

An additional mechanism of controlling gene transcription is by chromatin. In general, chromatin modification affects the accessibility of the transcription factor binding sites. This can be accomplished by modifying the DNA molecule itself by methylation of cytosines, which is generally repressive, or by histone modifications, such as acetylation, which generally promotes transcription (Natoli and Andrau 2012; Thurman et al. 2012).

As mentioned previously, genes are regulated by a collection of transcription factors (TFs) that modulate transcription from a promoter. These TFs act by binding to DNA sequences (called transcription factor binding sites or TFBS) near the controlled promoter and modulating recruitment of Pol II. These TFs bind TFBS, and function like a “logic switch”, turning on and off genes at important times (Jeziorska et al. 2009; Hobert 2008; Levine and Tjian 2003) Although many TF/TFBS interactions approximate Boolean logic, they are not truly Boolean and are often better described by three component logic (eg. low, medium, and high concentrations rather than on or off) (Teif 2010). A collection of these transcription factors that defines an expression pattern is referred to as a cis-regulatory module CRM) (Okkema and Krause 2005; Reinke et al. 2013; Jeziorska et al. 2009). A single gene may contain multiple CRMs which provide fine control of gene expression under multiple conditions (Jeziorska et al. 2009). A collection of interacting genes function together to form a gene regulatory network (GRN) (Figure 1-5). By combining collections of activators and repressors GRNs are able to implement multiple types of logic, including traditional Boolean AND, OR and NOT logic. Feedback loops are also very important. This can be positive feedback, where a gene either directly or indirectly activates itself. These can be important for state changes, once the gene is activated it stays on, and for amplification of the signal. Alternatively, negative feedback is possible. This involves a gene inhibiting itself. This can be used to keep the amount of a gene product within certain limits or to create an oscillating signal (Davidson and Levine 2008). A final type of genetic circuit that is common in development is the double negative circuit. This involves a global repressor that represses a certain collection of genes throughout a large region, and a more locally expressed repressor that represses the global repressor, thereby resulting in the activation of the collection of genes in the smaller region (Figure 1-5c) (Davidson and Levine 2008; Davidson 2009).

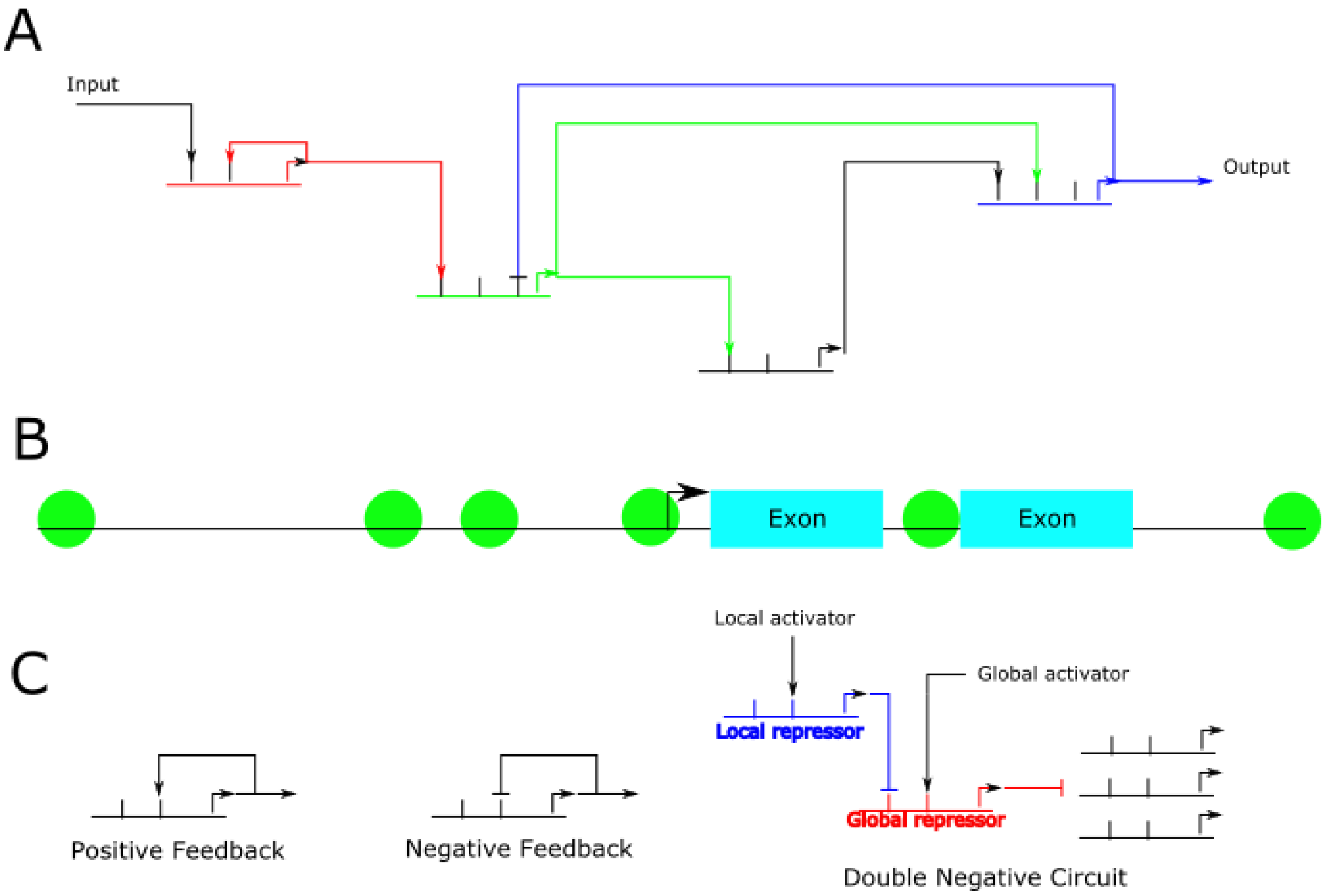

**Figure 1-5. Gene regulatory networks and cis-regulatory modules.**
A. Schematic diagram of hypothetical gene regulatory network (GRN). Red gene senses some input and self activates and activates green gene. Green gene activates black gene. Black and green gene activates blue gene which then represses the green gene expression and produces some output. All types of logic are possible with different configurations. B. Example gene containing cis-regulatory modules (green circles, CRM). Multiple CRMs can regulate the same gene. Each CRM can contain multiple TFBS. C. Examples of GRN circuit types.

### 1.2.2. Transcriptional regulation in *Caenorhabditis elegans*

Transcriptional regulation has been well studied in *C. elegans* (Chen et al. 2013; Reinke et al. 2013; Okkema and Krause 2005; Grishkevich et al. 2011). It is fairly typical for eukaryotes and thus is similar to transcription described above with a few additions. Some *C. elegans* core promoters contain T-blocks, which are blocks of thymine residues repeated approximately ten base pairs apart which have been shown to positively affect transcription. Large numbers of T-blocks are associated with higher transcription, for example genes with six or more T-blocks have fivefold higher expression (Grishkevich et al. 2011). An additional complication in *C. elegans* is transplicing (Bektesh and Hirsh 1988). Upwards of 70% of all genes in *C. elegans* have a twenty-two nucleotide splicing leader spliced onto the 5' end of their messenger RNAs. This splicing typically occurs at a TTnCAG upstream of the start codon. Because the 5' region of the mRNA is removed, the true TSS is often difficult to identify (Bektesh and Hirsh 1988; Chen et al. 2013; Grishkevich et al. 2011).

*C. elegans* has been used previously to uncover several cis-regulatory modules including those that control the genes required for determination of several neuronal fates (Wenick and Hobert 2004; Etchberger et al. 2007; Lanjuin and Sengupta 2004) and pharyngeal development (Ao et al. 2004; Gaudet and Mango 2002; Gaudet et al. 2004). As specialization of the cilium is one result of this fate determination, it is possible for some overlap with cilia gene regulation. For example, FKH-2 has been shown to be required for cilium specialization in ASH neurons (Mukhopadhyay et al. 2007). This collection of previous work will aid in determining the regulation mechanisms controlling cilia in *C. elegans*, which will be generally applicable to humans due to the similarity in regulatory mechanisms.

### 1.2.3. Importance of Regulation

Regulation of genes is immensely important as mis-regulation can have as dramatic an effect as mutation of the protein it encodes. Correct temporal and special expression of genes is necessary for the function many processes including development (Davidson and Erwin 2006; Li and Davidson 2009). Even small changes can have visible effects. For example, blonde hair in Europeans is the result of a single nucleotide polymorphism in a lymphoid enhancer-binding factor 1 (LEF1) binding site in the *KITLG* gene (Guenther et al. 2014). In addition, mis-regulation of genes can cause disease. For example, bare lymphocyte syndrome and cone-rod dystrophy are both caused by mutations in transcription factors (*RFX5* and *CRX* respectively) (Reith and Mach 2001; Freund et al. 1997). Also, many cancers are also caused by inappropriate regulation of genes, for example p53.

## 1.3. Transcriptional regulation of ciliary genes

### 1.3.1. Regulation of ciliary genes in humans and model organisms

In animals, cilia are regulated by a very well conserved class of transcription factors, Regulatory Factor X (RFX) (Emery et al. 1996; Chu et al. 2010). Many species have multiple RFX genes including two in *Drosophila melanogaster*, eight in humans, and one in *C. elegans* (Aftab et al. 2008; Swoboda et al. 2000; Chu et al. 2010; Emery et al. 1996; Reith et al. 1990). Based on phylogenetic analysis, RFX proteins can be placed into three groups. Group 1, consists of human RFX 1-3, share several protein domains in addition to the RFX DNA binding domain. Group 2, containing human RFX4 and 6, *Drosophila* RFX and *C. elegans daf-19*, lack the activation domain present in the first group but are otherwise conserved. Finally, the third group contains human RFX 5 and 7 and *Drosophila* RFX1 and 2, in which only the DNA binding domain is conserved (Chu et al. 2010). In humans, RFX3 is primarily involved in ciliogenesis, with RFX2 and 4 having additional cilia roles in the brain and testis (Choksi et al. 2014). RFX5 has roles in regulating MHC class II genes which are important for immune function and mutation of this gene can cause bare lymphocyte syndrome (Steimle et al. 1995). Interestingly, the role of RFX5 in the immune system may be related to the function of RFX in ciliogenesis as intraflagellar transport is important for the formation of the immune synapse (Finetti et al. 2011; Baldari and Rosenbaum 2010).

RFX functions by binding to an approximately fourteen base pair DNA sequence called the X-box (Reith et al. 1990; Emery et al. 1996) which is found in the promoter region of the regulated gene. This sequence consists of an imperfect inverted repeat of six base pair binding sites separated by a one to three base pair spacer (e.g. GTNRCC-$N_{1-3}$-RGYAAC ) (Swoboda et al. 2000; Chen et al. 2006; Efimenko et al. 2005; Blacque et al. 2005; Chu et al. 2012; Burghoorn et al. 2012; Emery et al. 1996; Newton et al. 2012). Some genes are regulated by more than one X-box which appears to fine-tune expression of the particular gene (Chu et al. 2012).

Evidence in several species suggests that X-box/RFX works in conjunction with other elements to specify ciliary expression. In vertebrates, the forkhead box transcription factor FoxJ1 specifies motile cilia formation. Mice lacking FoxJ1 show a lack of motile cilia both on single ciliated cells and multi-ciliated cells (Chen et al. 1998; Brody et al. 2000). The role of FoxJ1 in ciliogenesis of motile cilia is conserved in other vertebrates including *Xenopus* and zebrafish (Stubbs et al. 2008; Yu et al. 2008). FoxJ1 functions by binding to a short motif similar to TRTTTA (Badis et al. 2009; Nakagawa et al. 2013) (Figure 1-6).

In *Drosophila*, a different forkhead box transcription factor, *FD3F,* is required for specialisation of the chordotonal neurons (Newton et al. 2012). Although *FD3F* is only distantly related to FoxJ1, it is interesting to note that *Drosophila* lack motile cilia with the exception of the chordotonal neurons which maintain some motile features suggesting that this system maybe highly derived from the FoxJ1 system seen in vertebrates. *FD3F* functions by binding to a Fox motif, RYMAAYA, present in promoters (Newton et al. 2012).

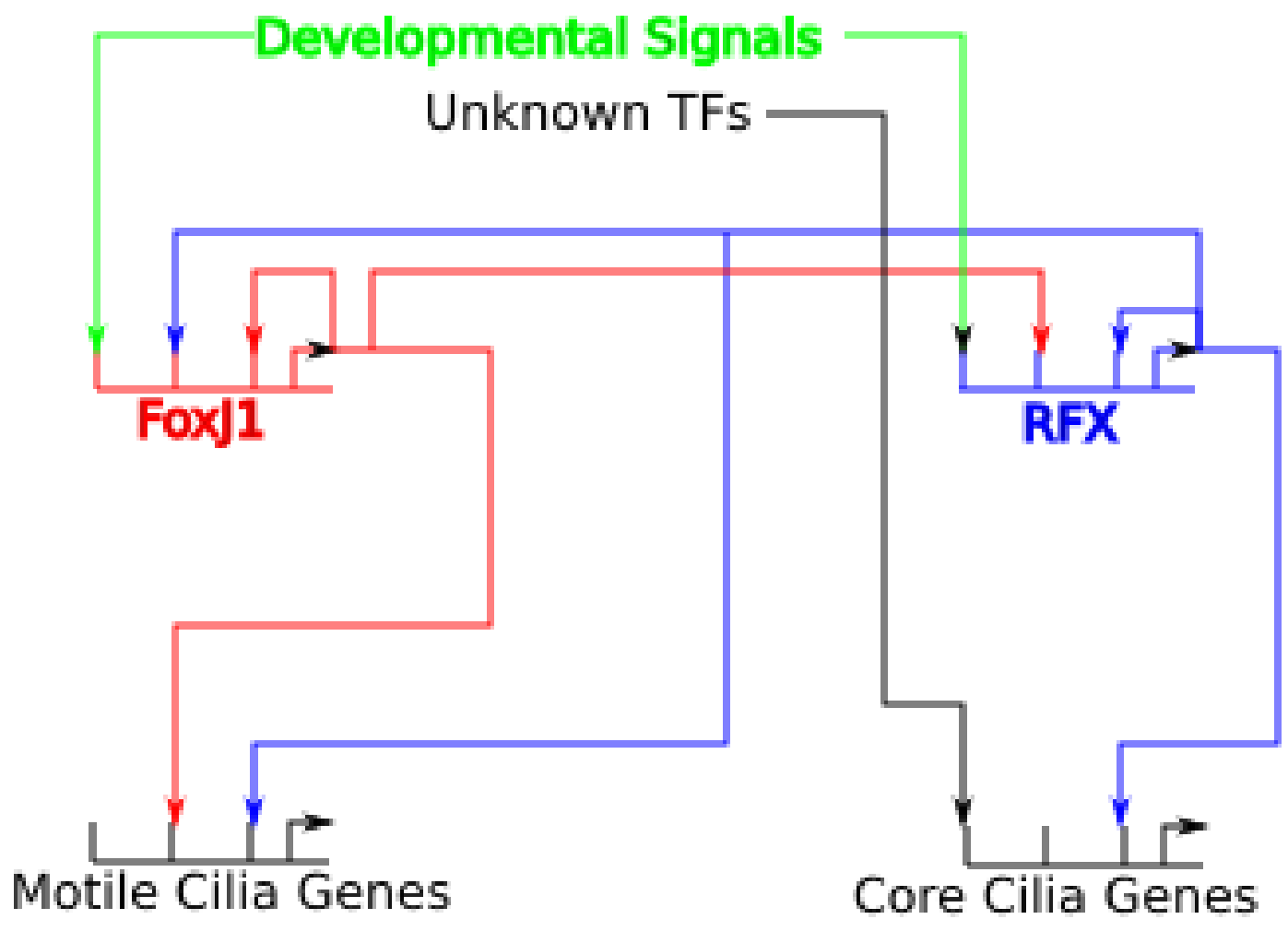

**Figure 1-6. RFX and FoxJ1 regulate motile cilia genes in vertebrates.** RFX and FoxJ1 are controlled by various developmental signals as well as regulate each other. FoxJ1 and RFX together activate motile cilia genes, whereas RFX and other unknown TFs activate core cilia genes.

### 1.3.2. Regulation of ciliary genes in *Caenorhabdits elegans*

*C. elegans* is a well established model organism, first established by Sydney Brenner (Brenner 1974). In recent years, the toolkit of techniques has been expanded (Kaletta and Hengartner 2006; Xu and Kim 2011). Additionally, *C. elegans* is part of the modENCODE project, which seeks to identify functional DNA elements (Gerstein et al. 2010). This organism is small and easily maintained as well as being transparent which makes them ideal for reporter gene studies. Additionally, *C. elegans* has many closely related species that have been sequenced, allowing comparative genomics (Kiontke and Fitch 2005; Kiontke et al. 2011; Stein et al. 2003). Worms are also easy to transform by microinjection (Praitis and Maduro 2011), and single copy insertions via Mos transposase are also possible (Frøkjaer-Jensen et al. 2008; Frøkjær-Jensen et al. 2012). Recently CRISPR genome editing tools have also become available (Frøkjær-Jensen 2013).

*C. elegans* is a convenient model for studies of cilia as it has only 60 ciliated cells, all of which are sensory neurons. These cells are organised in three main clusters, the labial and amphid neurons in the head, and the phasmid neurons in the tail (Inglis et al. 2007). These sensory neurons all have specific functions and cilia specialisations. Many of the amphid cilia are rod shaped, possessing either a single rod (ASE, ASG, ASH, ASI, ASJ, and ASK neurons) or dual rod shape cilia (ADF, ADL neurons). These neurons are involved in either chemosensation or mechanosensation. The phasmid neurons PHA and PHB also possess single rod shaped cilia and are involved in chemorepulsion. The wing-shaped cilia of the AWA, AWB, and AWC neurons sense volatile odorants. Another specialised cilium is found on the AFD neuron, which contains a small, finger-like cilium surrounded by villi, which is used for sensing temperature and magnetic fields (Bae and Barr 2008; Inglis et al. 2007; Vidal-Gadea et al. 2015). There are a number of observable phenotypes associated with cilia dysfunction including dye-filling defective (Dyf), osmotic avoidance abnormal (Osm), and chemotaxis defective (Che) (Starich et al. 1995). In terms of genetics, more than forty cilia genes have been identified in *C. elegans* (Swoboda et al. 2000; Blacque et al. 2005; Efimenko et al. 2005; Chen et al. 2006), and all known cilia genes are regulated by a single RFX gene, *daf-19* (Swoboda et al. 2000).

As with other animals, *daf-19* regulates most, if not all, cilia genes. To date more than forty *daf-19* targets have been identified (Table A-1) (Swoboda et al. 2000; Efimenko et al. 2005; Chen et al. 2006; Blacque et al. 2005). Daf-19 functions by binding an X-box sequence very similar to that of other animals (GTHNYY AT RRNAAC) although genes which show more restricted ciliary expression often contain more degenerate X-boxes (Efimenko et al. 2005).

An example of a *daf-19* target gene is *dyf-5*.The DYF-5 protein, encoded by this gene, is a mitogen-activated protein kinase (MAP kinase) that is involved with intraflagellar trafficking. It is expressed exclusively in ciliated neurons and is found in all sixty ciliated neurons in *C. elegans*(Burghoorn et al. 2007, 2012; Chen et al. 2006). The role of this protein is not entirely clear but it appears to play a role in docking and undocking kinesin motors to IFT particles. Normally, kinesin-II is restricted to the middle segment of cilia, however in *dyf-5* loss of function mutants kinesin-II is found throughout the cilium. Additionally, the other kinesin motor, OSM-3, seems to be affected as it is not attached to IFT particles and its speed is reduced in these mutants. The retrograde dynein motor may also be affected as at least six proteins accumulate at the tip of cilia in *dyf-5* mutants (Burghoorn et al. 2007). The phenotype of a null mutation of this gene is dye-filling defective, in which the 6 amphid neurons (ASI, ADL, ASK, AWB, ASH, and ASJ) and two phasmid neurons (PHA and PHB) that are exposed to the environment fail to take up a fluorescent dye. This is suggestive of cilia dysfunction although not conclusive as channel cell defects can block exposure of the cilia without disrupting them (Inglis et al. 2007). It was previously mapped to chromosome I of *C. elegans* (Starich et al. 1995) and the molecular identity was identified by our lab by using a computational approach to identify X-box containing genes (Chen et al. 2006). Two orthologs of note are *lf4* from *Chlamydomonas*, mutants of which cause long flagella (Berman et al. 2003), and Male-associated kinase (*MAK*) in humans (Tucker et al. 2011). Mutations in *MAK* are associated with the ciliopathy retinitis pigmentosa. This gene was identified by looking for targets of the Cone Rod Homeobox, a transcription factor associated with the disease (Ozgül et al. 2011). This gene is the main topic of this thesis.

Another interesting *daf-19* target gene is *peli-1*. *peli-1* belongs to the Pellino family of U3 ubiquitin ligases. It was identified in our lab as having two functional X-box motifs. Both the proximal and distal X-boxes are able to drive expression of a reporter gene in ciliated neurons with proximal motif stronger than the distal motif. Interestingly, when the distal motif is deleted, expression increases suggesting that the proximal motif is primarily responsible for cilia expression while the distal motif modulates expression of this gene (Chu et al. 2012)

Finally, an example of a *daf-19* target gene showing a restricted expression pattern is *nhr-44*. This gene belongs to the nuclear hormone receptor family of transcription factors. Although its role in cilia is currently unknown, its expression is X-box dependent. Whereas, *dyf-5* is expressed in all sixty ciliated neurons including amphids, phasmids and labial neurons, *nhr-44* is expressed solely in the amphid neurons (Burghoorn et al. 2012).

As X-box seems to specify ciliary expression, experiments to insert the X-box into non-cilary promoters were undertaken. However, from these results it appears that the X-box does not function by itself (Chen lab unpublished). Some factors working with the X-box/*daf-19* have been identified in *C. elegans*. The transcription factor *fkh-2* was identified to have a role in specialising cilia in AWB neurons based on its expression pattern; it is expressed post-embryonically almost exclusively in AWB neurons. Because of this, it was speculated that it may play a role in specifying the unique cilia morphology of the AWB neuron, which consists of a branched cilia with each branch showing a different length. Consistent with this, *fkh-2* mutants were found to have AWB specific cilia defects. It was determined that *fkh-2* regulates the kinesin-II subunit gene, *kap-1*, which is important during anterograde IFT. *fkh-2* is *daf-19* dependent but no X-box has been identified in its promoter (Mukhopadhyay et al. 2007). Interestingly, *fkh-2* is a forkhead domain transcription factor like FoxJ1 and *FD3F*; however, *fkh-2* appears to only be a distant relative of this family. The binding site of *fkh-2* was not reported but it has strong homology to yeast *fkh-1 and fkh-2* (www.wormbase.org) which were shown to bind to the sequence GTAAACA (Kato et al. 2004)

Experiments involving swapping X-box sequences among cilia genes suggested the involvement of neighbouring regions, which led to the discovery of a new cilary DNA motif, the C-box. This motif is approximately ten based pairs long consisting almost exclusively of cytosines and thymines. It is found primarily in genes showing pan-ciliary expression and is usually found in multiple copies within a promoter, most often flanking the X-box (Burghoorn et al. 2012). It was shown to be very important in ciliary expression, however, the X-box and C-boxes alone were unable to recapitulate full ciliary expression which suggests the involvement of other elements (Burghoorn et al. 2012).

## 1.4. Identification of regulatory sequences

There are a number of methods for identifying regulatory sequences. Computational methods can be used to identify known motifs. This process involves using sequences of known examples of a particular motif to generate a “seed” which is then used to search the genome. This “seed” is often either a hidden markov model or a position weight matrix (Finn et al. 2011; Liefooghe et al. 2006). For *de novo* motif identification, phylogenetic footprinting is often used. This involves alignment of orthologous promoters looking for regions of high homology. The assumption is that functional DNA sequences will be evolutionarily constrained; therefore they should show more similarity than adjacent sequences. Care must be taken to use species with appropriate evolutionary distance as species that are too close will not show enough divergence to identify motifs whereas species that are too distant may use different regulatory mechanisms (Katara et al. 2012; Nam et al. 2010; Bredrup et al. 2011; Janky and van Helden 2008). A similar method can be used to compare promoters from the same species that show the same expression pattern (Smith et al. 2005). Genes that share an expression pattern are often regulated by the same transcription factors.

Alternatively, molecular methods can be used to identify regulatory sequences. When working from a known transcription factor, chromatin immunoprecipitation (ChIP) (Wang et al. 2012; Weirauch et al. 2014) or protein binding microarrays (Narasimhan et al. 2015) can be used to identify sequences that are bound by the transcription factor. These sequences can then be used as a “seed” for the previously described computational method or in the case of ChIP; the indentified sequence can be mapped directly to the genome. If the transcription factor is unknown, “promoter bashing” can be done to obtain a small promoter region that is still able to regulate the gene of interest. Then methods such as DNase footprinting (Neph et al. 2012) or linker scanning mutagenesis (Etchberger et al. 2007; Nokes et al. 2009) can be used to identify the nucleotides involved. Yeast-1-hybrid and electrophoretic mobility shift (EMSA) assays can be used to show interaction between a DNA sequence and a transcription factor of interest (Walhout 2011; Hellman and Fried 2007).

## 1.5. Research objectives

This thesis aims to identify the cis-regulatory module controlling *dyf-5* expression in *C. elegans*. This will be achieved by addressing three main aims. The first is to use computational tools to identify motifs shared by related species. Second, the smallest promoter region necessary for correct expression will be identified. Finally, this small promoter region will be dissected with molecular tools to uncover the identity of regulatory motifs and the interaction amongst these regulatory motifs.

# Chapter 2. Computational discovery of *dyf-5* regulatory elements

## 2.1. Overview

X-box/RFX regulation of cilia is well conserved among metazoans (Chu et al. 2010). Given that cis-regulatory modules are often well conserved as well (Cameron et al. 2005), it is reasonable to hypothesise that the entire X-box cis-regulatory module is conserved. This hypothesis leads to two testable predictions. First, an orthologous promoter from a related species should drive expression in a similar manner to the native promoter. Second, if the X-box cis-regulatory module is conserved, orthologous promoters should contain islands of sequence conservation that correspond to transcription factor binding sites. These sites can then be identified using computational methods such as phylogenetic footprinting (Katara et al. 2012; Nam et al. 2010; Bredrup et al. 2011; Janky and van Helden 2008).

The approach used was to address both of these predictions. First, promoters orthologous to *dyf-5* were identified from species closely related of *C. elegans* and assayed for expression in *C. elegans*. Next, the sequences of these promoters was analysed computationally to identify conserved motifs.

## 2.2. Materials and methods

### 2.2.1. Determination of orthologous promoters of *dyf-5* in nematodes

Orthologs of *C. elegans dyf-5* were identified using genBlastG (She et al. 2011). The long isoform sequence of DYF-5 was used as query and the WS230 version of each genome used as target. All sequences were obtained from WormBase

(www.wormbase.org). Promoters of these genes were obtained by determining the sequence approximately two thousand base pairs upstream of the start codon or the entire intergenic region if it was smaller than two thousand base pairs. Promoter regions used are displayed in Table 2-1.

**Table 2-1. Orthologous promoter regions.**

| Species | Genome Version | Promoter Coordinates | Length |
|---|---|---|---|
| *C. elegans* | WS230 | I: 9,357,080..9,359,123 | 2043 |
| *C. briggsae* | WS230 | chrI: 2,694,376..2,695,979 | 1603 |
| *C. brenneri* | WS230 | Cbre_Contig15: 1,142,808..1,140,458 | 2350 |
| *C. remanei* | WS230 | Crem_Contig68: 348,121..349,884 | 1763 |
| *C. japonica* | WS230 | Cjap_Contig17533: 117,979..119,947 | 1968 |
| *C. species 5* | WS230 | Csp5_scaffold_01078: 2,665..4,616 | 1951 |
| *C. species 11* | WS230 | Scaffold616: 43,849..46,057 | 2208 |
| *C. angaria* | WS230 | Can_chrRNAPATHr22180: 4,007..2,205 | 1802 |
| *P. pacificus* | WS230 | Ppa_Contig95: 444,085..447,018 | 2933 |

### 2.2.2. Generation of constructs

Promoter regions were PCR amplified using primers in Table 2-2. *C. briggsae dyf-5* promoter was fused to *tdTomtato* by cloning into plasmid vector VH23.05 using SalI and BamHI. *C. elegans dyf-5* promoter was fused to *tdTomato* by PCR fusion (Hobert 2002) using nested primers and *tdTomato* amplified from plasmid VH23.05. This was performed by Zhaozhao Qin. All other promoters were fused to GFP by PCR fusion (Hobert 2002) using the nested primers listed in Table 2-2. GFP was amplified from plasmid PD95.75.

For the H-box deletion construct, the 5' and 3' halves of the promoter were amplified using the forward and reverse primers in Table 2-2. The primers are designed to introduce a four base pair deletion into the H-box. The two promoter halves were then joined by PCR fusion using the nested primers. Finally, the H-box deletion promoter was fused to GFP using the *C. elegans pdyf-5* and GFP nested primers.

For the H-box deletion Mos single copy insertion construct, the previously described H-box deletion construct was amplified with the H-box MosSCI primers.

These primers introduce restriction endonuclease cut sites to the construct. This construct was then cloned into plasmid CFJ151 with AflII and SbfI restriction enzymes.

**Table 2-2. List of PCR Primers used in Chapter 2.**

| Primer Set | Forward Primer | Nested Primer | Reverse Primer |
|---|---|---|---|
| *C. elegans pdyf-5 tdTomato* | ttcgaaaagtcttgaagttggtc | gcctgcaaatttgtcatacatac | tgacctcctcgcccttgctcaccatggcttcttgcccttatattttc |
| *tdTomato* | atggtgagcaagggcgag | ttacttgtacagctcgtccatg | tgacagcggcccctattatt |
| *C. elegans pdyf-5 GFP* | gcctgcaaatttgtcatacat | tttcaattcgaaaaacagcttc | tgaaaagttcttctcctttactcatggcttcttgcccttatattttc |
| *C. briggsae pdyf-5* | ccatttatttattggctgtcca | N/A | ttccaggatcctgttcaatgtagtttatgtagtctttgtag |
| *C. Brenneri pdyf-5* | tggtcagttggcttacaagaa | tcaaccaagacagccccttta | tgaaaagttcttctcctttactcatcaggtatctgaaaattgtagagtgg |
| *C. remanei pdyf-5* | gaaatgtcttgatgaaacttcacg | tgtaatgcggaagtgaaacaa | tgaaaagttcttctcctttactcatcatctcttctttttcgaatttcttg |
| *C. japonica pdyf-5* | tcccgtgaataaccccataa | ctcctgctctctttcggttg | tgaaaagttcttctcctttactcatctgcaagtacgaggcgtgag |
| *C. species 5 pdyf-5* | ttgctgcctagggtaagctc | ggcgagtttcagatggaaag | tgaaaagttcttctcctttactcatttcggtcgttcactttttcg |
| *C. species 11 pdyf-5* | tcgttgactctaggttactgtatcttg | cgaaagactcggaaatgagc | tgaaaagttcttctcctttactcattttttgtaggctaataaccagtatga |
| *C. angaria pdyf-5* | ccaaattcatccccacaatc | acatcaatttcgcgtcaaga | tgaaaagttcttctcctttactcattttcagtcaaattttattttcacgtag |
| *P. pacificus pdyf-5* | ttgttgctaagcgcggaaat | agtcaggagtgttcgccagt | tgaaaagttcttctcctttactcatcgatcgatcaaggctcctac |
| *GFP* | atgagtaaaggagaagaacttttcactgg | ggaaacagttatgtttggtatattggg | aagggcccgtacggccgactagtagg |
| *ΔH-box 5’* | gcctgcaaatttgtcatacat | tttcaattcgaaaaacagcttc | catgctatgcactttcggtagatagagaaactaag |
| *ΔH-box 3’* | cttagtttctctatctaccgaaagtgcatagcatg | tgaaaagttcttctcctttactcatggcttcttgcccttatattttc | tgaaaagttcttctcctttactcatggcttcttgcccttatattttc |
| *ΔH-box MosSCI* | ttccacttaagtttcaattcgaaaaacagcttc | N/A | ggataacctgcaggccagacgtgcg |

### 2.2.3. Generation of strains

Constructs were micro-injected using a method similar to Mello *et al.* (Mello et al. 1991). A DNA mixture containing 50ng/µl of construct DNA and 100ng/µl of CEH361 plasmid was injected into *dpy-5(e907)* worms. The plasmid contains a wild-type copy of *dpy-5* which can rescue the *Dpy* phenotype and therefore functions as a selectable marker. F1 worms displaying wild-type phenotype were selected after four days at 20ºC. After another four days wild-type F2 worms were isolated and observed.

The MosSCI strains were generated using a method similar to Frøkjær-Jensen *et al.* (Frøkjær-Jensen et al. 2012; Robert et al. 2009). Briefly, previously generated H-box deletion construct was inserted into plasmid CFJ151. A DNA mixture containing 50ng/µl of CFJ151 with insert, 50ng/µl JL43.1, 5ng/µl GH8, 5ng/µl CFJ104, and 2.5 ng/µl CFJ90 was injected into JNC1021 worms. F1 worms displaying wild-type phenotype were selected after four days at 20ºC. After another four days wild-type F2 worms were isolated and observed for *mCherry* expression. Worms with wild-type phenotype lacking *mCherry* expression were then grown for 3 generations to ensure homozygousity and observed by confocal microscopy.

Strains generated and used are listed in Table 2-3.

**Table 2-3. List of strains used in Chapter 2.**

| Strain | Sex | Source | Genotype | Notes |
|---|---|---|---|---|
| CB907 | Hermaphrodites | CGC | *dpy-5(e907)I;* | |
| JNC212 | Hermaphrodites | Injection Extrachromasomal | *dpy-5(e907)I; dotEx [Pr cbr-dyf-5::tdTomato + dpy-5(+)]* | *Caenorhabditis briggsae dyf-5* promoter (1731bp upstream of start codon) fused to tdTomato |
| JNC145 | Male/Female | CGC | *Caenorhabditis brenneri* | Same as CGC strain PB2801 |
| JNC146 | Hermaphrodites | Injection Extrachromasomal | *dpy-5(e907)I; dotEx [Pr CBN-mks-1::GFP + dpy-5(+)]* | *Caenorhabditis brenneri mks-1* promoter (1335bp upstream of start codon) fused to GFP |
| JNC147 | Male/Female | CGC | *Caenorhabditis remanei* | Same as CGC strain PB4641 |
| JNC148 | Hermaphrodites | CGC | *Pristionchus pacificus* | Same as CGC strain PS312 |
| JNC149 | Hermaphrodites | Injection Extrachromasomal | *dpy-5(e907)I; dotEx [Pr cbr-dyf-5::tdTomato + dpy-5(+)]* | *Caenorhabditis briggsae dyf-5* promoter (1731bp upstream of start codon) fused to tdTomato |
| JNC150 | Hermaphrodites | Injection Extrachromasomal | *dpy-5(e907)I; dotEx [Pr cbr-dyf-5::tdTomato + dpy-5(+)]* | *Caenorhabditis briggsae dyf-5* promoter (1731bp upstream of start codon) fused to tdTomato |
| JNC151 | Hermaphrodites | Injection Extrachromasomal | *dpy-5(e907)I; dotEx [Pr cbr-dyf-5::tdTomato + dpy-5(+)]* | *Caenorhabditis briggsae dyf-5* promoter (1731bp upstream of start codon) fused to tdTomato |
| JNC152 | Hermaphrodites | Injection Extrachromasomal | *dpy-5(e907)I; dotEx [Pr CBN-mks-1::GFP + dpy-5(+)]* | *Caenorhabditis brenneri mks-1* promoter (1335bp upstream of start codon) fused to GFP |
| JNC153 | Hermaphrodites | CGC | *Caenorhabditis species 11* | Same as CGC JU1373 |
| JNC154 | Male/Female | CGC | *Caenorhabditis angaria* | Same as CGC PS1010 |
| JNC500 | Hermaphrodites | Injection Extrachromasomal | *dpy-5(e907)I; dotEX [Pr Cre-mks-1::GFP + dpy-5(+)]* | *Caenorhabditis remanei mks-1* promoter (2452bp upstream of start codon) fused to GFP |
| JNC501 | Hermaphrodites | Injection Extrachromasomal | *dpy-5(e907)I; dotEX [Pr Cre-mks-1::GFP + dpy-5(+)]* | *Caenorhabditis remanei mks-1* promoter (2452bp upstream of start codon) fused to GFP |
| JNC502 | Hermaphrodites | Injection Extrachromasomal | *dpy-5(e907)I; dotEX [Pr Cre-mks-1::GFP + dpy-5(+)]* | *Caenorhabditis remanei mks-1* promoter (2452bp upstream of start codon) fused to GFP |
| JNC503 | Hermaphrodites | Injection | *dpy-5(e907)I; dotEX [Pr Cre-dyf-* | *Caenorhabditis remanei dyf-5* promoter (1785bp upstream of start |
| | | Extrachromasomal | *5::GFP + dpy-5(+)]* | codon) fused to GFP |
| JNC504 | Hermaphrodites | Injection Extrachromasomal | *dpy-5(e907)I; dotEX [Pr Cre-dyf-5::GFP + dpy-5(+)]* | *Caenorhabditis remanei dyf-5* promoter (1785bp upstream of start codon) fused to GFP |
| JNC505 | Hermaphrodites | Injection Extrachromasomal | *dpy-5(e907)I; dotEX [Pr Cre-dyf-5::GFP + dpy-5(+)]* | *Caenorhabditis remanei dyf-5* promoter (1785bp upstream of start codon) fused to GFP |
| JNC506 | Hermaphrodites | Injection Extrachromasomal | *dpy-5(e907)I; dotEX [Pr Cre-dyf-5::GFP + dpy-5(+)]* | *Caenorhabditis remanei dyf-5* promoter (1785bp upstream of start codon) fused to GFP |
| JNC507 | Hermaphrodites | Injection Extrachromasomal | *dpy-5(e907)I; dotEX [Pr Can-dyf-5::GFP + dpy-5(+)]* | *Caenorhabditis angaria dyf-5* promoter (1907bp upstream of start codon) fused to GFP |
| JNC508 | Hermaphrodites | Injection Extrachromasomal | *dpy-5(e907)I; dotEX [Pr Can-dyf-5::GFP + dpy-5(+)]* | *Caenorhabditis angaria dyf-5* promoter (1907bp upstream of start codon) fused to GFP |
| JNC509 | Male/Female | CGC | *Caenorhabditis japonica* | Same as CGC DF5081 |
| JNC510 | Hermaphrodites | Injection Extrachromasomal | *dpy-5(e907)I; dotEX [Pr Csp11-dyf-5::GFP + dpy-5(+)]* | *Caenorhabditis species 11 dyf-5* promoter (2209bp upstream of start codon) fused to GFP |
| JNC511 | Hermaphrodites | Injection Extrachromasomal | *dpy-5(e907)I; dotEX [Pr Csp11-dyf-5::GFP + dpy-5(+)]* | *Caenorhabditis species 11 dyf-5* promoter (2209bp upstream of start codon) fused to GFP |
| JNC512 | Hermaphrodites | Injection Extrachromasomal | *dpy-5(e907)I; dotEX [Pr Csp11-dyf-5::GFP + dpy-5(+)]* | Caenorhabditis species 11 *dyf-5* promoter (2209bp upstream of start codon) fused to GFP |
| JNC513 | Hermaphrodites | Injection Extrachromasomal | *dpy-5(e907)I; dotEX [Pr Csp11-dyf-5::GFP + dpy-5(+)]* | *Caenorhabditis species 11 dyf-5* promoter (2209bp upstream of start codon) fused to GFP |
| JNC514 | Hermaphrodites | Injection Extrachromasomal | *dpy-5(e907)I; dotEX [Pr Csp5-dyf-5::GFP + dpy-5(+)]* | *Caenorhabditis species 5 dyf-5* promoter (1953bp upstream of start codon) fused to GFP |
| JNC515 | Male/Female | CGC | *Caenorhabditis species 5* | Same as CGC JU1201 |
| JNC523 | Hermaphrodites | Injection Extrachromasomal | *dpy-5(e907)I; dotEX [Pr dyf-5::GFP + dpy-5(+)]* | *dyf-5* promoter (1929bp upstream of start codon) fused to GFP (injected at 50ng/ul) |
| JNC524 | Hermaphrodites | Injection Extrachromasomal | *dpy-5(e907)I; dotEX [Pr dyf-5::GFP + dpy-5(+)]* | *dyf-5* promoter (1929bp upstream of start codon) fused to GFP (injected at 50ng/ul) |
| JNC525 | Hermaphrodites | Injection | *dpy-5(e907)I; dotEX [Pr dyf-* | *dyf-5*ΔH promoter (1929bp upstream of start codon with 4bp deletion |
| | | Extrachromasomal | *5ΔH::GFP + dpy-5(+)]* | 392bp upstream of start codon) fused to GFP (Injected at 100ng/ul) |
| JNC526 | Hermaphrodites | Injection Extrachromasomal | *dpy-5(e907)I; dotEX [Pr dyf-5ΔH::GFP + dpy-5(+)]* | *dyf-5*ΔH promoter (1929bp upstream of start codon with 4bp deletion 392bp upstream of start codon) fused to GFP (Injected at 100ng/ul) |
| JNC527 | Hermaphrodites | Injection Extrachromasomal | *dpy-5(e907)I; dotEX [Pr dyf-5ΔH::GFP + dpy-5(+)]* | *dyf-5*ΔH promoter (1929bp upstream of start codon with 4bp deletion 392bp upstream of start codon) fused to GFP (injected at 50ng/ul) |
| JNC1021 | Hermaphrodites | CGC | *ttTi5605 II; unc-119(ed3) III; oxEx1578* | Same as EG6699 |

### 2.2.4. Visualization of strains

Worms were fixed on glass slides with 2% agarose pads and M9 solution containing 2%(w/v) sodium azide. M9 solution contains 0.3%(w/v) $KH_2PO_4$, 0.6%(w/v)$Na_2HPO_4$, 0.5%(w/v) NaCl, and 0.01%(w/v) $MgSO_4$. Slides were then observed on Zeiss spinning disc confocal microscope with 40x oil immersion lens.

### 2.2.5. Detection of known motifs

Known motifs (X-box and C-box) were identified in a similar manner to Chen *et al.* (Chen et al. 2006). This involves using the HMMER software package (ver. 1.8.5) (Finn et al. 2011). Briefly, this involves aligning sequences of known motifs with ClustalW (Higgins et al. 1996), submitting this alignment to hmmb to create a hidden Markov model (HMM), and finally submitting this hmm to hmmfs to identify motifs. For X-box, the known motifs were the thirty-one used by Chu *et al.* (Chu et al. 2012) plus the two reported in that paper. For C-box, the validated C-boxes reported by Burghoorn *et al.* were used (Burghoorn et al. 2012). The motifs identified in each promoter were visualised with a perl script written by Christian Frech.

### 2.2.6. *de novo* detection of regulatory motifs

For *de novo* detection of regulatory motifs the software XXmotif was used (Hartmann et al. 2013). This software consists of a webserver to uploading files (http://xxmotif.genzentrum.lmu.de/). The input set is the collection of *dyf-5* promoters that were shown to be functional in *C. elegans*. The similarity threshold for merging motifs was set to high, default settings were used for everything else.

## 2.3. Results

### 2.3.1. Determination of orthologous promoters of *dyf-5* in nematodes

In order to characterise promoters orthologous to the *dyf-5* promoter, *dyf-5* orthologs from other species must first be identified. To accomplish this the software genBlastG was used (She et al. 2011). This program is a homology-based gene finder. The protein sequence for the long isoform of DYF-5 protein was obtained from WormBase (www.wormbase.org) and used to query the eight nematode genomes. Using this approach, *dyf-5* orthologs for all eight species were identified. For *C. briggsae, C. species 5 (C. sinica), C. remanei, C. species 11 (C. tropicalis), C. angaria,* and *P. pacificus*, only a single ortholog was identified. For the remaining two species *C. brenneri* and *C. japonica,* there were two potential orthologs identified. The two orthologs are likely due to heterozygosity of this region in the sequenced strain, as both species have male and female sexes (Barrière et al. 2009). Due to ambiguities in the genomic sequences of these species they were excluded from further analysis. For the remaining species, PCR primers were designed to amplify either an approximately two kilobase promoter region upstream of the start codon or the entire intergenic region if it was smaller than two kilobases. This has previously been reported and an appropriate promoter region (Okkema and Krause 2005). PCR bands of the correct size were isolated for all species.

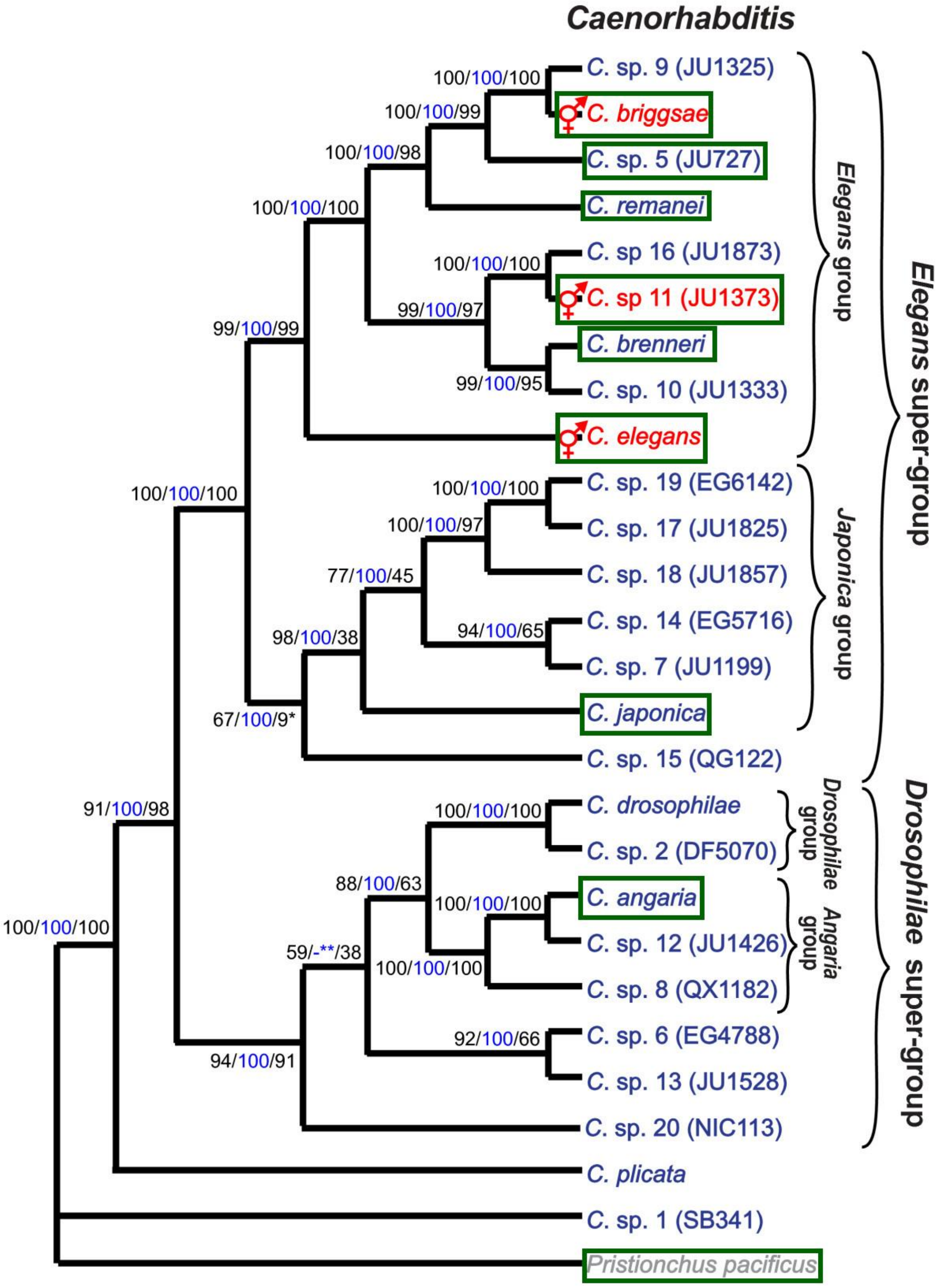
Caenorhabditis
C. sp. 9 (JU1325)
C. briggsae
C. sp. 5 (JU727)
C. remanei
C. sp 16 (JU1873)
C. sp 11 (JU1373)
C. brenneri
C. sp. 10 (JU1333)
C. elegans
C. sp. 19 (EG6142)
C. sp. 17 (JU1825)
C. sp. 18 (JU1857)
C. sp. 14 (EG5716)
C. sp. 7 (JU1199)
C. japonica
C. sp. 15 (QG122)
C. drosophilae
C. sp. 2 (DF5070)
C. angaria
C. sp. 12 (JU1426)
C. sp. 8 (QX1182)
C. sp. 6 (EG4788)
C. sp. 13 (JU1528)
C. sp. 20 (NIC113)
C. plicata
C. sp. 1 (SB341)
Pristionchus pacificus
Elegans group
Japonica group
Drosophilae group
Angaria group
Elegans super-group
Drosophilae super-group
100/100/100
100/100/99
100/100/98
100/100/100
100/100/100
99/100/97
99/100/95
99/100/99
100/100/100
100/100/100
100/100/97
77/100/45
94/100/65
98/100/38
67/100/9*
91/100/98
100/100/100
88/100/63
100/100/100
59/-**/38
100/100/100
92/100/66
94/100/91
100/100/100

**Figure 2-1. Phylogeny of *Caenorhabditis* species**

Phylogeny showing evolutionary relationships of *Caenorhabditis* species. Species in red are hermaphroditic. Green boxes were added to denote the species used in this study. Image adapted from: (Kiontke et al. 2011). Used under creative commons license (http://creativecommons.org/licenses/by/2.0).

## 2.3.2. Search for known motifs in *dyf-5* ortholog promoters

To determine if previously identified motifs were present in these promoters, analysis by HMMER (version 1.8.5) was used (Finn et al. 2011). This software takes alignments of known motifs to build a hidden Markov model which is then used to find motifs that match the model in the given sequences. It was expected that this approach would identify X-boxes and C-boxes in all orthologs. For X-boxes, thirty-three previously identified X-boxes were used as the seed (Chu et al. 2012). For C-boxes, the validated C-boxes identified by Burghoorn *et al.* were used (Burghoorn et al. 2012).

As predicted, an X-box was found in each sequence as well as at least one C-box (fig. 2-2). Given that not all previously predicted C-boxes were found even for *C. elegans*, this suggests that there may be additional C-boxes present.

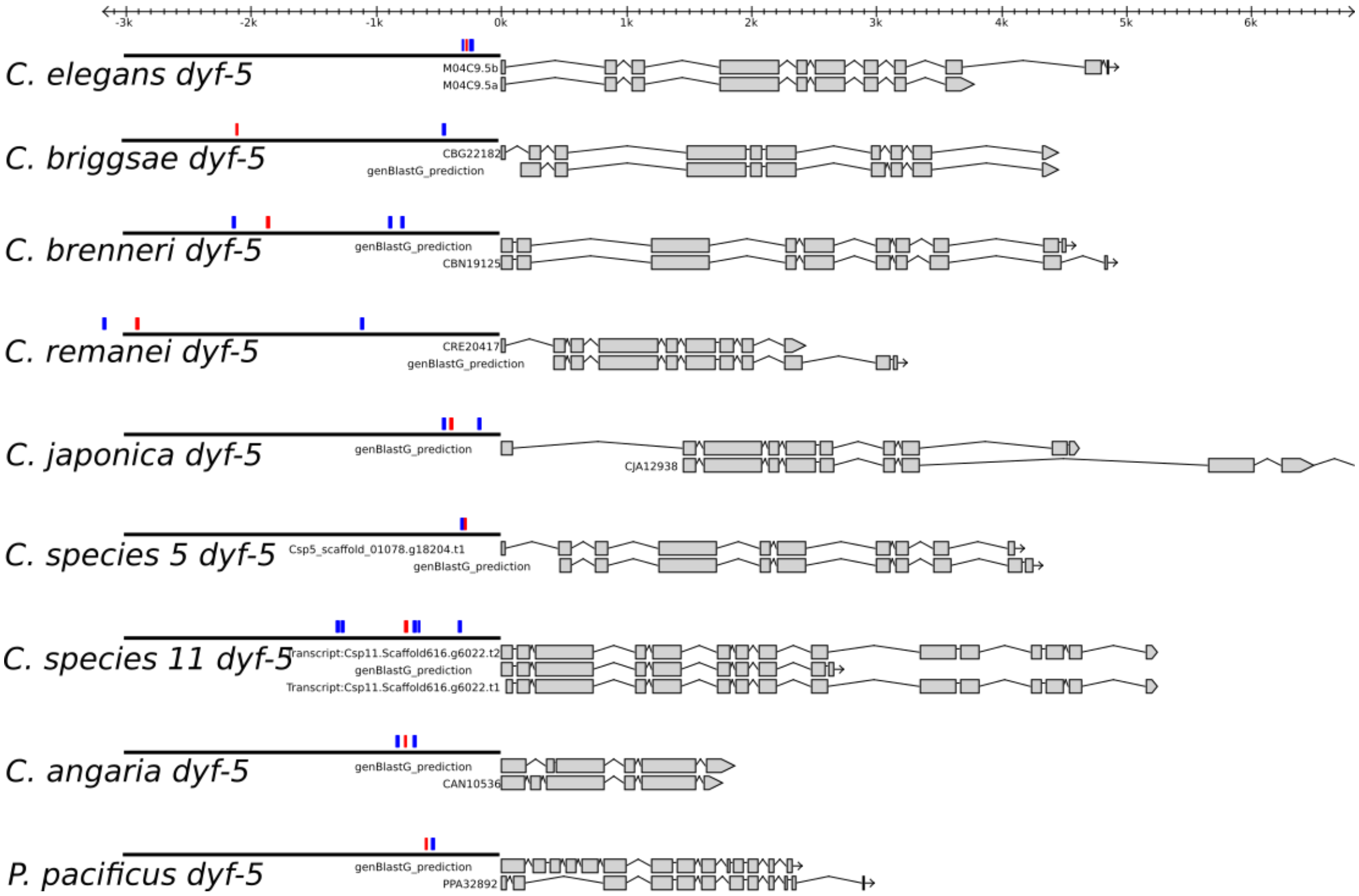

**Figure 2-2. Prediction of known motifs in orthologous promoters.**
Motifs predicted in each promoter using HMMER. X-boxes are depicted in red and predicted C-boxes are blue. The annotated gene model is shown next to the gene model predicted by GenBlastG. An X-box was identified in each promoter along with at least one C-box.

### 2.3.3. Conservation of expression in *Caenorhabditis*

Next, we wished to determine if these promoters were able to correctly drive expression when introduced into *C. elegans*. If the orthologous promoters are able to drive expression similar to native *dyf-5* promoter it strongly suggests that the promoters contain conserved functional elements. This may not always be the case as similar patterns of expression may be driven by unrelated factors (Zhao et al. 2005) or co-evolution between transcription factors and their binding sites can preserve function while changing the factors involved (Barrière et al. 2012). However, given that at least one component of the cis-regulatory module appears to be conserved, the X-box, it is reasonable to assume other components may be conserved as well.

To test the expression driven by the various promoters, each of the previously isolated promoters were fused to a fluorescent reported gene. The *C. briggsae dyf-5* promoter was fused to *tdTomato* by cloning it into a plasmid VH23.05. The rest of the promoters were fused to GFP by PCR fusion (Hobert 2002). These constructs were then injected into *C. elegans* to create extra-chromosomal array strains expressing the reporter gene. These strains were then photographed by confocal microscopy (Figure 2-3). With the exception of *C. species 11* and *P. pacificus*, all strains showed expression of similar pattern and intensity to that driven by the *C. elegans dyf-5* promoter, which showed expression exclusively in the sixty ciliated neurons as previously reported (Burghoorn et al. 2007, 2012; Chen et al. 2006). *C. species 11* showed the same pattern of expression but was substantially weaker than the others. *P. pacificus* showed no detectable expression. From these results, it was concluded that all the promoters could contain similar elements with the exception of *C. species 11,* which may be missing an element, and *P. pacificus*, which appears to be too divergent.

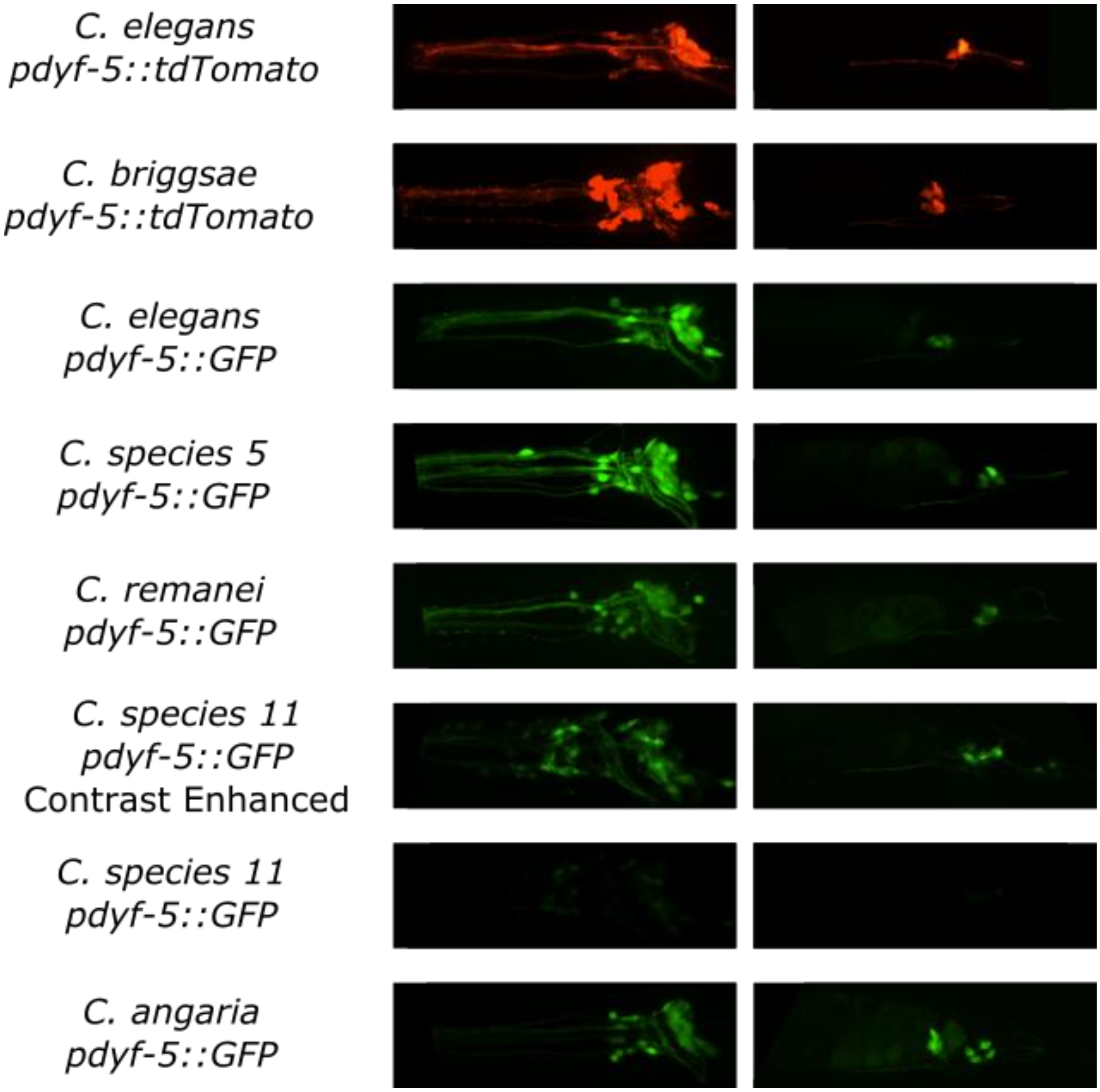

**Figure 2-3. Expression patterns of orthologous *dyf-5* promoters in *C. elegans*.** The promoter of the *dyf-5* ortholog from each species was fused to GFP by PCR fusion and injected into *C. elegans*. The left-hand column shows the head of the worm containing the labial and amphid neurons. The right-hand column shows the tail of the worm containg the phasmid neurons. *C.* briggsae promoter was fused to tdTomato via cloning. All species appear to show expression in all sixty ciliated neurons. All intensities were similar by visual comparison with the exception of *C. species 11* which is show with and without contrast enhancement. *P. pacificus* showed no expression (not shown).

## 2.3.4. de novo discovery of motifs using XXmotif

To uncover the functional motifs responsible for the expression of the identified promoters, the software XXmotif was used (Hartmann et al. 2013). This software uses a position-weight matrix based algorithm that identifies motifs that are over-represented in the dataset. Because of this, it is able to identify motifs shared among the submitted sequences as well as motifs repeated multiple times within a single sequence. Therefore, motifs were selected that had a low E value in addition to being common to the majority of promoters, thus reducing the number of repetitive motifs.

Using this approach, the promoters previously shown to be functional were submitted. The software was able to identify three high quality motifs that were shared by all six promoters (Figure 2-4). These included the known X-box, motifs consistent with C-boxes, and a putative new element termed the H-box for its similarity to homeodomain transcription factor binding sites (Figure 2-5). Interestingly, this software was only able to detect one of the two C-boxes previously predicted by Burghoorn *et al.* (Burghoorn et al. 2012). Transplice sites were also detected in these promoters (data not shown).

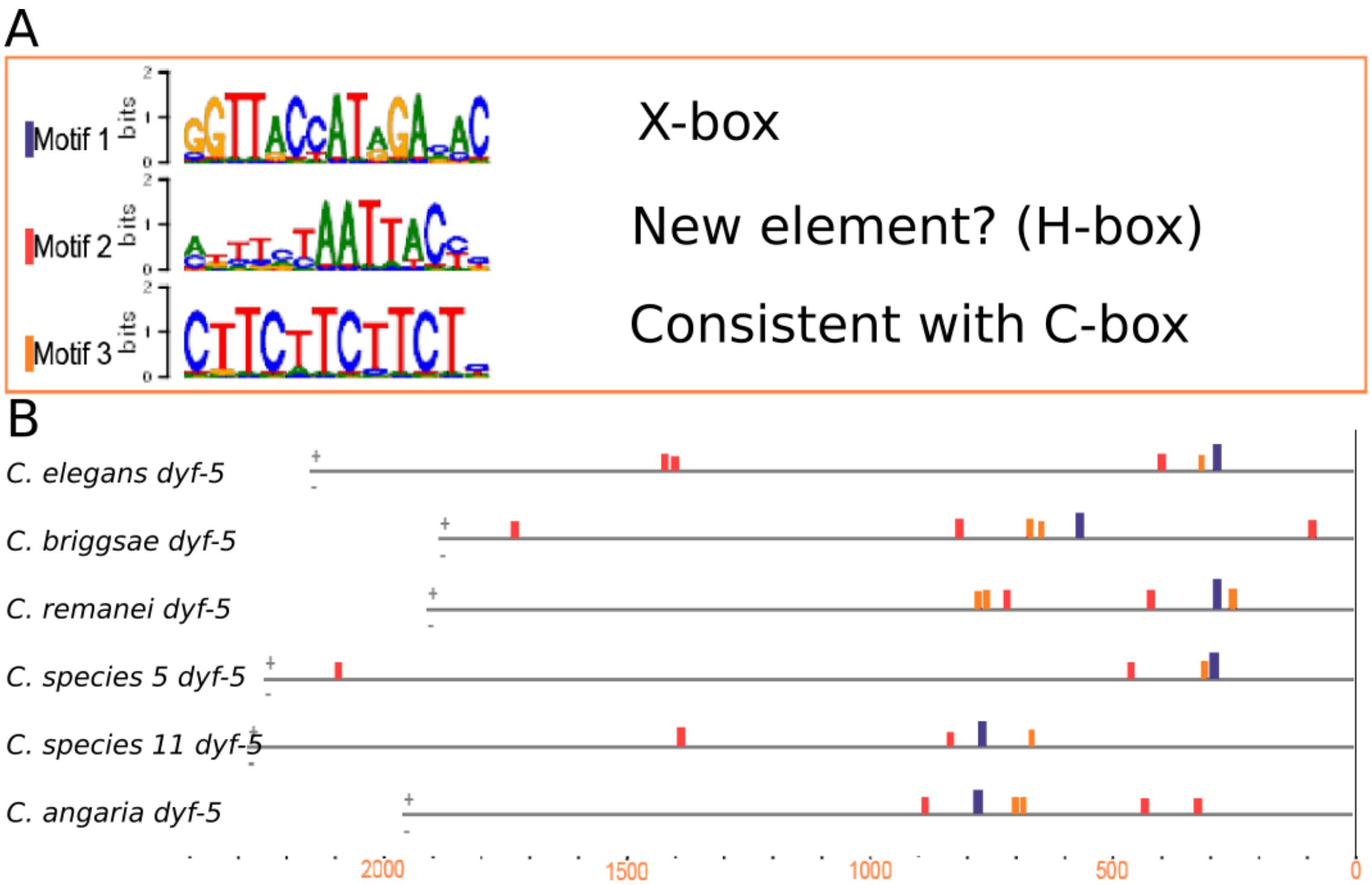

**Figure 2-4. *De novo* detection of regulatory motifs using XXmotif.**
The orthologous *dyf-5* promoters previously shown to be functional were submitted to XXmotif. (A) The best three hits are shown. Motif 1 corresponds to the X-box, Motif 3 is consistent with a C-box motif. Motif 2 represents a potential new motif. (B) Distribution of identified motifs in the submitted sequences. All sequences count backwards from the predicted translation start site indicated by the vertical line on the right hand side.

A) Motif Sequences

| Sequence | Strand | Start | Site P-value | -20bp | Motif | +20bp |
|---|---|---|---|---|---|---|
| *C. elegans dyf-5* | + | 187 | 6.6697e-06 | CGAGGCTCTTAGTTTCTCTA | TCTAATTACC | GAAAGTGCATAGCATGCTCA |
| *C. briggsae dyf-5* | + | 55 | 6.6697e-06 | TATGCTTTTTCTGAACCCTC | TCTAATTACC | TGCTCTCCTTCGAAAAGTGC |
| *C. remanei dyf-5* | + | 166 | 1.2696e-05 | ATACTTCTGACCACCTTCTT | CCTAATTACC | TCTTCATTCCCTAGTTTCAA |
| *C. remanei dyf-5* | + | 203 | 4.5226e-05 | TCCCTAGTTTCAAAGTCAAA | TCTACTTACC | ACTACATTCTCTTTCTCTCT |
| *C. species 5 dyf-5* | + | 131 | 1.2696e-05 | CTCAAGCACTGCTCCATCTT | CCTAATTACC | ATTCAAAGTGCATCTACCTT |
| *C. species 11 dyf-5* | + | 172 | 5.8613e-05 | CATTTCATATGATCTATTTG | GGTAACTACC | ACTTGGGGAACGTTATAATT |
| *C. species 11 dyf-5* | + | 236 | 2.9315e-05 | TTTCCAGTGCCCTTTCCAAA | TGCAATTACC | CTCGTGTTTCGCCGGAAAGG |
| *C. angaria dyf-5* | + | 193 | 3.7042e-05 | CACCGATTTTGGTTCTCAAT | TCTAATTACT | TTTATTTTTGCATTCGCTTT |

B) Top hit from Jaspar and TRANSFAC

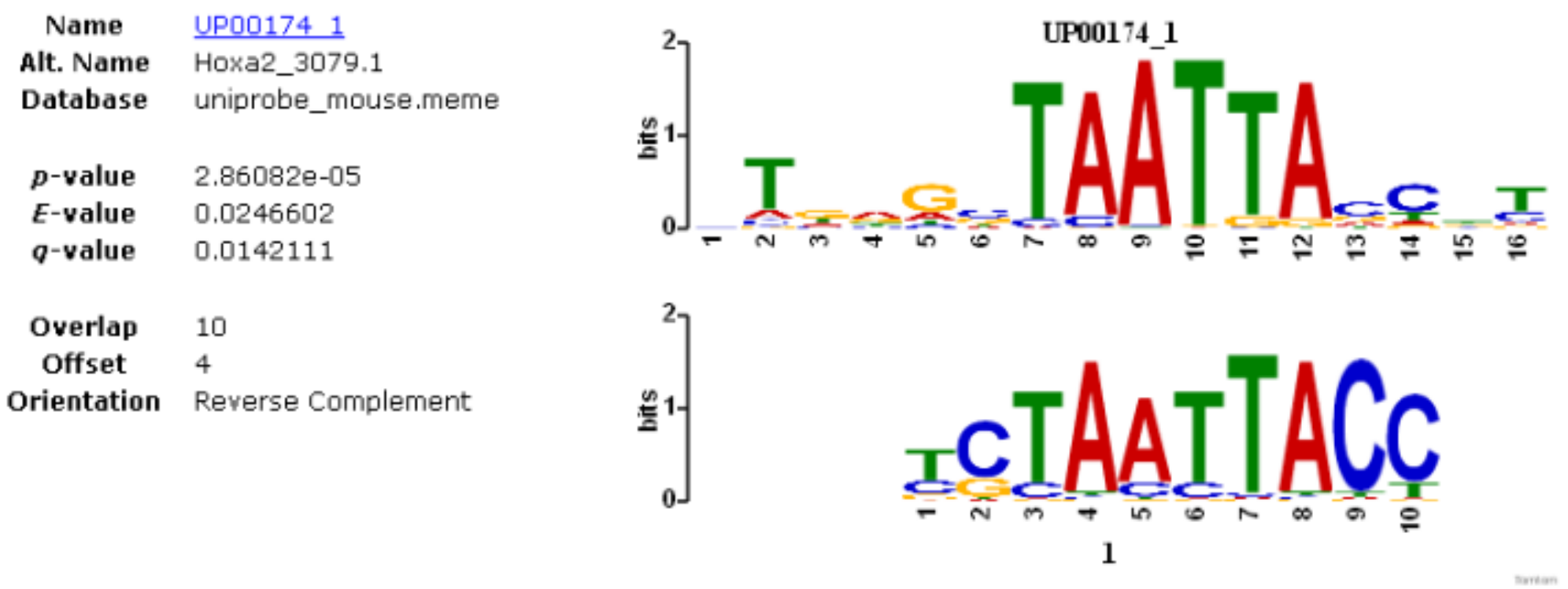

**Figure 2-5. Further analysis of new motif.**
XXmotif analysis was performed on the orthologous promoter sequences. Sequences were shortened to just 300bp upstream and 100bp downstream of X-box to reduce search space. The new motif was again detected. (A) Sequences of detected motif from each submitted promoter. (B) Search for similar motifs in JASPAR and TRANSFAC revealed strong similarity to homodomain binding sites.

### 2.3.5. H-box deletion

To determine if the H-box motif is functional, constructs were created with the H-box deleted from the *dyf-5* promoter fused to GFP. If the H-box is functional, deletion of the H-box should result in a different pattern or intensity of expression when compared to the wild type promoter.

Results showed a slight reduction of expression when injected as an extra-chromosomal array (Figure 2-6). The reduction could have been the result of copy number variations in the array. To test for this a single copy insertion strain was generated. This experiment showed no significant difference between the wild type and H-box deleted strains (Figure 2-7). From this result it can be concluded that the H-box is non-functional. Highly conserved motifs without function have been reported previously (Sleumer et al. 2009).

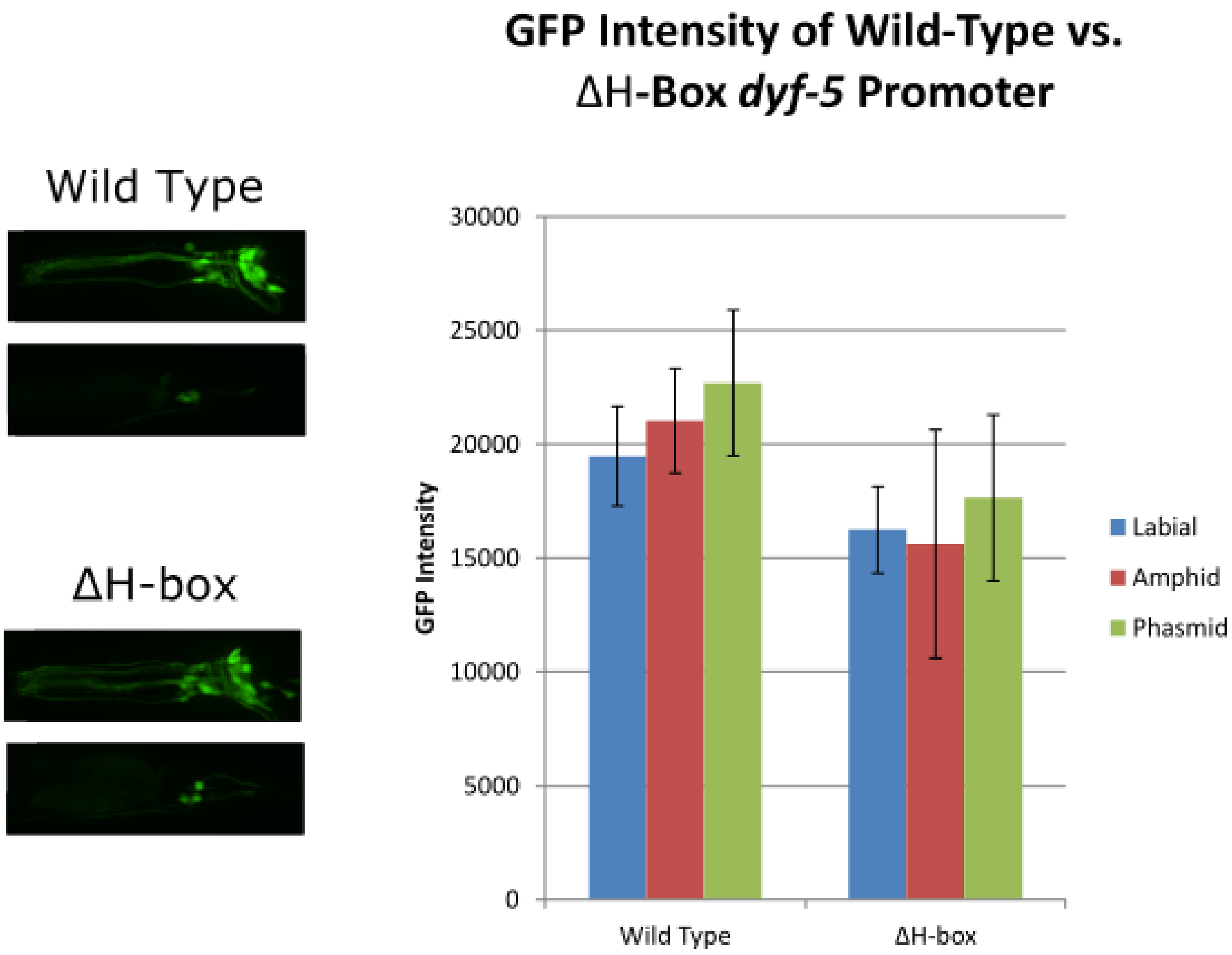

**Figure 2-6. Effect of H-box deletion on expression of extra-chromosomal array strains.**

*C. elegans* strain with either the full length wild-type *dyf-5* promoter fused to GFP or the *dyf-5* promoter with H-box deleted fused to GFP were created. (Left) Confocal images of expression. Pattern and intensity of expression is very similar (Right) Intensity of GFP determined by analysing confocal images (N=10)

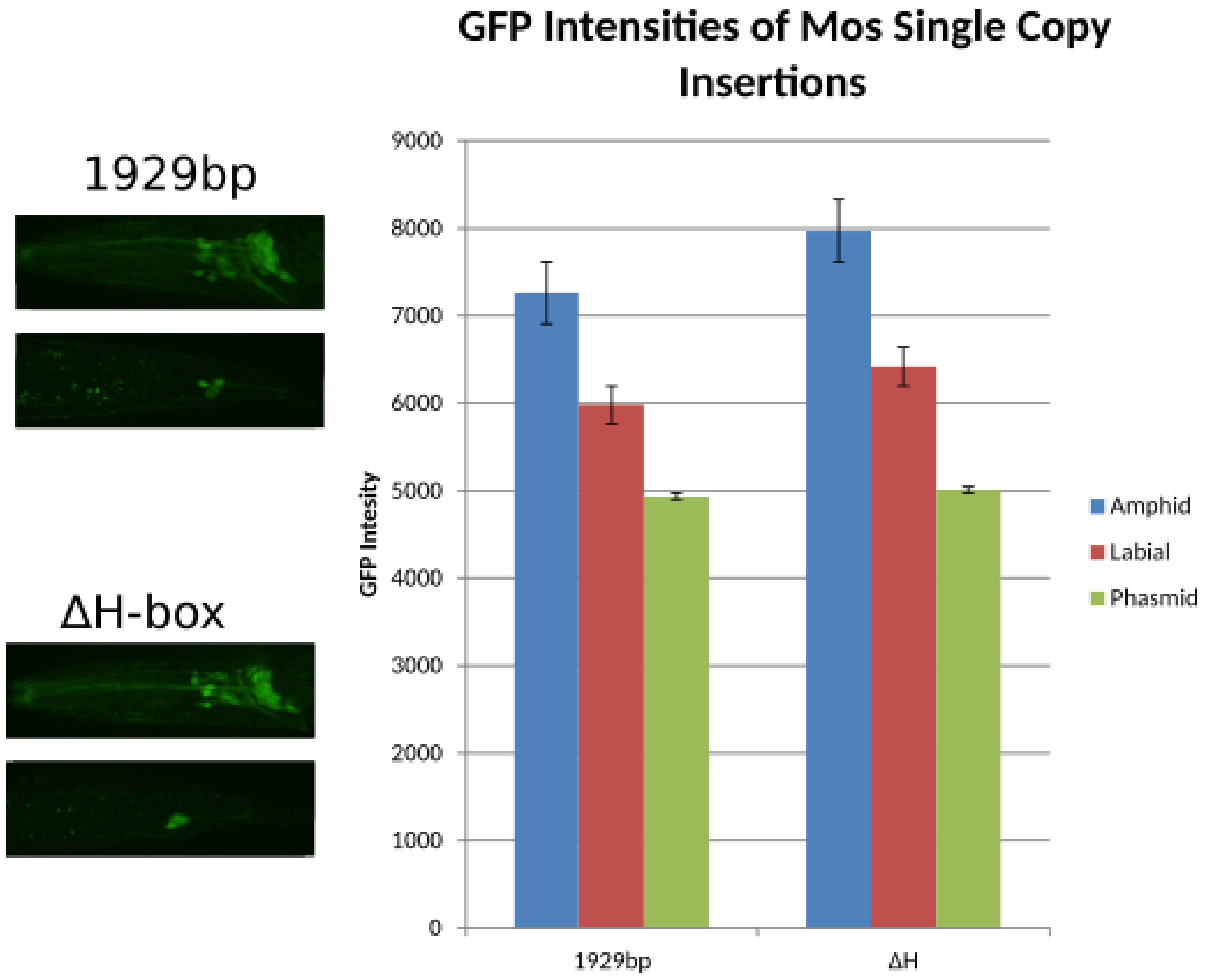

**Figure 2-7. Mos Single-copy insertions of H-box deletion.**
Single copy insertion strains of both full length wild-type (1929bp) *dyf-5* promoter fused to GFP and *dyf-5* promoter with H-box deleted fused to GFP were created. (Left) Confocal images showing expression. Expression pattern and intensity is very similar. (Right) GFP intensity of strains determined by analysing confocal images. No significant change was observed (N=9).

## 2.4. Discussion

The results of the reporter gene assay clearly show conservation of function of the *dyf-5* orthologous promoters. This suggests that expression of *dyf-5* is well conserved within the *Caenorhabditis* genus. In retrospect, transgenic worms containing the *C. elegans dyf-5* promoter fused to one colour of reporter gene (e.g. *tdTomato*) and the *dyf-5* promoter from another species fused to another colour of reporter gene (e.g. *GFP*) may have yielded more information. A direct comparison of the expression patterns could then have been made to ensure that the promoter does in fact drive expression in all sixty cells. However, by visual comparison the patterns do seem to be well conserved. The exact composition of motifs may not be conserved however as evidenced by the reduced expression of the *C. species 11* promoter GFP fusion. However, it cannot be ruled out that only the particular promoter region used is lacking the motif and that a larger region may be able to drive expression at the same level as the other promoters. Additionally, it is unclear if this CRM is conserved outside of the *Caenorhabditis* genus. The lack of expression observed from the *P. pacificus* promoter GFP fusion suggests that this is true but it cannot be ruled out that the promoter region selected simply is not the true promoter for this gene. Shared expression is not definitive proof that the regulatory elements are conserved, however, as the same pattern of expression could be driven by different factors (Zhao et al. 2005). For example, co-evolution between transcription factors and their binding sites can preserve expression pattern while changing factors (Barrière et al. 2012).

Computational comparison of these promoters was able to identify several conserved motifs. This confirms that these promoters are expressed using mostly the same factors. Detection of the X-box and some C-boxes in each sequence served as an internal control indicating that this method does indeed work. Since the method only analyses conservation, the H-box cannot be considered a true false positive. The sequence is indeed conserved; however, the reason for its conservation is not obvious. It is possible that this sequence serves some function other than as a binding site for transcription factors. Similar cases of cryptic conservation have been reported where the most conserved motif is not functional (Sleumer et al. 2009).

It is interesting to note that the software did not detect all C-boxes known to be present in the *C. elegans* promoter. XXmotif was able to identify only one C-box in *C. elegans* whereas two were reported by Burghoorn *et al.* when comparing genes with pan-ciliary expression (Burghoorn et al. 2012). In fact, a third functional C-box is present (Chapter 4). The reason for these false negatives is unclear. Potentially, the sequence of these C-boxes may be too divergent for XXmotif to detect. Alternatively, the threshold for calling the motif may simply be too high resulting in these motifs being filtered out of the final dataset. Either way, it raises the possibility that motifs exist that were not detected by this software. This possibility will be addressed in the next chapter.

# Chapter 3. Identification of *dyf-5* minimal promoter

## 3.1. Overview

Computational techniques did not identify any new regulatory elements. This suggests one of two things: either there are no additional elements or there are other regulatory elements but they are poorly conserved at sequence level therefore computational comparison was unable to identify them. Since XXmotif was unable to identify all the known C-boxes, the second hypothesis seems plausible.

To identify motifs, progressive promoter truncations were made. Removing important elements is predicted to alter expression. This technique has been successfully used previously. For example, the excretory cell specifying element Ex1 was identified in the *pgp-12* promoter (Zhao et al. 2005), and to indentify elements responsible for AFD neuron specific expression of *gcy-8* and *gcy-18* (Kagoshima and Kohara 2015). The goal is to identify the smallest region that can drive correct expression. This was termed the minimal promoter. Previously, the term minimal promoter was used synonymously with core promoter, for example the Δ*pes-10* minimal promoter contains only the part of the promoter thought to be needed to properly initiate transcription (Seydoux and Fire 1994). More recently, however, it has come to mean the smallest region required for basal activity, which includes promoter transcription initiation as well tissue-specific TFBS (Reidling et al. 2002; Reidling and Said 2003; Sawata et al. 2004). This later definition is the one I will be using.

## 3.2. Materials and methods

### 3.2.1. Generation of Constructs

Constructs were constructed in a similar manner to Chapter 2. Briefly, promoters were amplified with forward and reverse primers and fused to GFP by PCR fusion using nested primers. For MosSCI constructs, GFP fusion constructs were amplified using the MosSCI primers and cloned into plasmid CFJ151. For the rescue construct, minimal promoter GFP fusion construct was amplified with forward and reverse primers and fused to *dyf-5* using nested primers. The plasmid promoter truncations were created by Jun Wang by cloning promoters into PD95.75.

**Table 3-1. List of PCR Primers used in Chapter 3.**

| Primer Set | Forward Primer | Nested Primer | Reverse Primer |
|---|---|---|---|
| *C. elegans pdyf-5 (1929bp) GFP* | gcctgcaaatttgtcatacat | tttcaattcgaaaaacagcttc | tgaaaagttcttctcctttactcatggcttcttgcccttatattttc |
| *C. elegans pdyf-5 (1595bp) GFP* | gcctgcaaatttgtcatacat | gaggcactaaatgccgagtg | tgaaaagttcttctcctttactcatggcttcttgcccttatattttc |
| *C. elegans pdyf-5 (1002bp) GFP* | gcctgcaaatttgtcatacat | ctttgggcaaggtttttgtg | tgaaaagttcttctcctttactcatggcttcttgcccttatattttc |
| *C. elegans pdyf-5 (898bp) GFP* | gcctgcaaatttgtcatacat | aaaaatccaggagaacaatattcc | tgaaaagttcttctcctttactcatggcttcttgcccttatattttc |
| *C. elegans pdyf-5 (790bp) GFP* | gcctgcaaatttgtcatacat | acgtttttcattgcatgaatttt | tgaaaagttcttctcctttactcatggcttcttgcccttatattttc |
| *C. elegans pdyf-5 (324bp) GFP* | gcctgcaaatttgtcatacat | tcatctcgtcttcttcttgtgc | tgaaaagttcttctcctttactcatggcttcttgcccttatattttc |
| *C. elegans pdyf-5 (299bp) GFP* | gcctgcaaatttgtcatacat | ttccacttaaggccgtttgctcttggttac | tgaaaagttcttctcctttactcatggcttcttgcccttatattttc |
| *C. elegans pdyf-5 (285bp) GFP* | gcctgcaaatttgtcatacat | ttccacttaaggttaccatagaaactgtctgttacacc | tgaaaagttcttctcctttactcatggcttcttgcccttatattttc |
| *C. elegans pdyf-5 (minimal) GFP* | gcctgcaaatttgtcatacat | ttccacttaagtcatctcgtcttcttcttgtgc | tgaaaagttcttctcctttactcatgagtgagccatgagaggaaag |
| *GFP* | atgagtaaaggagaagaacttttcactgg | ggaaacagttatgtttggtatattggg | aagggcccgtacggccgactagtagg |
| *C. elegans pdyf-5 (minimal) GFP dyf-5* | ttccacttaagtcatctcgtcttcttcttgtgc | ttccacttaagtcatctcgtcttcttcttgtgc | caagtttaacagccgatgacattttgtatagttcatccatgccatgtgta |
| *dyf-5* | atgtcatcggctgttaaacttg | ttttgccacaattcactatatca | cccgaaaattgacatttgct |
| *C. elegans pdyf-5 (1929bp)* | ttccacttaagtttcaattcgaaaaacagcttc | N/A | ggataacctgcaggccagacgtgcg |
| *GFP MosSCI* | | | |
| *C. elegans pdyf-5 (324bp) GFP MosSCI* | ttccacttaagtcatctcgtcttcttcttgtgc | N/A | ggataacctgcaggccagacgtgcg |
| *C. elegans pdyf-5 (299bp) GFP MosSCI* | ttccacttaaggccgtttgctcttggttac | N/A | ggataacctgcaggccagacgtgcg |
| *C. elegans pdyf-5 (285bp) GFP MosSCI* | ttccacttaaggttaccatagaaactgtctgttacacc | N/A | ggataacctgcaggccagacgtgcg |
| *C. elegans pdyf-5 (minimal) GFP MosSCI* | ttccacttaagtcatctcgtcttcttcttgtgc | N/A | ggataacctgcaggccagacgtgcg |

### 3.2.2. Generation of Strains

Constructs were generated similar to Chapter 2. Briefly, constructs were micro-injected with a DNA mixture contain 50ng/µl of construct DNA and 100ng/µl of CEH361 into *dpy-5(e907)* worms. F1 worms displaying wild-type phenotype were selected after four days at 20ºC. After another four days wild-type F2 worms were isolated and observed.

The MosSCI strains were generated by micro-injecting JNC1021 worms with a DNA mixture containing 50ng/µl of CFJ151 with insert, 50ng/µl JL43.1, 5ng/µl GH8, 5ng/µl CFJ104, and 2.5 ng/µl CFJ90. F1 worms displaying wild-type phenotype were selected after four days at 20ºC. After another four days wild-type F2 worms were isolated and observed for *mCherry* expression. Worms with wild-type phenotype lacking *mCherry* expression were then grown for 3 generations to ensure homozygousity and observed by confocal microscopy.

For the *dyf-5* rescue strains, 50ng/µl of construct DNA and 100ng/µl of *rol-6* marker plasmid RF-4 DNA were injected into JNC528 worms. F1 worms displaying the roller phenotype were selected after four days at 20ºC. After another four days, roller F2 worms were isolated and observed. Strains generated and used are listed in Table 3-2.

**Table 3-2. List of strains used in Chapter 3.**

| Strain | Sex | Source | Genotype | Notes |
|---|---|---|---|---|
| JNC516 | Hermaphrodites | Injection Extrachromasomal | *dpy-5(e907)I; dotEX [Pr dyf-5::GFP + dpy-5(+)]* | *dyf-5* promoter (285bp upstream of start codon) fused to GFP (Same as BC7732) |
| JNC517 | Hermaphrodites | Injection Extrachromasomal | *dpy-5(e907)I; dotEX [Pr dyf-5::GFP + dpy-5(+)]* | *dyf-5* promoter (1929bp upstream of start codon) fused to *GFP (Same as BC7696)* |
| JNC518 | Hermaphrodites | Injection Extrachromasomal | *dpy-5(e907)I; dotEX [Pr dyf-5::GFP + dpy-5(+)]* | *dyf-5* promoter (898bp upstream of start codon) fused to GFP (Same as BC7841) |
| JNC519 | Hermaphrodites | Injection Extrachromasomal | *dpy-5(e907)I; dotEX [Pr dyf-5::GFP + dpy-5(+)]* | *dyf-5* promoter (790bp upstream of start codon) fused to GFP (Same as BC7840) |
| JNC520 | Hermaphrodites | Injection Extrachromasomal | *dpy-5(e907)I; dotEX [Pr dyf-5::GFP + dpy-5(+)]* | *dyf-5* promoter (324bp upstream of start codon) fused to GFP (Same as BC7833) |
| JNC521 | Hermaphrodites | Injection Extrachromasomal | *dpy-5(e907)I; dotEX [Pr dyf-5::GFP + dpy-5(+)]* | *dyf-5* promoter (299bp upstream of start codon) fused to GFP (Same as BC7801) |
| JNC522 | Hermaphrodites | Injection Extrachromasomal | *dpy-5(e907)I; dotEX [Pr dyf-5::GFP + dpy-5(+)]* | *dyf-5* promoter (1929bp upstream of start codon) fused to GFP (injected at 50ng/ul) |
| JNC523 | Hermaphrodites | Injection Extrachromasomal | *dpy-5(e907)I; dotEX [Pr dyf-5::GFP + dpy-5(+)]* | *dyf-5* promoter (1929bp upstream of start codon) fused to GFP (injected at 50ng/ul) |
| JNC524 | Hermaphrodites | Injection Extrachromasomal | *dpy-5(e907)I; dotEX [Pr dyf-5::GFP + dpy-5(+)]* | *dyf-5* promoter (1929bp upstream of start codon) fused to GFP (injected at 50ng/ul) |
| JNC525 | Hermaphrodites | Injection Extrachromasomal | *dpy-5(e907)I; dotEX [Pr dyf-5ΔH::GFP + dpy-5(+)]* | *dyf-5*ΔH promoter (1929bp upstream of start codon with 4bp deletion 392bp upstream of start codon) fused to GFP (Injected at 100ng/ul) |
| JNC526 | Hermaphrodites | Injection Extrachromasomal | *dpy-5(e907)I; dotEX [Pr dyf-5ΔH::GFP + dpy-5(+)]* | *dyf-5*ΔH promoter (1929bp upstream of start codon with 4bp deletion 392bp upstream of start codon) fused to GFP (Injected at 100ng/ul) |
| JNC527 | Hermaphrodites | Injection Extrachromasomal | *dpy-5(e907)I; dotEX [Pr dyf-5ΔH::GFP + dpy-5(+)]* | *dyf-5*ΔH promoter (1929bp upstream of start codon with 4bp deletion 392bp upstream of start codon) fused to GFP (injected at 50ng/ul) |
| JNC528 | Hermaphrodites | CGC | *dyf-5(mn400)* | Same as CGC SP1745 |
| JNC529 | Hermaphrodites | CGC | *M04C9.8(ok1170)* | Same as CGC RB1146 |
| JNC530 | Hermaphrodites | Injection Extrachromasomal | *dpy-5(e907)I; dotEX [Pr dyf-5::GFP + dpy-5(+)]* | Minimal *dyf-5* promoter (324bp upstream of start codon to 204bp upstream of start codon [120bp promoter]) fused to GFP (injected at 50ng/ul) |
| JNC530 | Hermaphrodites | Injection Extrachromasomal | *dpy-5(e907)I; dotEX [Pr dyf-5::GFP + dpy-5(+)]* | Minimal *dyf-5* promoter (324bp upstream of start codon to 204bp upstream of start codon [120bp promoter]) fused to GFP (injected at 50ng/ul) |
| JNC531 | Hermaphrodites | Injection Extrachromasomal | *dpy-5(e907)I; dotEX [Pr dyf-5::GFP + dpy-5(+)]* | Minimal *dyf-5* promoter (324bp upstream of start codon to 204bp upstream of start codon [120bp promoter]) fused to GFP (injected at 50ng/ul) |
| JNC532 | Hermaphrodites | Injection Extrachromasomal | *dpy-5(e907)I; dotEX [Pr dyf-5::GFP + dpy-5(+)]* | Minimal *dyf-5* promoter (324bp upstream of start codon to 204bp upstream of start codon [120bp promoter]) fused to GFP (injected at 50ng/ul) |
| JNC533 | Hermaphrodites | Injection Extrachromasomal | *dpy-5(e907)I; dotEX [Pr dyf-5::GFP + dpy-5(+)]* | *dyf-5* promoter (1595bp upstream of start codon) fused to GFP (injected at 50ng/ul, linear construct) |
| JNC534 | Hermaphrodites | Injection Extrachromasomal | *dpy-5(e907)I; dotEX [Pr dyf-5::GFP + dpy-5(+)]* | *dyf-5* promoter (1002bp upstream of start codon) fused to GFP (injected at 50ng/ul, linear construct) |
| JNC535 | Hermaphrodites | Injection Extrachromasomal | *dpy-5(e907)I; dotEX [Pr dyf-5::GFP + dpy-5(+)]* | *dyf-5* promoter (898bp upstream of start codon) fused to GFP (injected at 50ng/ul, linear construct) |
| JNC536 | Hermaphrodites | Injection Extrachromasomal | *dpy-5(e907)I; dotEX [Pr dyf-5::GFP + dpy-5(+)]* | *dyf-5* promoter (790bp upstream of start codon) fused to GFP (injected at 50ng/ul, linear construct) |
| JNC537 | Hermaphrodites | Injection Extrachromasomal | *dpy-5(e907)I; dotEX [Pr dyf-5::GFP + dpy-5(+)]* | *dyf-5* promoter (324bp upstream of start codon) fused to GFP (injected at 50ng/ul, linear construct) |
| JNC538 | Hermaphrodites | Injection Extrachromasomal | *dpy-5(e907)I; dotEX [Pr dyf-5::GFP + dpy-5(+)]* | *dyf-5* promoter (299bp upstream of start codon) fused to GFP (injected at 50ng/ul, linear construct) |
| JNC539 | Hermaphrodites | Injection Extrachromasomal | *dpy-5(e907)I; dotEX [Pr dyf-5::GFP + dpy-5(+)]* | *dyf-5* promoter (299bp upstream of start codon) fused to GFP (injected at 50ng/ul, linear construct) (1-3) |
| JNC540 | Hermaphrodites | Injection Extrachromasomal | *dpy-5(e907)I; dotEX [Pr dyf-5::GFP + dpy-5(+)]* | *dyf-5* promoter (299bp upstream of start codon) fused to GFP (injected at 50ng/ul, linear construct) (4-2) |
| JNC541 | Hermaphrodites | Injection Extrachromasomal | *dpy-5(e907)I; dotEX [Pr dyf-5::GFP + dpy-5(+)]* | *dyf-5* promoter (299bp upstream of start codon) fused to GFP (injected at 50ng/ul, linear construct) (4-12) |
| JNC542 | Hermaphrodites | Injection Extrachromasomal | *dpy-5(e907)I; dotEX [Pr dyf-5::GFP + dpy-5(+)]* | *dyf-5* promoter (299bp upstream of start codon) fused to GFP (injected at 50ng/ul, linear construct) (4-19) |
| JNC543 | Hermaphrodites | Injection Extrachromasomal | *dpy-5(e907)I; dotEX [Pr dyf-5::GFP + dpy-5(+)]* | *dyf-5* promoter (285bp upstream of start codon) fused to GFP (injected at 50ng/ul, linear construct) (3-11) |
| JNC544 | Hermaphrodites | Injection Extrachromasomal | *dpy-5(e907)I; dotEX [Pr dyf-5::GFP + dpy-5(+)]* | *dyf-5* promoter (285bp upstream of start codon) fused to GFP (injected at 50ng/ul, linear construct) (4-18) |
| JNC545 | Hermaphrodites | Injection Extrachromasomal | *dyf-5(mn400); dotEX [Pr dyf-5(min)::GFP::dyf-5]* | Minimal *dyf-5* promoter (324bp upstream of start codon to 204bp upstream of start codon [120bp promoter]) fused to GFP fused to *DYF-5* protein coding region |
| JNC546 | Hermaphrodites | MosSCI | *ttTi5605 II; unc-119(ed3) III; [Pr dyf-5::GFP + Cbr-unc-119]* | *dyf-5* promoter (1929bp upstream of start codon) fused to GFP inserted at mos site (1-4) |
| JNC547 | Hermaphrodites | MosSCI | *ttTi5605 II; unc-119(ed3) III; [Pr dyf-5::GFP + Cbr-unc-119]* | *dyf-5* promoter (1929bp upstream of start codon) fused to GFP inserted at mos site (5-1) |
| JNC548 | Hermaphrodites | MosSCI | *ttTi5605 II; unc-119(ed3) III; [Pr dyf-5::GFP + Cbr-unc-119]* | *dyf-5*ΔH promoter (1929bp upstream of start codon with 4bp deletion 392bp upstream of start codon) fused to GFP inserted at mos site (1-1) |
| JNC549 | Hermaphrodites | MosSCI | *ttTi5605 II; unc-119(ed3) III; [Pr dyf-5::GFP + Cbr-unc-119]* | *dyf-5*ΔH promoter (1929bp upstream of start codon with 4bp deletion 392bp upstream of start codon) fused to GFP inserted at mos site (1-3) |
| JNC550 | Hermaphrodites | MosSCI | *ttTi5605 II; unc-119(ed3) III; [Pr dyf-5::GFP + Cbr-unc-119]* | *dyf-5* promoter (324bp upstream of start codon) fused to GFP inserted at mos site (8-5-1) |
| JNC551 | Hermaphrodites | MosSCI | *ttTi5605 II; unc-119(ed3) III; [Pr dyf-5::GFP + Cbr-unc-119]* | *dyf-5* promoter (324bp upstream of start codon) fused to GFP inserted at mos site (8-5-2) |
| JNC552 | Hermaphrodites | MosSCI | *ttTi5605 II; unc-119(ed3) III; [Pr dyf-5::GFP + Cbr-unc-119]* | *dyf-5* promoter (299bp upstream of start codon) fused to GFP inserted at mos site |
| JNC553 | Hermaphrodites | MosSCI | *ttTi5605 II; unc-119(ed3) III; [Pr dyf-5::GFP + Cbr-unc-119]* | *dyf-5* promoter (285bp upstream of start codon) fused to GFP inserted at mos site (2-1) |
| JNC554 | Hermaphrodites | MosSCI | *ttTi5605 II; unc-119(ed3) III; [Pr dyf-5::GFP + Cbr-unc-119]* | *dyf-5* promoter (285bp upstream of start codon) fused to GFP inserted at mos site (2-2) |
| JNC555 | Hermaphrodites | MosSCI | *ttTi5605 II; unc-119(ed3) III; [Pr dyf-5::GFP + Cbr-unc-119]* | Minimal *dyf-5* promoter (324bp upstream of start codon to 204bp upstream of start codon [120bp promoter]) fused to GFP inserted at mos site |
| JNC556 | Hermaphrodites | MosSCI | *ttTi5605 II; unc-119(ed3) III; [Pr dyf-5::GFP + Cbr-unc-119]* | Minimal *dyf-5* promoter + transplice (324bp upstream of start codon to 204bp upstream of start codon [120bp promoter] with transplice site inserted) fused to GFP inserted at mos site (3-2-1) |
| JNC557 | Hermaphrodites | MosSCI | *ttTi5605 II; unc-119(ed3) III; [Pr dyf-5::GFP + Cbr-unc-119]* | Minimal *dyf-5* promoter + transplice (324bp upstream of start codon to 204bp upstream of start codon [120bp promoter] with transplice site inserted) fused to GFP inserted at mos site (3-2-2) |
| JNC1021 | Hermaphrodites | CGC | *ttTi5605 II; unc-119(ed3) III; oxEx1578* | Same as EG6699 |

### 3.2.3. Visualization of strains

Strains were visualised the same as Chapter 2. Worms were fixed on glass slides with 2% agarose pads and M9 solution containing 2%(w/v) sodium azide. M9 solution contains 0.3%(w/v) $KH_2PO_4$, 0.6%(w/v)$Na_2HPO_4$, 0.5%(w/v) NaCl, and 0.01%(w/v) $MgSO_4$. Slides were then observed on Zeiss spinning disc confocal microscope with 40x oil immersion lens.

## 3.3. Results

### 3.3.1. Promoter truncations (plasmid)

Because the computational approach was unable to find any new elements, a molecular approach was attempted. A student in our lab, Jun Wang, had previously created a series of *dyf-5* promoter truncations. These were created by inserting the various promoters into the GFP containing plasmid PD95.75 and injecting them into *C. elegans* as extra-chromosomal arrays. If a functional motif is removed by the truncation, a change in expression (either pattern or intensity) is expected.

The results show a dramatic reduction between 1929bp full length promoter and the 898bp promoter. There is also a further reduction between 324bp and 299bp. Interestingly, this region contains one of the C-boxes reported by Burghoorn *et al.* and detected by XXmotif analysis (Burghoorn et al. 2012). A final reduction of expression is observed between 299bp and 285bp. 285bp is the 5’ end of the X-box so it is possible the linkers are interfering with the function of the X-box (Figure 3-1).

Unfortunately, there were no constructs available with promoter lengths between 1929bp and 898bp. In addition, these constructs were significantly weaker than constructs produced previously using PCR Fusion. Plasmid constructs have previously been shown to produce weaker expression (Etchberger and Hobert 2008).

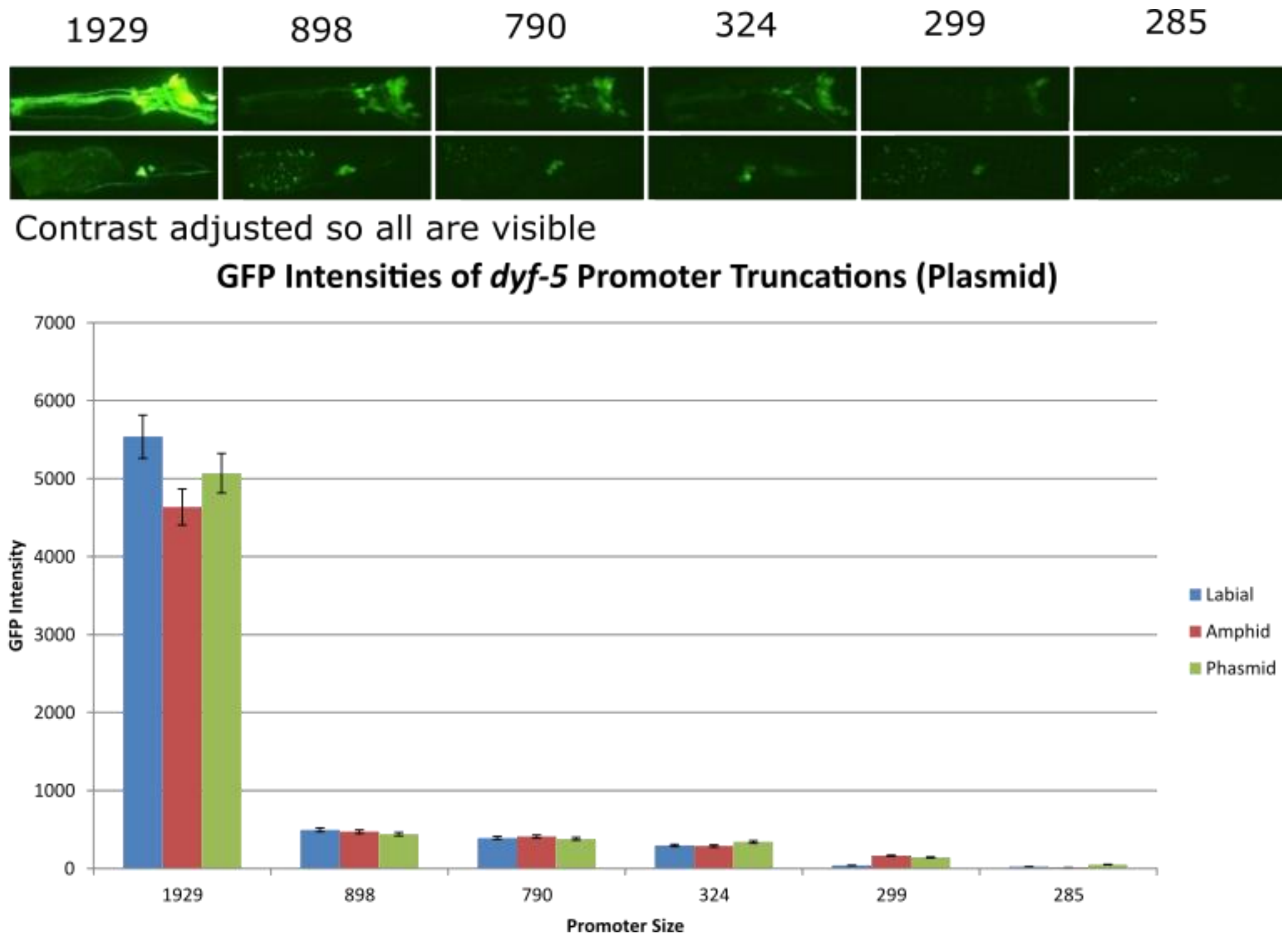

**Figure 3-1. Promoter truncations of *dyf-5* promoter fused to GFP via cloning.** (Top) Confocal images of all truncations. Contrast has been adjusted so expression of all constructs is visible and direct comparison is possible. A very sharp decrease in expression intensity is observed when promoter is truncated from 1929bp to 898bp. Further reduction is observed after 324bp truncations (Bottom) GFP intensities of each strain determined by confocal microscopy. Background expression has been removed (N=3).

### 3.3.2. Promoter truncations (linear)

In order to fill in the gap of promoter truncations between 1929bp and 898bp, a new series of constructs was produced. These were created using PCR fusion as this technique is simpler and the constructs show better expression (Etchberger and Hobert 2008). It was expected that these constructs would recapitulate the results previously observed and provide more resolution between 1929bp and 898bp.

Interestingly, the length dependent reduction of expression observed previously was not present in this new dataset. Expression was relatively similar amongst all the constructs, with 1929bp and 324bp showing very similar expression (Figure 3-2, Figure 3-3). 299bp was slightly weaker and this can be explained by the loss of a C-box.

Reduced expression of 285bp is likely the result of linkers interfering with the function of the X-box. Expression of constructs with intermediate length is somewhat higher. Although all constructs were injected and the same concentration variations in copy number amongst extra-chromosomal arrays are possible and likely explain the variability in expression.

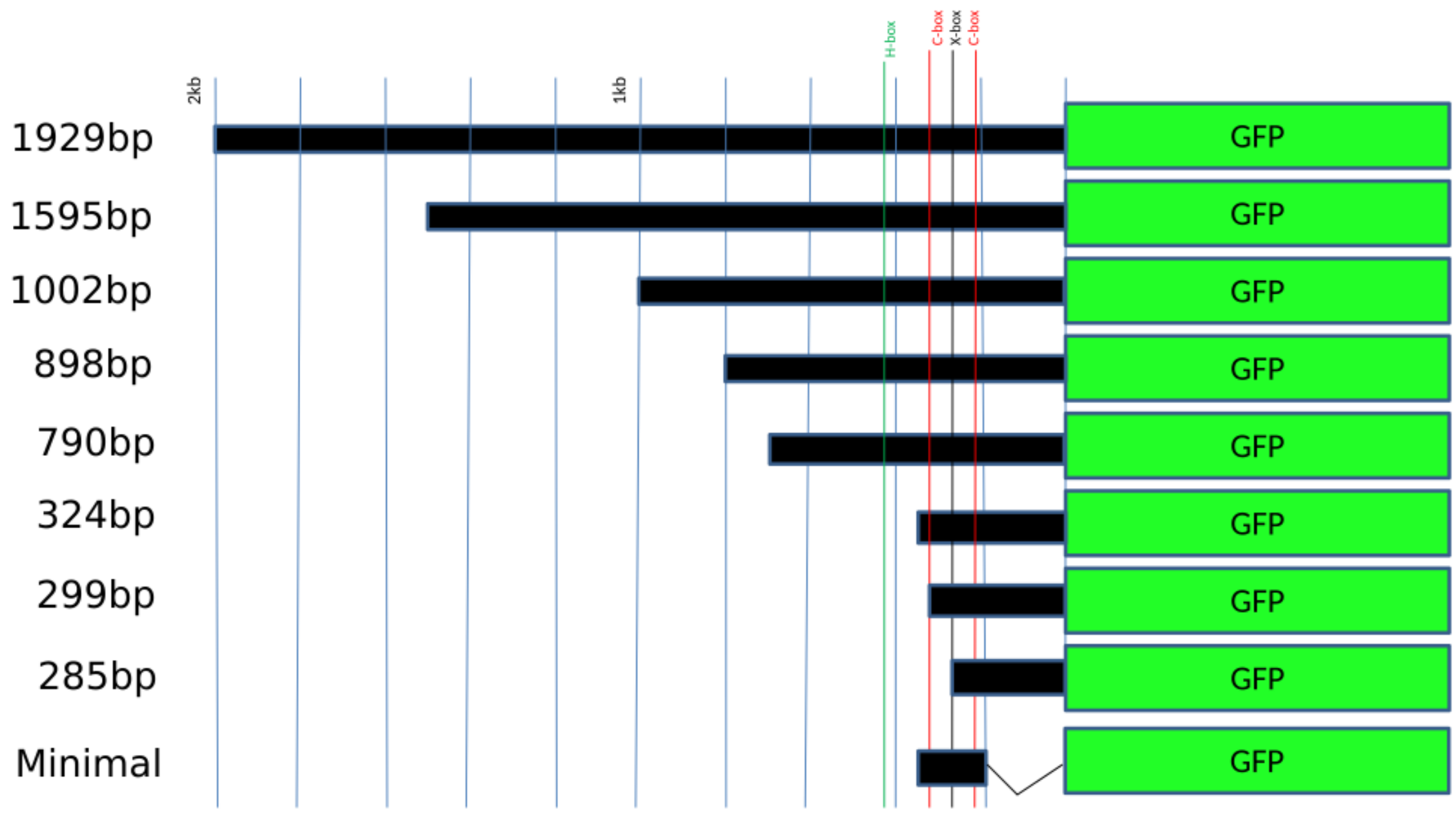

**Figure 3-2. Schematic representation of promoter truncation constructs.** Promoter of *dyf-5* was truncated to various lengths and compared to full length *dyf-5* promoter (1929bp). Minimal promoter has 3’ 200bp removed. 1929, 898, 324, 299 and 285 were also previously created as plasmid constructs.

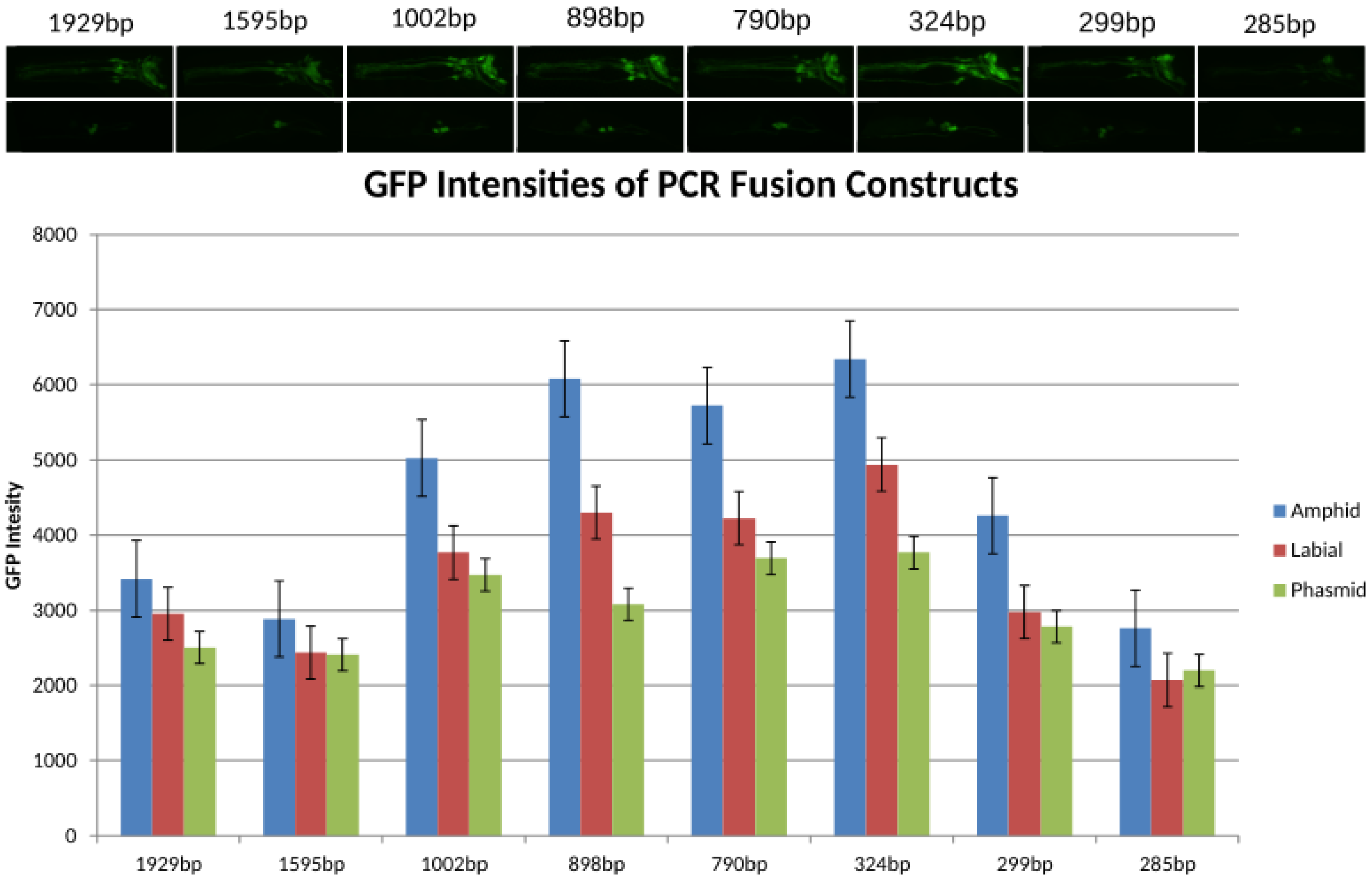

**Figure 3-3. Promoter truncations of *dyf-5* promoter fused to PCR fusion.**
(Top) Confocal images of all truncations. Pattern of length dependent expression observed in plasmid constructs is absent. Higher expression in the midsize constructs is likely the result of copy number variations in extra-chromosomal constructs. (Bottom) GFP intensities of each strain determined by cofocal microscopy. No clear length dependent expression observed (N=9).

### 3.3.3. Mos single copy insertions (truncation)

To confirm that the previous result is due to copy number variations of the extra-chromosomal arrays and not cryptic enhancers, single copy insertions of 1929bp, 324bp, 299bp, and 285bp promoter truncations were performed. It was expected that in the absence of additional enhancers, the 1929bp and 324bp promoters would show the same level of expression, whereas the 299bp and 285bp promoters would be weaker as a result of the missing C-box.

The results were as expected. The 1929bp and 324bp promoters showed virtually identical expression whereas the others were slightly decreased. The 285bp promoter showed a larger decrease in expression which is consistent with the hypothesis that the linkers are interfering with the function of the X-box. From these results, it can be concluded that the expression of the PCR fusion constructs is more similar to the single copy constructs and can thus be considered to be most like the endogenous situation. The length dependent decrease in expression of the plasmid constructs could be the result of unusual chromatin structure formed by the foreign plasmid DNA. Future studies should be aware of this possibility when investigating expression.

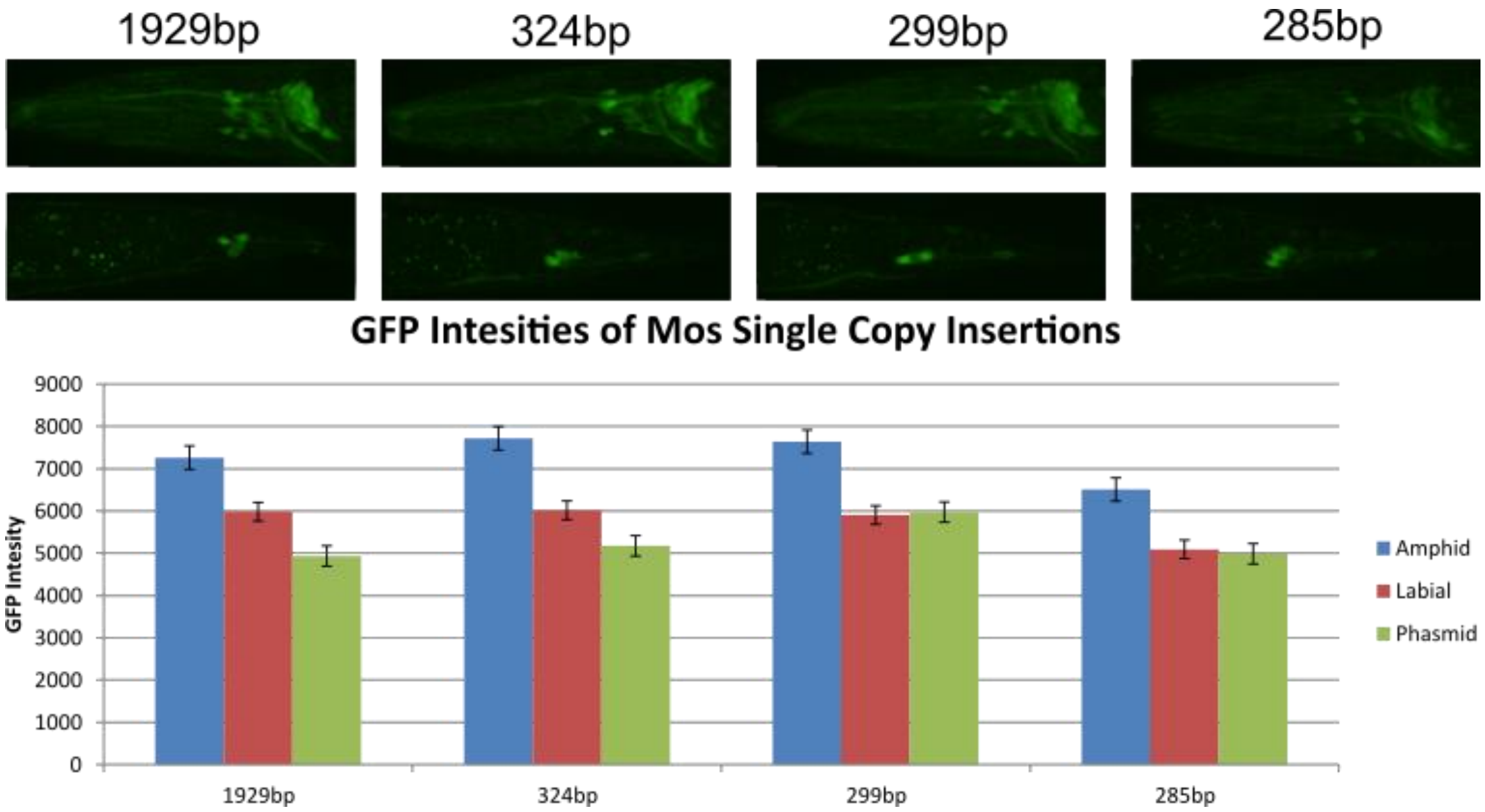

**Figure 3-4. Mos single-copy insertions of *dyf-5* promoter truncations.**
(Top) Confocal images of all single copy truncations. Pattern of expression is maintained. No significant reduction of expression observed until promoter is reduced below 299bp. (Bottom) GFP intensities of each strain determined by confocal microscopy. No significant changes in expression observed (N=6).

### 3.3.4. Minimal promoter constructs

The previous experiments revealed that there are no promoter elements upstream of 324bp. It was then necessary to determine if there were any elements downstream of the X-box. To that end, a deletion of the 205bp upstream of the start codon was generated. This region was chosen because this 120bp contains only the X-box and C-boxes reported by Burghoorn *et al.* and confirmed by my experiments. If this region can drive expression similar to the full length promoter this would suggest there are no other elements present. This region was termed the minimal promoter.

When injected as an extra-chromosomal array, the minimal promoter drives the same pattern of expression as the full length promoter with nearly identical intensity (Figure 3-5). This suggests that there are no important promoter elements downstream of this region.

Because extra-chromosomal arrays can have variable copy numbers, single copy insertions of these constructs was also performed. In this case, the pattern was again the same but the intensity of the minimal promoter was somewhat weaker than the full length promoter (Figure 3-6).

Since no difference in expression was observed between these promoters when expressed from an extra-chromosomal array, it was hypothesised that it was not a transcription factor binding site that was missing but some portion of the basal promoter apparatus which was limiting the efficiency of transcription. Another construct was produced including a proper transplice site into the minimal promoter. The transplice site improved expression of the minimal promoter but did not bring it back to the levels of the full length promoter. This suggests that something else is still missing, likely some other part of the basal promoter. There are several possibilities such as the proper transcription start site or T-blocks (Reinke et al. 2013).

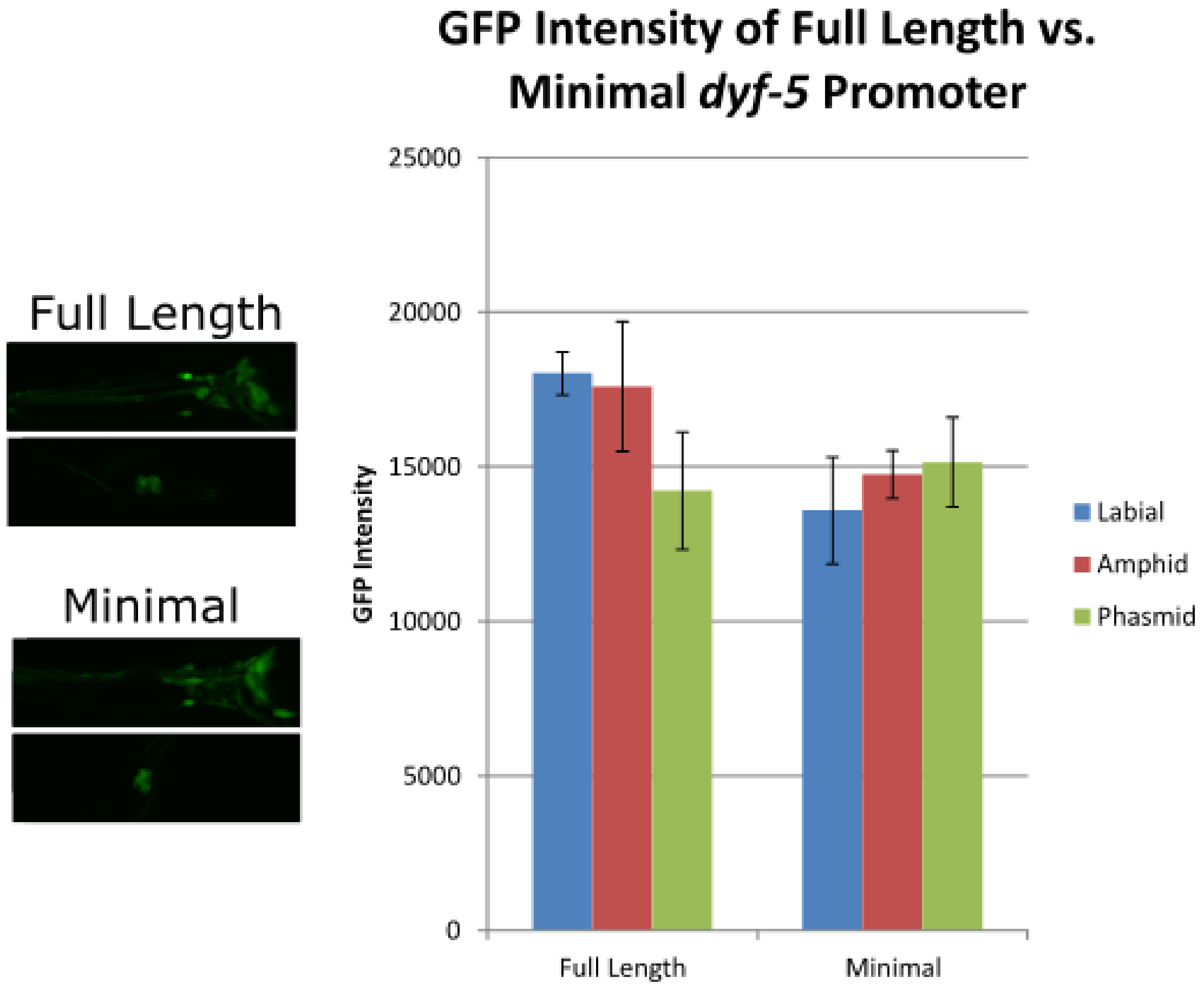

**Figure 3-5. Expression of minimal *dyf-5* promoter.**
*C. elegans* strains with either the full length wild-type *dyf-5* promoter fused to GFP or the 120bp minimal *dyf-5* promoter fused to GFP were created. (Left) Confocal images of expression. Pattern and intensity appears very similar. (Right) Intensity of GFP determined by analysing confocal images (N=5).

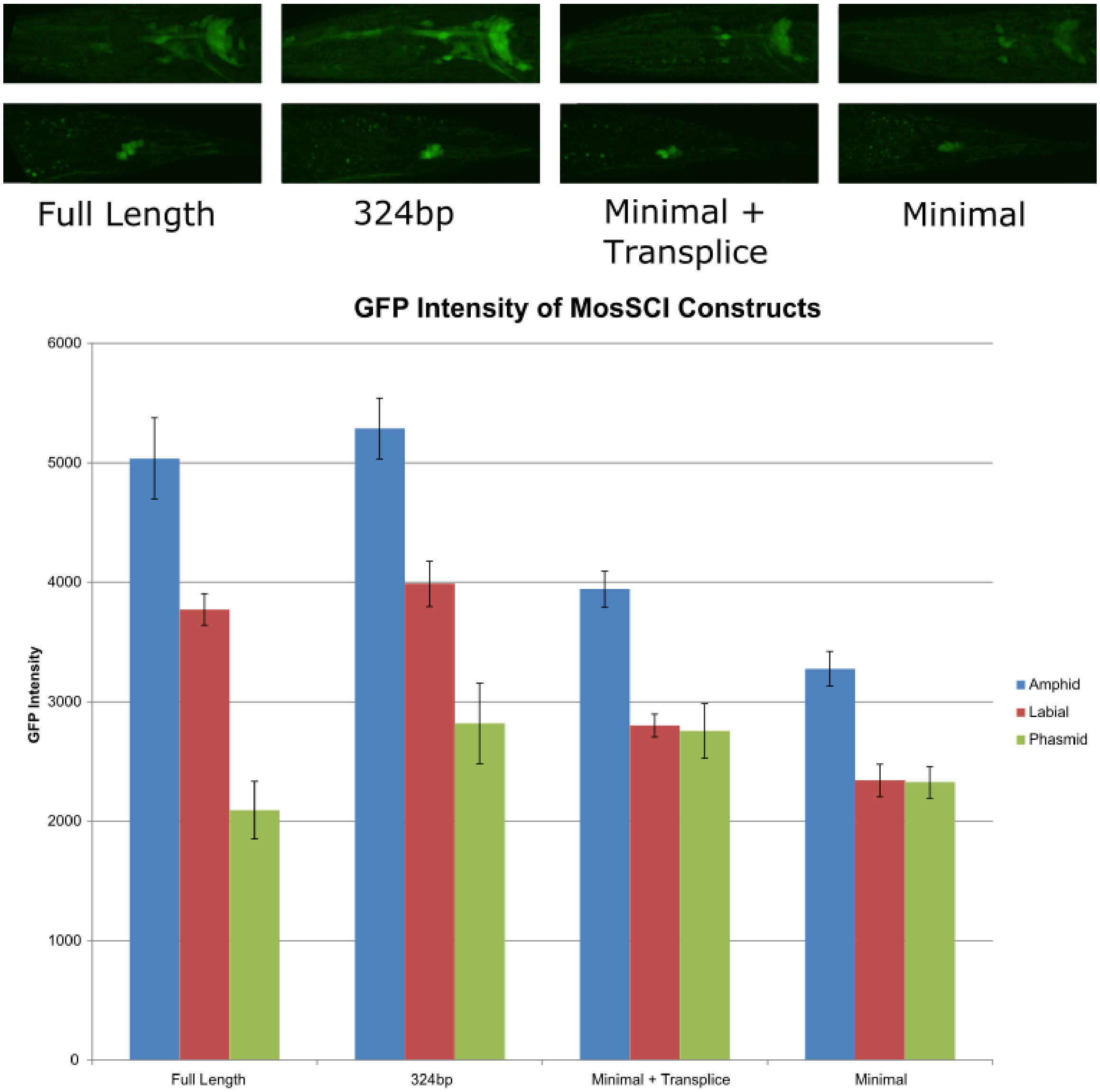

**Figure 3-6. Expression of single copy minimal *dyf-5* promoter.**
*C. elegans* strains with single copy insertions of either the full length wild-type *dyf-5* promoter fused to GFP or the 120bp minimal *dyf-5* promoter fused to GFP were created. Full length (1929bp) and 324bp promoters show very similar expression. Minimal promoter is visible weaker however pattern of expression is maintained. Addition of a transplice site to the promoter restores some expression. Intensity values determined by confocal microscopy (N=6).

### 3.3.5. Rescue constructs

All the previous experiments have shown that the minimal promoter is largely able to express in the same manner as the full length promoter. The next question is to ascertain if this promoter can rescue a *dyf-5* mutant when driving the DYF-5 protein. To do this, a construct with the minimal promoter driving the *dyf-5* gene with an N-terminal GFP fusion was created. This construct was injected into a *dyf-5(mn400)* mutant along with a *rol-6* marker gene. These worms were then dye-filled with DiI along with N2 and mutant worms. The construct clearly rescues the mutant phenotype (Figure 3-7). This suggests that the 120bp minimal promoter contains all the major element controlling *dyf-5* expression.

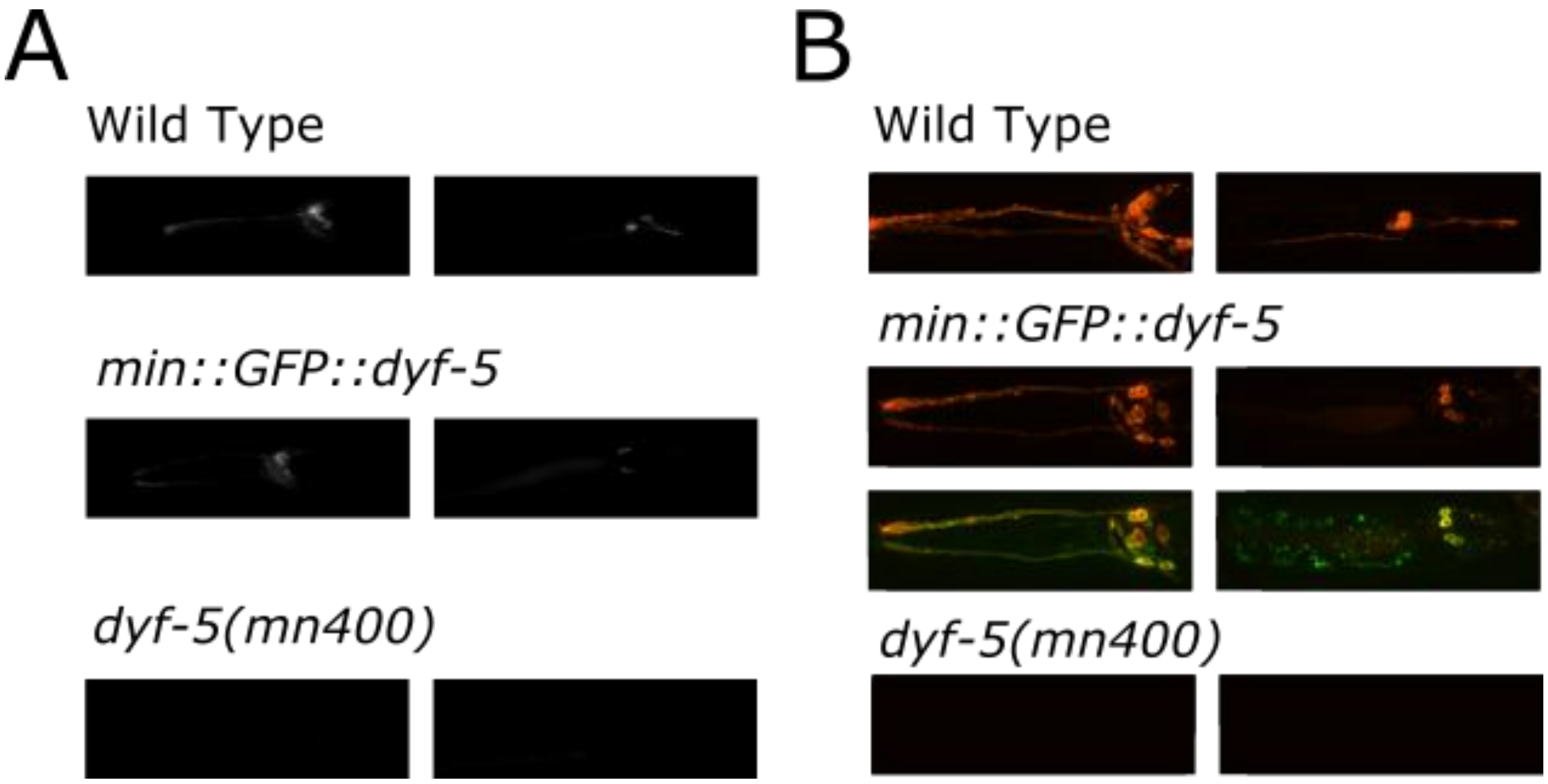

**Figure 3-7. *dyf-5* minimal promoter rescue.**
The minimal *dyf-5* promoter was fused to a GFP::*DYF-5* translational fusion construct. This was injected into a *dyf-5* mutant (mn400). Dye-filling with DiI was performed on N2, rescue construct, and mutant. (A) compound microscope images (B) Confocal images.

## 3.4. Discussion

In general, the results from the promoter truncations agree with that of the bioinformatics; there is no evidence of motifs outside of a small region surrounding the X-box. Interestingly, the plasmid and PCR fusion based promoter truncations show vastly different results. The plasmid truncations show a very clear length dependent effect whereas the linear constructs do not. It has been previously shown that plasmid based constructs show lower expression levels than PCR fusion based constructs. The reason for this is unclear, although it was speculated that packaging of the vector DNA may occlude regulatory elements (Etchberger and Hobert 2008). My results support this hypothesis, as one would expect the repressive effect of the vector DNA to be greater the closer it is to the regulatory elements. Given that the important elements are approximately 300 base pairs upstream of the start codon, it appears that the repressive effects can extend at least 500 base pairs as the 898 base pair promoter showed repression. The results of the single copy insertion construct also suggest that the vector free constructs more closely resemble the native situation, as no length dependent repression was observed in these strains. As previously mentioned, double labelling with a "control" promoter driving one colour and the "test" promoter driving a different colour of reporter gene may have helped to confirm the expression pattern. Although the pattern appears the same upon visual inspection, small differences cannot be ruled out. Since no differences in expression pattern were observed expression intensity was the main parameter for comparison.

The minimal promoter appears to contain all the major elements necessary for expression. Although, the expression of the minimal promoter was weaker than that of the full length promoter, the full expression pattern was observed and full rescue of the mutant phenotype was observed. Only rescue of the dye-filling defective phenotype was tested, however, it is likely that other cilia defective-related phenotypes could also be rescued, for example chemotaxis defects. As no conserved sequences were observed in the 200 base pairs removed to create the minimal promoter, it seems logical that the difference of expression is the result of the removal of basal promoter elements rather than portions of the X-box CRM. There is support for this hypothesis as adding a transplice site to the minimal promoter improves expression. Other potential elements still missing include a native transcription start site and T-blocks (Grishkevich et al. 2011).

As the *C. elegans dyf-5* minimal promoter was achieved by hypothesising that the computationally detected motifs (the X-box and C-boxes) were the only motifs required for function of the promoter, it is possible that the same could be done to identify minimal promoters from related species. This would include identifying the X-box and C-boxes in these orthologous promoters and amplifying the region containing them.

While it seems likely that all the functional elements are contained within the minimal promoter, it is possible that this may not be the case. The region downstream of the minimal promoter may contain elements aside from basal promoter elements that could contribute to the reduced expression. Additionally, it is possible elements reside upstream of the minimal promoter. Although no conserved elements were detected and a change in expression was not observed when this upstream region was removed it is still possible that functional elements that are redundant or function under a condition not tested are present. For example, shadow enhancers have been reported that are redundant with other motifs and function to improve robustness of expression under conditions of stress (Hobert 2010; Barolo 2012). Since all my experiments were performed in the relatively stress-free environment of the lab it is possible that they were missed.

# Chapter 4. Molecular dissection of *dyf-5* minimal promoter

## 4.1. Overview

It was previously shown that only the 120bp minimal promoter of *dyf-5* was necessary to drive correct expression. This suggests that all necessary motifs are present within this promoter. The question of what functional motifs are present within the promoter remains. In addition, how the motifs act together to drive correct expression must be addressed. From the computational analysis, only X-boxes and C-boxes are predicted to reside in the minimal promoter. Therefore, the first step is to sequentially mutate these motifs and observe the effect on expression. Once this is accomplished, further dissection of the remaining region will determine what other motifs are present if there are any.

Since the minimal promoter is only 120bp it becomes possible to have synthetic promoters synthesised biochemically. This allows far greater control of the final sequence than is possible otherwise.

## 4.2. Materials and methods

### 4.2.1. Generation of constructs

Mutation constructs were generated by having 80-mers corresponding to each half of the promoter synthesised by Eurofins (www.operon.com). The left oligo is in the forward direction and contains twenty base pair overlap with the right oligo. The right oligo is in the reverse direction and contains twenty base pair overlap with GFP. The constructs were then generated by PCR fusion using the left oligo and the nested GFP primer as primers and GFP and the right oligo diluted to 0.1µM as template. The primers used are listed in Table 4-1. Using various combinations of these primers the promoters listed in Table 4-2 were created. Promoters also schematically represented in Figure 4-1.

**Table 4-1. Primers used in Chapter 4.**

| Primer Name | Sequence |
|---|---|
| minL | tcatctcgtcttcttcttgtgctccgccgtttgctcttggttaccatagaaactgtctgttacacccttttctcttcttc |
| minR | agttcttctcctttactcatgagtgagccatgagaggaaagactaaaagagaagaagcatgaagaagagaaaagggtgta |
| minL_xbox | tcatctcgtcttcttcttgtgctccgccgtttgctcttggttaggatagaaactgtctgttacacccttttctcttcttc |
| minL_cbox | tcatctcgtgttgttgttgtgctccgccgtttgctcttggttaccatagaaactgtctgttacacccttttctcttcttc |
| minR_cbox | agttcttctcctttactcatgagtgagccatgagaggaaagactaaaacacaacaaccatgaagaagagaaaagggtgta |
| minLC_cbox | tcatctcgtgttgttgttgtgctccgccgtttgctcttggttaccatagaaactgtctgttacaccgttttgtgttgttg |
| minRC_cbox | agttcttctcctttactcatgagtgagccatgagaggaaagactaaaacacaacaaccatcaacaacacaaaacggtgta |
| minC_cboxL+ | tcatctcgtcttcttcttgtgctccgccgtttgctcttggttaccatagaaactgtctgttacaccgttttgtgttgttg |
| minC_cboxR+ | agttcttctcctttactcatgagtgagccatgagaggaaagactaaaagagaagaagcatcaacaacacaaaacggtgta |
| Scramble_1_L | tatcgttggtgatcctgcgcattttgttttacgctcttggttaccatagaaactgtctgttttgtcggtccctttttac |
| Scramble_1_R | agttcttctcctttactcatgcagcgcagccctcaaaaaaggctaaaagggcaaaacaaggtaaaaaagggaccgacaaa |
| Scramble_2_L | ttctgcgccgctagtcggtcgtctcttttggtgctcttggttaccatagaaactgtctgtggctttttcctcttttttt |
| Scramble_2_R | agttcttctcctttactcatcgaaccactgttcccggagaaactaaaagatgaatacagaaaaaaaaagaggaaaaagcc |
| minL_LC_strong | tcatctcgtgtagtagtagtgctccgccgtttgctcttggttaccatagaaactgtctgttacacccttatgtgatgatg |
| minR_RC_strong | agttcttctcctttactcatgagtgagccatgagaggaaagactaatacactactaccatcatcatcacataagggtgta |
| minR_syntrans | agttcttctcctttactcatggcttcttgcccttatatttactgaaaagagaagaagcatgaagaagagaaaagggtgta |
| minL_xbox_str | tcatctcgtcttcttcttgtgctccgccgtttgctcttgctcataaaatacggtgtctgttacacccttttctcttcttc |
| Scramble_3_L | tcttcctctatttatcagcttccccttttcggtgctcttggttaccatagaaactgtctgttgttctgattgttgtttggt |
| Scramble_3_R | agttcttctcctttactcatgagtgaggcaagcaccccaaacctaaaagcaacgtgcgaaaccaaacaacaatcagaaca |
| Scramble_4_L | ttgtttgtctgtccatggcgttgtgccttcaggctcttggttaccatagaaactgtctgttcgtactttttgcttccatt |
| Scramble_4_R | agttcttctcctttactcatgagtgagagaatcaagcccaaactaaaagcacgaacgaacaatggaagcaaaaagtacga |
| Scramble_neg_L | tcttcctctatttatcagcttccccttttcggtttgatagattgtcagagctacctgttcctgttctgattgttgtttggt |

**Table 4-2.** ***dyf-5*** **Promoter Mutation Constructs**

| Construct Name | Sequence |
|---|---|
| Minimal Promoter | tcatctcgtCTTCTTCTTgtgctccgccgtttgctcttgGTTACCATAGAAACtgtctgttacacccTTTTCT<br>CTTCTTCatgCTTCTTCTCTTTTagtctttcctctcatggctcactc |
| X-box | tcatctcgtCTTCTTCTTgtgctccgccgtttgctcttgGTTA**GG**ATAGAAACtgtctgttacacccTTTTCT<br>CTTCTTCatgCTTCTTCTCTTTTagtctttcctctcatggctcactc |
| X-box Strong | tcatctcgtCTTCTTCTTgtgctccgccgtttgctcttg**CTCATAAAATACGG**tgtctgttacacccTTTTCT<br>CTTCTTCatgCTTCTTCTCTTTTagtctttcctctcatggctcactc |
| Left C-box | tcatctcgt**G**TT**G**TT**G**TTgtgctccgccgtttgctcttgGTTACCATAGAAACtgtctgttacacccTTTTCT<br>CTTCTTCatgCTTCTTCTCTTTTagtctttcctctcatggctcactc |
| Right C-box | tcatctcgtCTTCTTCTTgtgctccgccgtttgctcttgGTTACCATAGAAACtgtctgttacacccTTTTCT<br>CTTCTTCatg**G**TT**G**TT**G**T**G**TTTTagtctttcctctcatggctcactc |
| Centre C-box | tcatctcgtCTTCTTCTTgtgctccgccgtttgctcttgGTTACCATAGAAACtgtctgttacaccgTTTT**G**T<br>**G**TT**G**TT**G**atgCTTCTTCTCTTTTagtctttcctctcatggctcactc |
| Left and Centre C-box | tcatctcgt**G**TT**G**TT**G**TTgtgctccgccgtttgctcttgGTTACCATAGAAACtgtctgttacacccTTTTCT<br>CTTCTTCatg**G**TT**G**TT**G**T**G**TTTTagtctttcctctcatggctcactc |
| Centre and Right C-box | tcatctcgtCTTCTTCTTgtgctccgccgtttgctcttgGTTACCATAGAAACtgtctgttacaccgTTTT**G**T<br>**G**TT**G**TT**G**atg**G**TT**G**TT**G**T**G**TTTTagtctttcctctcatggctcactc |
| Left and Right C-box | tcatctcgt**G**TT**G**TT**G**TTgtgctccgccgtttgctcttgGTTACCATAGAAACtgtctgttacaccgTTTT**G**T<br>**G**TT**G**TT**G**atgCTTCTTCTCTTTTagtctttcctctcatggctcactc |
| 3 C-box | tcatctcgt**G**TT**G**TT**G**TTgtgctccgccgtttgctcttgGTTACCATAGAAACtgtctgttacaccgTTTT**G**T<br>**G**TT**G**TT**G**atg**G**TT**G**TT**G**T**G**TTTTagtctttcctctcatggctcactc |
| 3 C-box Strong | tcatctcgt**G**T**AG**T**AG**T**A**gtgctccgccgtttgctcttgGTTACCATAGAAACtgtctgttacacccTT**A**T**G**<br>T**GA**T**GA**T**G**atg**G**T**AG**T**AG**T**G**T**A**TTagtctttcctctcatggctcactc |
| Scramble 1 | **tatcgttggtgatcctgcgcattttgttttac**gctcttgGTTACCATAGAAACtgtctgt**tttgtcggtcccttttta**<br>**ccttgttttgccc**ttttag**cctttttgagggctgcgctgc** |
| Scramble 2 | **ttctgcgccgctagtcggtcgtctcttttggt**gctcttgGTTACCATAGAAACtgtctgt**ggctttttcctctttttt**<br>**ttctgtattcatc**ttttag**tttctccgggaacagtggttcg** |
| Scramble 3 | **tcttcctctatttatcagcttccccttttcggt**gctcttgGTTACCATAGAAACtgtctgt**tgttctgattgttgtttggt**<br>**ttcgcacgttgc**ttttag**gtttggggtgcttgc**ctcactc |
| Scramble 4 | **ttgtttgtctgtccatggcgttgtgccttcag**gctcttgGTTACCATAGAAACtgtctgt**tcgtactttttgcttcca**<br>**ttgttcgttcgtgc**ttttag**tttgggcttgattct**ctcactc |
| Scramble Negative | **tcttcctctatttatcagcttccccttttcggtttgatagattgtcagagctacctgttcctgttctgattgttgtttggtttc**<br>**gcacgttgc**ttttag**gtttggggtgcttgc**ctcactc |
| Minimal + Transplice | tcatctcgtCTTCTTCTTgtgctccgccgtttgctcttgGTTACCATAGAAACtgtctgttacacccTTTTCT<br>CTTCTTCatgCTTCTTCTCTTTT**c**agt**aaatataagggcaagaagcc** |

X-box and C-boxes are capitalized. Changed bases are in bold.

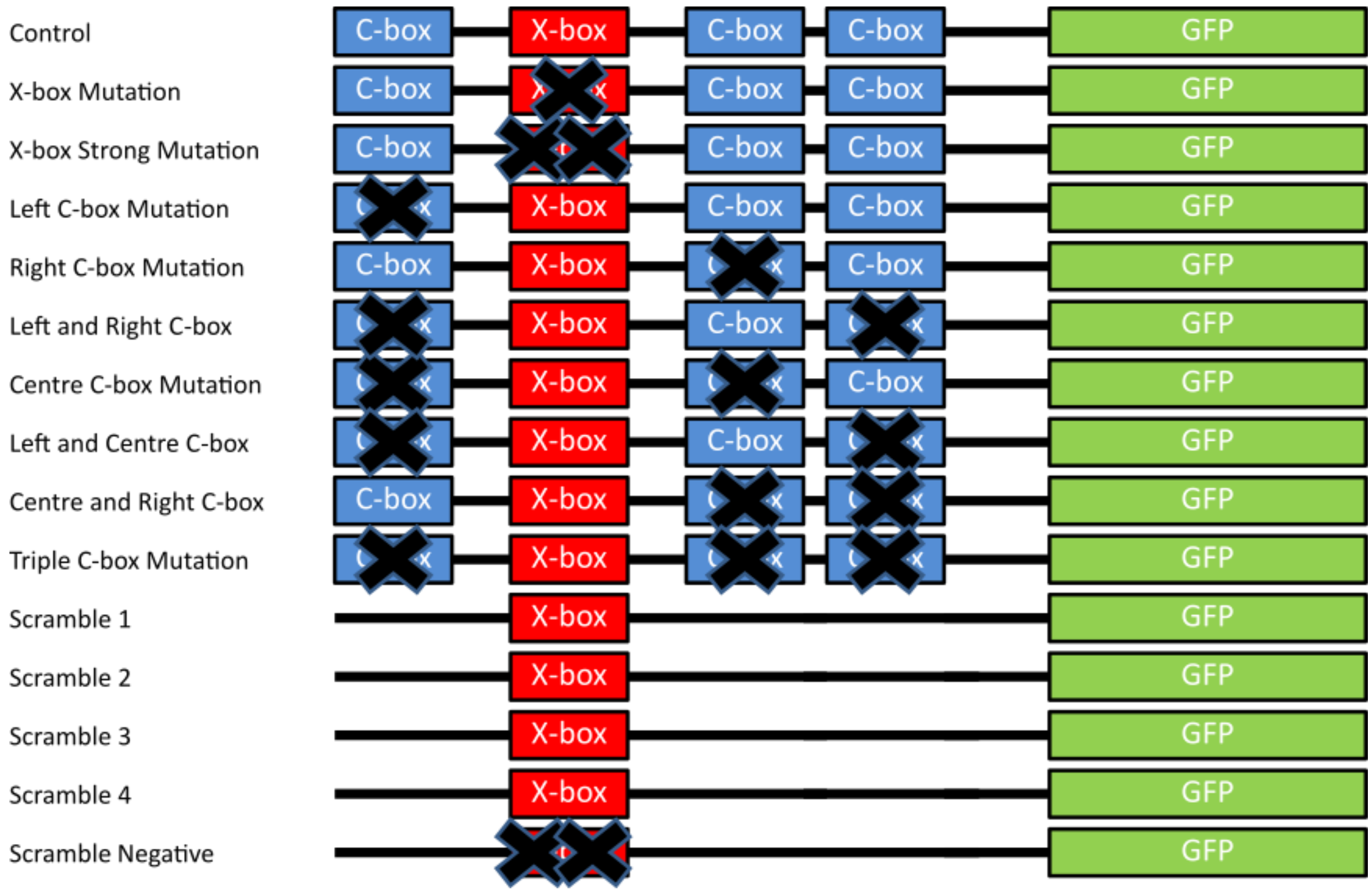

**Figure 4-1. Schematic representation of Chapter 4 constructs.**
X-box and C-boxes were sequentially mutated. Scramble mutations scrambled promoter but left X-box intact thus removing all C-boxes.

### 4.2.2. Generation of Strains

Strains were generated in the same manner as the extra-chromosomal array strains created in Chapters 2 and 3.

**Table 4-3. List of strains used in Chapter 4.**

| Strain | Sex | Source | Genotype | Notes |
|---|---|---|---|---|
| JNC558 | Hermaphrodites | Injection Extrachromasomal | *dpy-5(e907)I; dotEX [Pr dyf-5(min)::GFP + dpy-5(+)]* | Minimal *dyf-5* promoter (324bp upstream of start codon to 204bp upstream of start codon [120bp promoter]) fused to GFP (injected at 50ng/ul) |
| JNC559 | Hermaphrodites | Injection Extrachromasomal | *dpy-5(e907)I; dotEX [Pr dyf-5(min - xbox)::GFP + dpy-5(+)]* | Minimal *dyf-5* promoter (324bp upstream of start codon to 204bp upstream of start codon [120bp promoter]) with X-box mutated fused to GFP (injected at 50ng/ul) (1-3) |
| JNC560 | Hermaphrodites | Injection Extrachromasomal | *dpy-5(e907)I; dotEX [Pr dyf-5(min - xbox)::GFP + dpy-5(+)]* | Minimal *dyf-5* promoter (324bp upstream of start codon to 204bp upstream of start codon [120bp promoter]) with X-box mutated fused to GFP (injected at 50ng/ul) (2-1) |
| JNC561 | Hermaphrodites | Injection Extrachromasomal | *dpy-5(e907)I; dotEX [Pr dyf-5(min - xbox strong)::GFP + dpy-5(+)]* | Minimal *dyf-5* promoter (324bp upstream of start codon to 204bp upstream of start codon [120bp promoter]) with X-box mutated (strong mutation) fused to GFP (injected at 50ng/ul) (2-2) |
| JNC562 | Hermaphrodites | Injection Extrachromasomal | *dpy-5(e907)I; dotEX [Pr dyf-5(min - xbox strong)::GFP + dpy-5(+)]* | Minimal *dyf-5* promoter (324bp upstream of start codon to 204bp upstream of start codon [120bp promoter]) with X-box mutated (strong mutation) fused to GFP (injected at 50ng/ul) (4-10) |
| JNC563 | Hermaphrodites | Injection Extrachromasomal | *dpy-5(e907)I; dotEX [Pr dyf-5(min - L c-box)::GFP + dpy-5(+)]* | Minimal *dyf-5* promoter (324bp upstream of start codon to 204bp upstream of start codon [120bp promoter]) with Left C-box mutated fused to GFP (injected at 50ng/ul) (1-3) |
| JNC564 | Hermaphrodites | Injection Extrachromasomal | *dpy-5(e907)I; dotEX [Pr dyf-5(min - L c-box)::GFP + dpy-5(+)]* | Minimal *dyf-5* promoter (324bp upstream of start codon to 204bp upstream of start codon [120bp promoter]) with Left C-box mutated fused to GFP (injected at 50ng/ul) (5-7) |
| JNC565 | Hermaphrodites | Injection Extrachromasomal | *dpy-5(e907)I; dotEX [Pr dyf-5(min - R c-box)::GFP + dpy-5(+)]* | Minimal *dyf-5* promoter (324bp upstream of start codon to 204bp upstream of start codon [120bp promoter]) with Right C-box mutated fused to GFP (injected at 50ng/ul) |
| JNC566 | Hermaphrodites | Injection Extrachromasomal | *dpy-5(e907)I; dotEX [Pr dyf-5(min - C c-box)::GFP + dpy-5(+)]* | Minimal *dyf-5* promoter (324bp upstream of start codon to 204bp upstream of start codon [120bp promoter]) with Centre C-box mutated fused to GFP (injected at 50ng/ul) (1-1) |
| JNC567 | Hermaphrodites | Injection Extrachromasomal | *dpy-5(e907)I; dotEX [Pr dyf-5(min - C c-box)::GFP + dpy-5(+)]* | Minimal *dyf-5* promoter (324bp upstream of start codon to 204bp upstream of start codon [120bp promoter]) with Centre C-box mutated fused to GFP (injected at 50ng/ul) (5-13) |
| JNC568 | Hermaphrodites | Injection Extrachromasomal | *dpy-5(e907)I; dotEX [Pr dyf-5(min - L+R c-box)::GFP + dpy-5(+)]* | Minimal *dyf-5* promoter (324bp upstream of start codon to 204bp upstream of start codon [120bp promoter]) with Left and Right C-boxes mutated fused to GFP (injected at 50ng/ul) (2-2) |
| JNC569 | Hermaphrodites | Injection Extrachromasomal | *dpy-5(e907)I; dotEX [Pr dyf-5(min - L+R c-box)::GFP + dpy-5(+)]* | Minimal *dyf-5* promoter (324bp upstream of start codon to 204bp upstream of start codon [120bp promoter]) with Left and Right C-boxes mutated fused to GFP (injected at 50ng/ul) (2-14) |
| JNC570 | Hermaphrodites | Injection Extrachromasomal | *dpy-5(e907)I; dotEX [Pr dyf-5(min - L+C c-box)::GFP + dpy-5(+)]* | Minimal *dyf-5* promoter (324bp upstream of start codon to 204bp upstream of start codon [120bp promoter]) with Left and Centre C-boxes mutated fused to GFP (injected at 50ng/ul) (2-1) |
| JNC571 | Hermaphrodites | Injection Extrachromasomal | *dpy-5(e907)I; dotEX [Pr dyf-5(min - L+C c-box)::GFP + dpy-5(+)]* | Minimal *dyf-5* promoter (324bp upstream of start codon to 204bp upstream of start codon [120bp promoter]) with Left and Centre C-boxes mutated fused to GFP (injected at 50ng/ul) (2-2) |
| JNC572 | Hermaphrodites | Injection Extrachromasomal | *dpy-5(e907)I; dotEX [Pr dyf-5(min - R+C c-box)::GFP + dpy-5(+)]* | Minimal *dyf-5* promoter (324bp upstream of start codon to 204bp upstream of start codon [120bp promoter]) with Right and Centre C-boxes mutated fused to GFP (injected at 50ng/ul) (2-1) |
| JNC573 | Hermaphrodites | Injection Extrachromasomal | *dpy-5(e907)I; dotEX [Pr dyf-5(min - R+C c-box)::GFP + dpy-5(+)]* | Minimal *dyf-5* promoter (324bp upstream of start codon to 204bp upstream of start codon [120bp promoter]) with Right and Centre C-boxes mutated fused to GFP (injected at 50ng/ul) (5-10) |
| JNC574 | Hermaphrodites | Injection Extrachromasomal | *dpy-5(e907)I; dotEX [Pr dyf-5(min - 3 c-box)::GFP + dpy-5(+)]* | Minimal *dyf-5* promoter (324bp upstream of start codon to 204bp upstream of start codon [120bp promoter]) with Left, Right and Centre C-boxes mutated fused to GFP (injected at 50ng/ul) (3-9) |
| JNC575 | Hermaphrodites | Injection Extrachromasomal | *dpy-5(e907)I; dotEX [Pr dyf-5(min - 3 c-box)::GFP + dpy-5(+)]* | Minimal *dyf-5* promoter (324bp upstream of start codon to 204bp upstream of start codon [120bp promoter]) with Left, Right and Centre C-boxes mutated fused to GFP (injected at 50ng/ul) (6-12) |
| JNC576 | Hermaphrodites | Injection Extrachromasomal | *dpy-5(e907)I; dotEX [Pr dyf-5(min - 3 c-box strong)::GFP + dpy-5(+)]* | Minimal *dyf-5* promoter (324bp upstream of start codon to 204bp upstream of start codon [120bp promoter]) with Left, Right and Centre C-boxes mutated (strong mutation) fused to GFP (injected at 50ng/ul) |
| JNC577 | Hermaphrodites | Injection Extrachromasomal | *dpy-5(e907)I; dotEX [Pr dyf-5(min - 4 c-box)::GFP + dpy-5(+)]* | Minimal *dyf-5* promoter (324bp upstream of start codon to 204bp upstream of start codon [120bp promoter]) with Left, Right, Centre and putative 4th C-boxes mutated fused to GFP (injected at 50ng/ul) (1-3) |
| JNC578 | Hermaphrodites | Injection Extrachromasomal | *dpy-5(e907)I; dotEX [Pr dyf-5(min - 4 c-box)::GFP + dpy-5(+)]* | Minimal *dyf-5* promoter (324bp upstream of start codon to 204bp upstream of start codon [120bp promoter]) with Left, Right, Centre and putative 4th C-boxes mutated fused to GFP (injected at 50ng/ul) (2-1) |
| JNC579 | Hermaphrodites | Injection Extrachromasomal | *dpy-5(e907)I; dotEX [Pr dyf-5(min - scramble1)::GFP + dpy-5(+)]* | Scrambled Minimal *dyf-5* promoter (324bp upstream of start codon to 204bp upstream of start codon [120bp promoter]) with only X-box left intact (version 1) fused to GFP (injected at 50ng/ul) (1-13) |
| JNC580 | Hermaphrodites | Injection Extrachromasomal | *dpy-5(e907)I; dotEX [Pr dyf-5(min - scramble1)::GFP + dpy-5(+)]* | Scrambled Minimal *dyf-5* promoter (324bp upstream of start codon to 204bp upstream of start codon [120bp promoter]) with only X-box left intact (version 1) fused to GFP (injected at 50ng/ul) (2-2) |
| JNC581 | Hermaphrodites | Injection Extrachromasomal | *dpy-5(e907)I; dotEX [Pr dyf-5(min - scramble2)::GFP + dpy-5(+)]* | Scrambled Minimal *dyf-5* promoter (324bp upstream of start codon to 204bp upstream of start codon [120bp promoter]) with only X-box left intact (version 2) fused to GFP (injected at 50ng/ul) (3-1) |
| JNC582 | Hermaphrodites | Injection Extrachromasomal | *dpy-5(e907)I; dotEX [Pr dyf-5(min - scramble2)::GFP + dpy-5(+)]* | Scrambled Minimal *dyf-5* promoter (324bp upstream of start codon to 204bp upstream of start codon [120bp promoter]) with only X-box left intact (version 2) fused to GFP (injected at 50ng/ul) (6-4) |
| JNC583 | Hermaphrodites | Injection Extrachromasomal | *dpy-5(e907)I; dotEX [Pr dyf-5(min - scramble3)::GFP + dpy-5(+)]* | Scrambled Minimal *dyf-5* promoter (324bp upstream of start codon to 204bp upstream of start codon [120bp promoter]) with only X-box left intact (version 3) fused to GFP (injected at 50ng/ul) (4-3) |
| JNC584 | Hermaphrodites | Injection Extrachromasomal | *dpy-5(e907)I; dotEX [Pr dyf-5(min - scramble3)::GFP + dpy-5(+)]* | Scrambled Minimal *dyf-5* promoter (324bp upstream of start codon to 204bp upstream of start codon [120bp promoter]) with only X-box left intact (version 3) fused to GFP (injected at 50ng/ul) (5-7) |
| JNC585 | Hermaphrodites | Injection Extrachromasomal | *dpy-5(e907)I; dotEX [Pr dyf-5(min - scramble4)::GFP + dpy-5(+)]* | Scrambled Minimal *dyf-5* promoter (324bp upstream of start codon to 204bp upstream of start codon [120bp promoter]) with only X-box left intact (version 4) fused to GFP (injected at 50ng/ul) (1-7) |
| JNC586 | Hermaphrodites | Injection Extrachromasomal | *dpy-5(e907)I; dotEX [Pr dyf-5(min - scramble4)::GFP + dpy-5(+)]* | Scrambled Minimal *dyf-5* promoter (324bp upstream of start codon to 204bp upstream of start codon [120bp promoter]) with only X-box left intact (version 4) fused to GFP (injected at 50ng/ul) (3-4) |
| JNC587 | Hermaphrodites | Injection Extrachromasomal | *dpy-5(e907)I; dotEX [Pr dyf-5(min - scramble negative)::GFP* + *dpy-5(+)]* | Scrambled Minimal *dyf-5* promoter (324bp upstream of start codon to 204bp upstream of start codon [120bp promoter]) with only X-box also scrambled (negative control) fused to GFP (injected at 50ng/ul) |

## 4.3. Results

### 4.3.1. X-box necessary for expression

In order to determine the function of each motif found in the *dyf-5* promoter, constructs were made with deletions in each motif. The first of these constructs mutated the X-box. This motif was previously been shown to be essential for cilia gene expression (Burghoorn et al. 2012; Swoboda et al. 2000), therefore mutating it is expected to produce no expression.

Mutation of the X-box produced a very dramatic reduction of expression (Figure 4-2, Figure 4-3). However, when exposure and sensitivity is increased, ciliated neuron expression is still observed. Because only two nucleotides of the X-box were changed, it was hypothesised that this mutation was not strong enough to completely abolish expression. A second mutation, term X-box strong mutation, was created to test this hypothesis. This mutation was created by randomising the X-box sequence thereby leaving no part of it intact. This construct was weaker than the original X-box mutation but still retained some cilia expression.

From these results, it can be concluded that the X-box is very important for expression of *dyf-5* but is not solely responsible for the observed pattern of expression. It should be noted that the level of expression observed is very low and likely only visible because of the construct residing in a multi-copy extra-chromosomal array. For this reason, it is unlikely that this level of expression is meaningful to a single copy endogenous gene. It also explains why this expression has not been reported before.

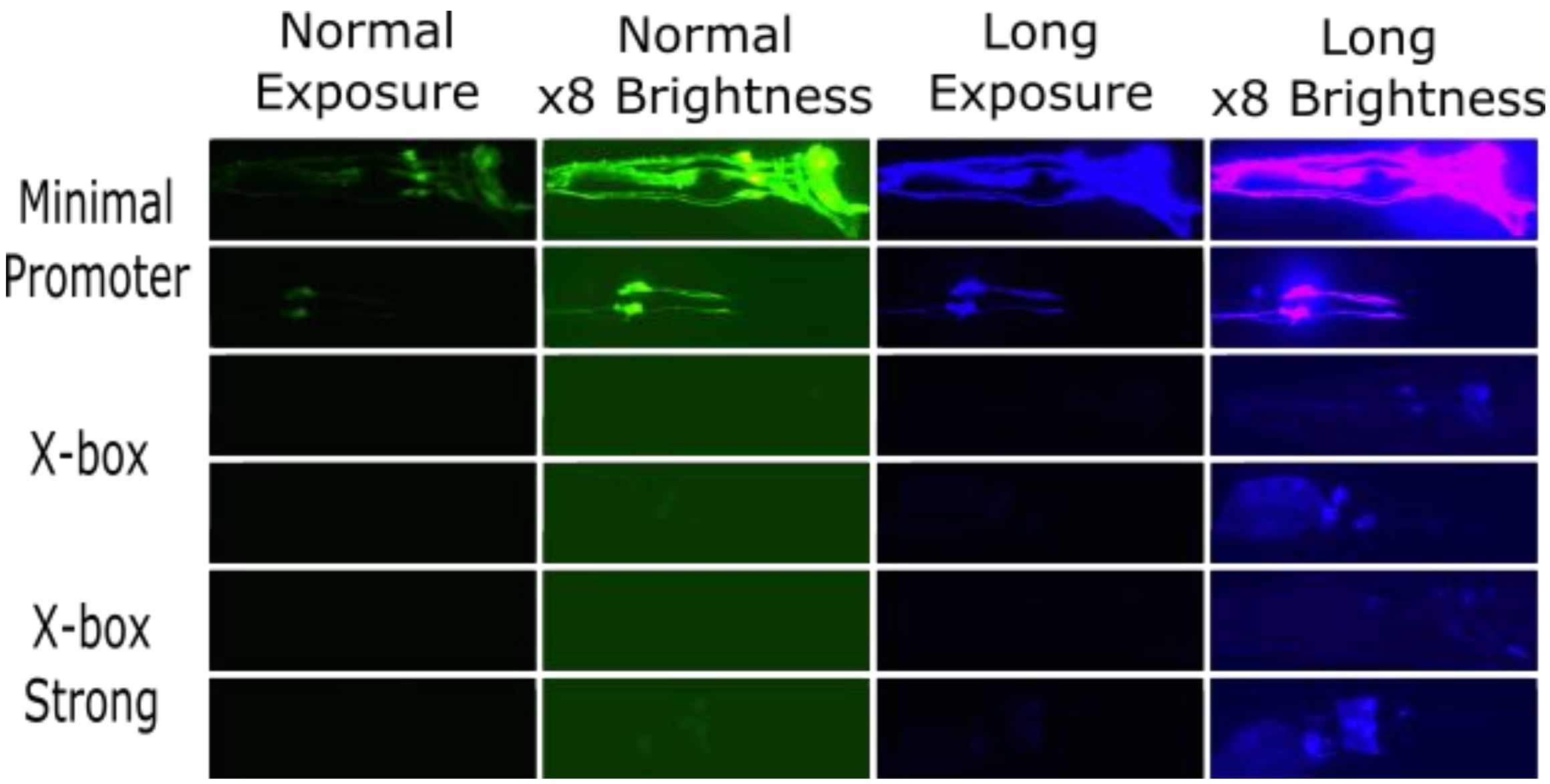

**Figure 4-2. Expression of X-box mutation strain.**
Confocal images of minimal promoter and X-box mutations. Normal exposure and 8x brightness (green). High exposure and 8x brightness (blue). Expression intensity is severely reduced in X-box mutants however expression pattern appears to be maintained.

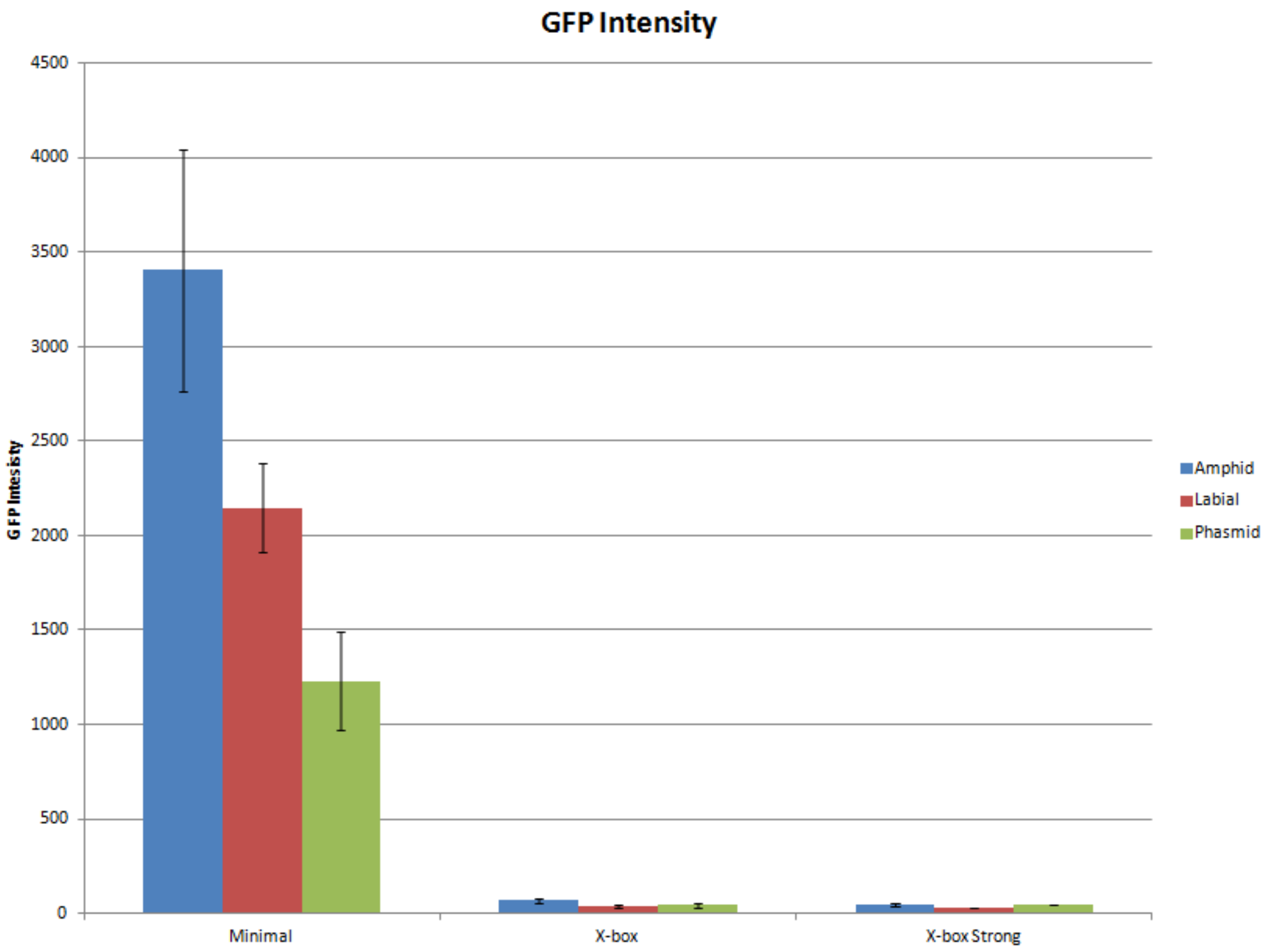

**Figure 4-3. GFP intensities of X-box mutation constructs**
Intensity values X-box mutant constructs obtained from confocal images. X-box mutants show only a fraction of wild-type expression (N=6).

### 4.3.2. Multiple C-boxes function with X-box to drive expression

Next, the role of the C-boxes was determined. By sequentially mutating each C-box and combination of C-boxes, the role they play in expression can be determined. The two C-boxes reported by Burghoorn *et al.* as well as a third putative C-box in between the X-box and the right C-box were tested. This third C-box was discovered by its sequence similarity to the right C-box; it is a mirror image. Due to the low complexity of the reported C-boxes it was suggested that this sequence may also be a functional C-box. It was expected that the three C-boxes would be partially redundant; mutating a single C-box would give a minor reduction in expression whereas mutating multiple C-boxes would give a more severe phenotype.

Mutating each C-box in isolation resulted in a reduction in expression. For the left and right C-boxes, this reduction was quite modest but the centre C-box mutation resulted in a much more dramatic reduction. It is unclear if this is the result of the greater importance of the centre C-box or a result of variability of the extra-chromosomal arrays.

Mutating pairs of C-boxes resulted in further reduction of expression. Mutation of the left and right C-boxes together resulted in expression similar to either the left or right mutated on their own. Contrary to the hypothesis, this suggests that these two C-boxes are not redundant and both are necessary for full expression. Mutation of the left and centre or right and centre C-boxes resulted in expression similar to the centre C-box on its own or somewhat weaker. This suggests that the centre C-box has greater importance than the other two. Finally, mutating all three C-boxes results in reduced expression. Interestingly, this expression is on par with the expression observed when the centre C-box was mutated on its own.

Mutation of all three C-boxes drastically reduced expression, however, there was still some cilia specific expression observed. Since it was previously suggested that the X-box does not function on its own (Burghoorn et al. 2012), two hypotheses were suggested. First, the C-box mutations may not represent null mutations thereby allowing some expression. Second, an additional undetected motif may be present. To test the first hypothesis, another construct was created where a stronger mutation was introduced to all three C-boxes. This construct expressed at a level similar to that of the original three C-box mutation. Since the expression was not further reduced, this suggests that the original mutation was strong enough to eliminate C-box function. The second hypothesis was tested in the next section.

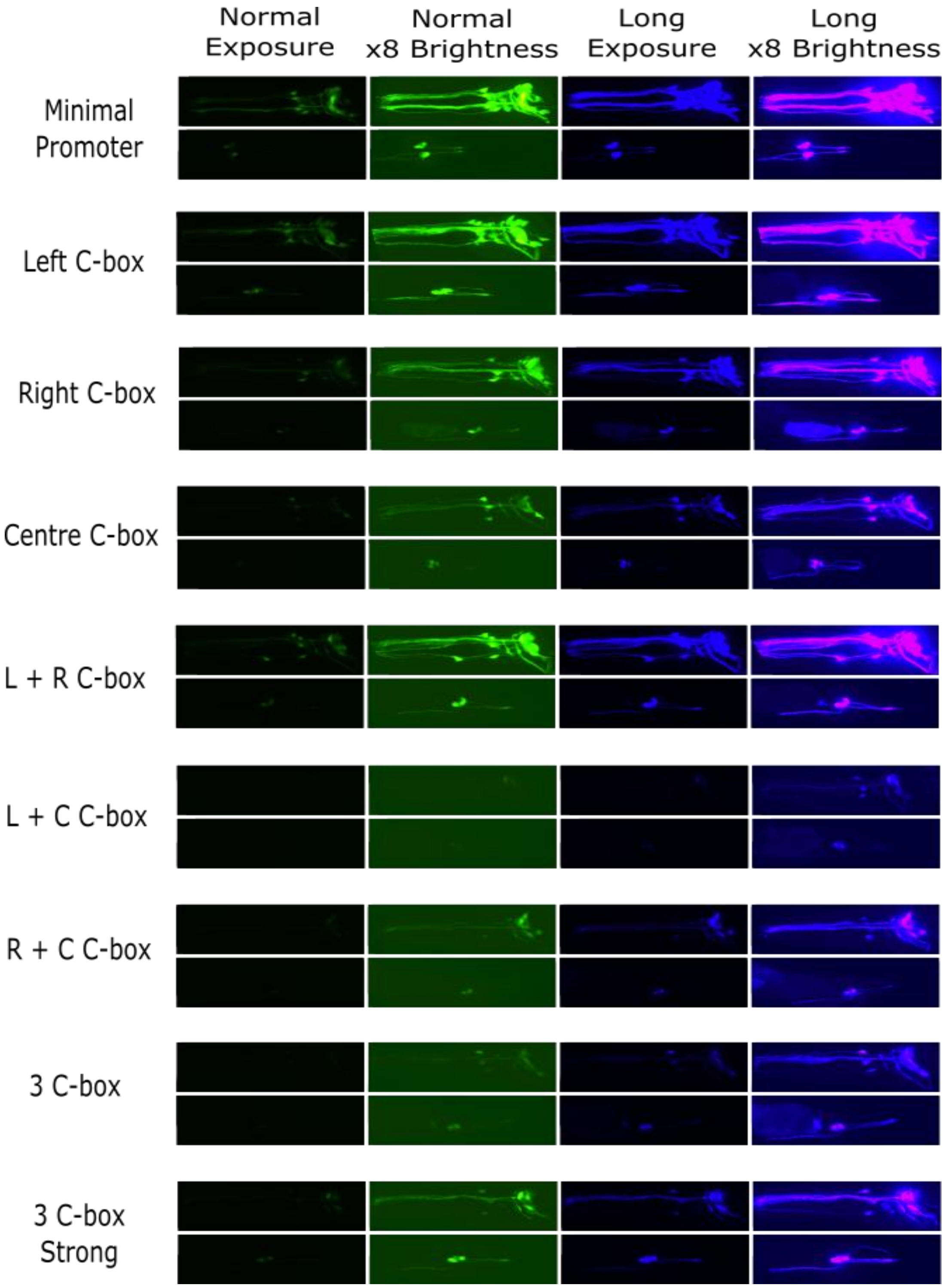

**Figure 4-4.** **Expression of C-box mutation strains.**

Normal exposure and 8x brightness (green). High exposure and 8x brightness (blue). The labels on the left indicate which C-box is mutated. L, R, and C refer to the left, right and centre C-boxes repectively. Left and right C-boxes only have a slight effect on expression. Any construct with a centre C-box mutation has a severe reduction in expression. Expression pattern is maintained in all constructs.

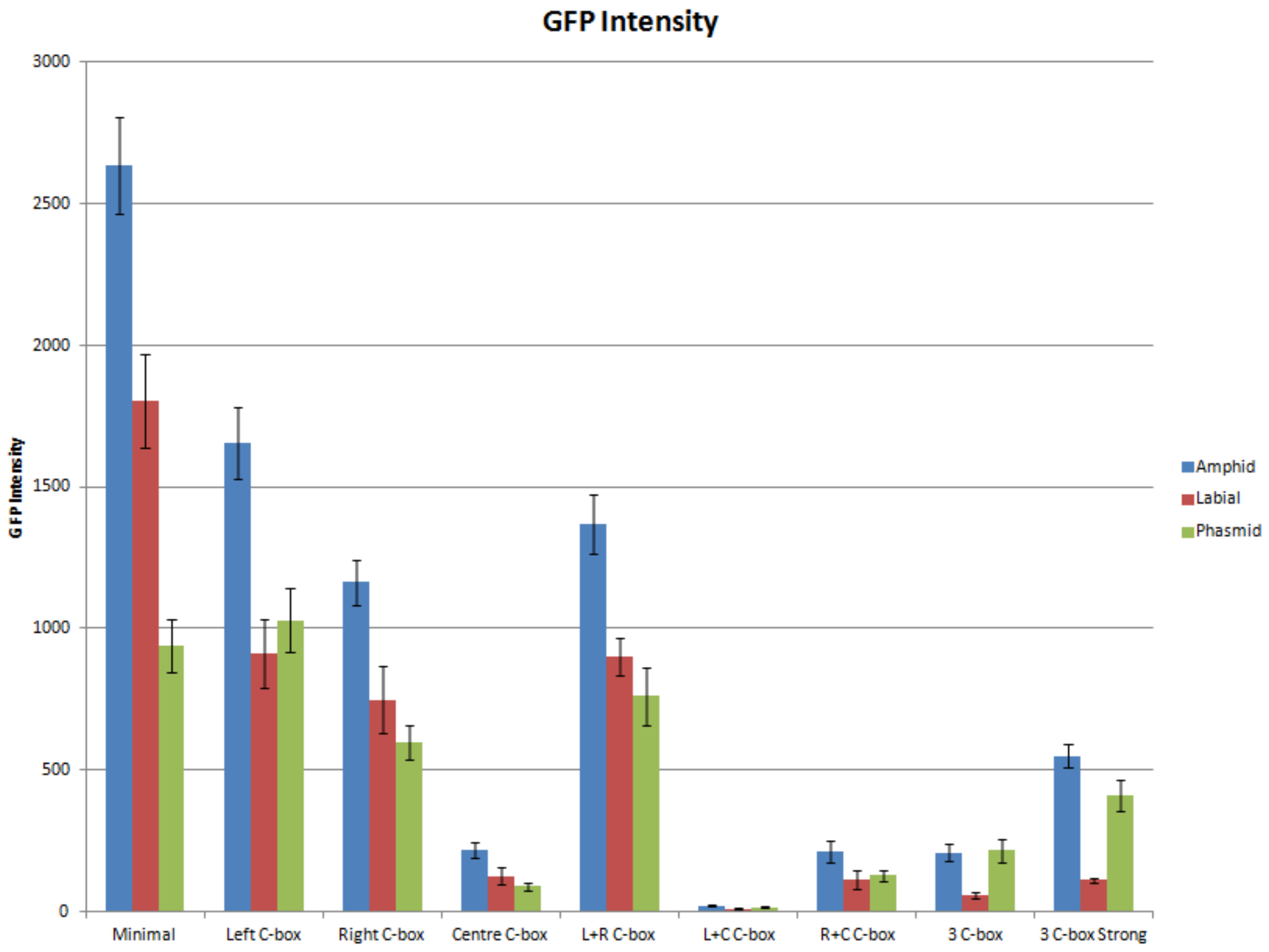

**Figure 4-5. GFP intensities of C-box mutation strains.**
Graph of intensities obtained by analysis of confocal images. Left and Right C-box mutants both result in similar reduction of expression. Centre C-box mutations are similar to constructs containing no C-boxes (N=5).

### 4.3.3. X-box sufficient for minimal expression

To test the hypothesis that an additional promoter element is present in the minimal promoter and responsible for the ciliated neuron expression, a series of "scrambled" constructs were produced. These constructs involve randomising the sequence of the minimal promoter while leaving the X-box intact. To reduce to possibility of reconstituting a C-box, the three C-box mutation construct was used as the starting sequence. Since these constructs were produced at random, four different constructs were produced to account for stochastic formation of functional enhancers or repressors. It was expected that these construct would show less expression than the three C-box mutation if an additional element was present. Alternatively, if the X-box is able to drive a low level of expression on its own, no change in expression would be observed.

Of the four constructs produced, three showed GFP expression similar to that of the three C-box mutation. This suggests that the X-box is able to drive some expression on its own, albeit at a fairly low level. Despite showing as similar level of expression, one of these construct does show a change in pattern, the absence of expression in labial neurons. It is possible that this promoter has formed a labial specific repressor. The final scramble mutation showed a complete absence of expression. This can be explained by the spurious formation of a repressor or change in sequence that results in the disruption of the transcription machinery.

Finally, to show that this low level of expression was driven by the X-box and not the result of some other part of the construct a negative control was produced. This construct is the same as the scramble 3 construct except the X-box has been randomised as well. If the X-box is responsible, no expression should be observed. This is exactly what was observed therefore the X-box is responsible for this expression.

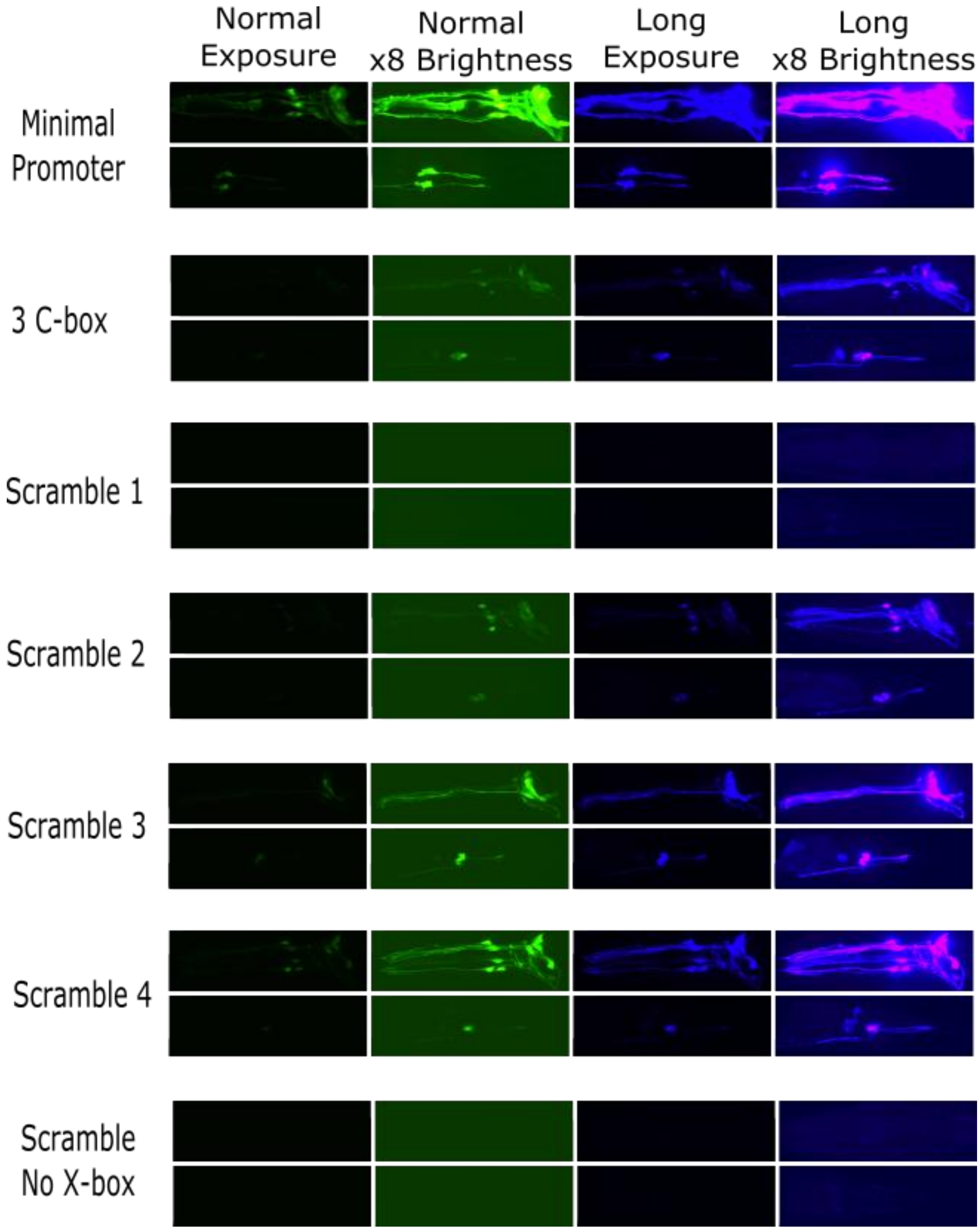

**Figure 4-6. Expression of Scramble constructs.**
Normal exposure and 8x brightness (green). High exposure and 8x brightness (blue). Three of four scamble mutations show similar expression intensity to the 3 C-box mutant construct. This likely represents X-box-only expression. Removal of X-box and C-boxes results in no expression. Scramble 1 shows no expression and Scramble 3 is missing expression in labial neurons.

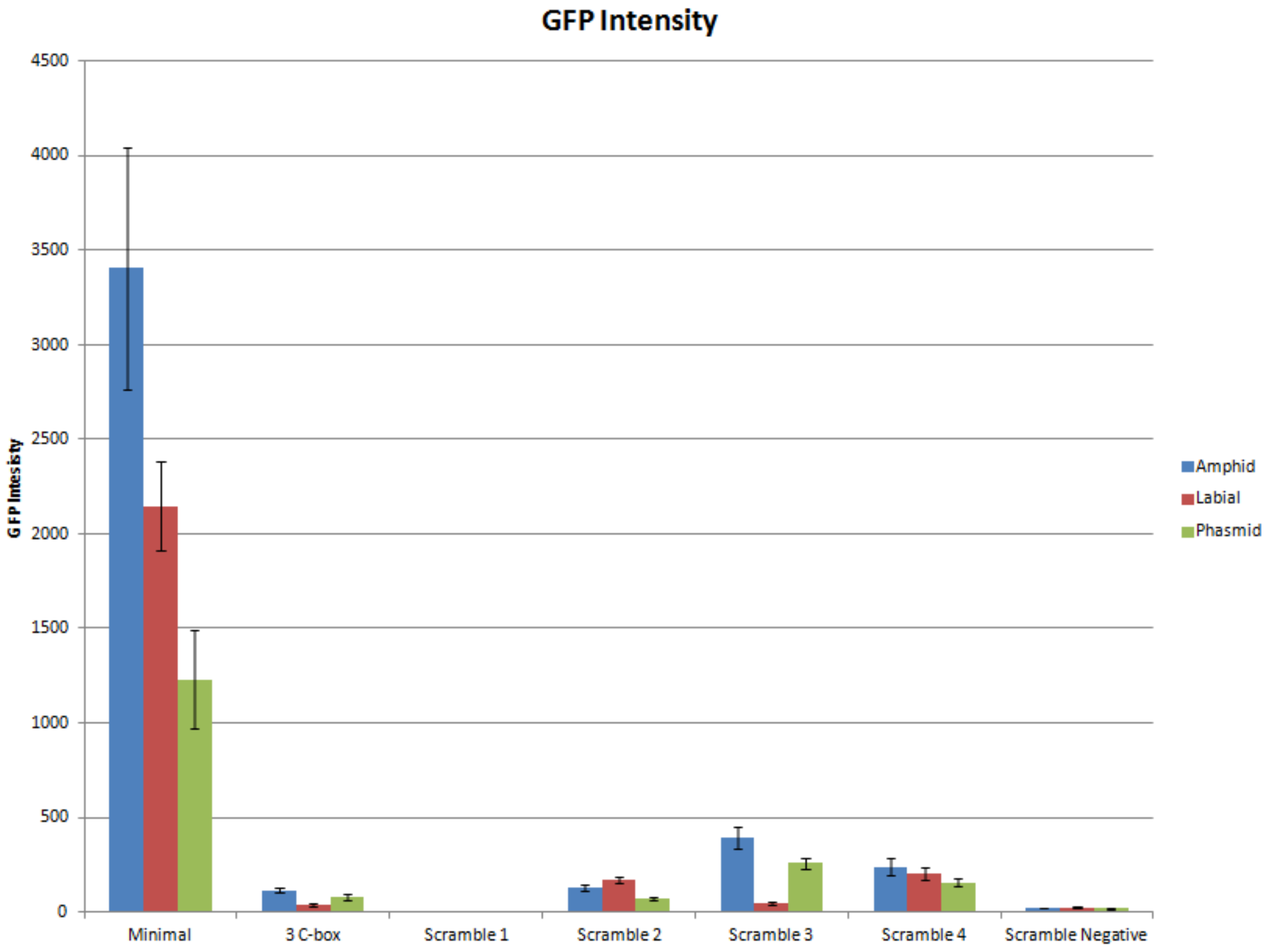

**Figure 4-7. GFP intensity of Scramble Constructs**
Graph of intensities obtained by analysis of confocal images. Intensity of expression in scamble constructs is similar to 3 C-box mutation construct (N=5).

## 4.4. Discussion

I have demonstrated that only the X-box and three C-boxes make up the X-box CRM for *dyf-5*. The X-box is the major regulator as mutating reduces expression quite dramatically. The three C-boxes all function together to increase expression. Interestingly, they aren't redundant as expected. All three are necessary for correct expression as mutating any one C-box results in reduced expression. Another interesting finding is that the three C-boxes are not equal in their effects; the centre C-box has a stronger effect. It is unclear what could account for this as the sequences of the three C-boxes are quite similar. The centre C-box is closer to the X-box however and this might account for its stronger effect. This also brings up the possibility that DAF-19 and the factor that binds the C-box interact physically. A double labelling experiment may have been useful for this work as well as the pattern of expression could be directly compared between “control” and “test” promoters. Some differences are apparent as the axons do not appear to contain *GFP* in many of the strains. This is most likely the result of lower concentrations of *GFP* in the cell as this is associated with the weaker promoters, so the *GFP* in the axons quickly drops below the detection threshold.

Both motifs appear to be able to drive the correct expression pattern on their own albeit weakly. This suggests the possibility that in genes with a more restricted pattern of cilia expression the X-box alone (or possibly C-box alone) would work in conjunction with another factor that shares the same restricted expression pattern.

# Chapter 5. Concluding remarks

## 5.1. Conclusions

The goal of this research was to study how *daf-19* works with other transcription factors by to identifying the cis-regulatory module controlling the broadly expressed cilia gene, *dyf-5*. To that end, I have identified a previously unreported C-box motif in the *dyf-5* promoter as well as demonstrated that *dyf-5* is regulated solely by the X-box and three C-boxes. There are of course some caveats. It is possible that the expression pattern of the altered promoters is not perfectly identical. Small, subtle differences may not have been visible to my visual inspection. A double reporter experiment could address this by allowing a direct comparison of the expression patterns in all cells. Additionally, it is possible that there are additional elements involved in the expression of *dyf-5* that were not detected. For example, shadow enhancers have been reported that drive expression in the same manner as the promoter proximal elements and therefore provide redundancy under stress (Barolo 2012; Hobert 2010). These could not be detected by my approach. Techniques, such as CRISPR, that can delete or mutate parts of the promoter while maintaining its genomic context would be required to detect them. Finally, this study only addresses the pan-ciliary expression of *dyf-5.* While, it is expected there are other genes that are expressed in a similar way, there is no guarantee the genes with similar expression patterns are all expressed this way.

I have also provided some new evidence on how C-boxes function. For example, the three C-boxes do not have an equal effect on expression and the C-box closest to the X-box was most important for expression. This suggests that distance between X-box and C-box is critical and hints that DAF-19 may physically interact with the protein that binds the C-box. Additionally, there were hints that the C-box alone may be able to specify ciliary expression, suggesting that, in conjunction with other factors, the C-box may be able to drive *daf-19* independent ciliary expression.

Although the exact regulatory machinery is likely not conserved between *C. elegans* and humans, this study gives us a new understanding of how ciliopathy genes may be regulated and opens up new avenues for exploring the question deeper. For example, ciliopathy genes in humans may be regulated in a similar manner albeit with different machinery. The more we understand cilia in one species the more able we can understand cilia and ciliopathies in general.

## 5.2. Future directions

### 5.2.1. Characterisation of the C-box

There are a number of experiments that could be performed to help characterise the C-box. First, the C-box is relatively difficult to identify computationally due to its low sequence complexity. It is possible that the sequence has more complexity than is readily apparent, as the exact sequence of a functional C-box has not been extensively studied. Therefore, experiments to mutate individual bases of the C-box could be undertaken to determine the specific pattern of bases necessary and possible protein contacts. Second, the C-box has not been looked at outside of its native context. It is quite possible that specific spacings and arrangements of C-boxes within a promoter are necessary for its function. This could be tested by moving C-boxes relative to each other and the X-box and measure the effect on transcription of a reporter gene. It will also be interesting to characterise the genes driven by C-box. The C-box seems to drive pan-ciliary expression and it will be interesting to see if all genes driven by C-box are pan-ciliary. It is certainly possible that a promoter containing both X-box and C-box could also contain cell specific repressors thus restricting expression of the gene. The lack of labial neuron expression in the Scramble 3 construct suggests this is possible. It will also be interesting to characterise the C-boxes from other species. My results strongly suggest other *Caenorhabditis* species make use of C-boxes but it is unclear if they are conserved in more distantly related species.

### 5.2.2. Identification of transcription factors

The most important next step in this research is to identify the protein that binds the C-box. Another student in our lab, Zhaozhao Qin, is working on this very problem. Her research focuses on a mutant screen seeking to identify mutants that effect expression of a reporter gene driven by a *dyf-5* promoter. To date, she has isolated a candidate mutation and putatively identified it as a transcription factor gene in the forkhead box family. There is still currently no evidence that this gene interacts with the C-box. However, this could be achieved through experiments such as EMSA.

The standard model of gene transcription hypothesises that the C-box is a TFBS and therefore must be bound by a TF. An intriguing alternative hypothesis is that the C-box is actually just a structural feature of the DNA that allows transcription to be activated solely by the X-box. For example, the sequence may facilitate DNA adopting a structure conducive to transcription, such as bending or unwinding the DNA, or more unusual conformation such as Z or H form DNA (Dai and Rothman-Denes 1999). Alternatively, the sequence may influence chromatin structure in the region like T-blocks which are hypothesised to exclude nucleosomes from the region (Grishkevich et al. 2011). The CT rich C-box may allow DNA to melt more easily permitting transcription. Two pieces of evidence hint at this possibility: the orientations of the C-boxes, and the complexity of the C-box. First, the centre C-box, which is closest to the X-box and almost immediately downstream, is most critical for expression whereas the others have only modest effect. This would make sense if transcription initiation happens in the vicinity of the centre C-box, due to its position this C-box would be required whereas the others would simply aid in the opening of the DNA helix. Second, the C-box has very low sequence complexity. This is not very ideal for a TFBS because the sequence could be formed quite readily by random chance and cause off target binding. Interestingly, polypurine/polypyrimidine sequences have been previously shown to function as cis-acting transcriptional regulators although most appear to be repressive rather than activating (Brahmachari et al. 1997). There is at least one exception, however (Zahedi et al. 1999). It has been suggested these polypurine/polypyrimidine sequences function by forming a triple helix structure by forming Hoogsteen base pairs which can then potentially be bound by proteins (Duca et al. 2008; Buske et al. 2011). In order for these to form and mirror image repeat of the polypurine/polypyrimidine must be nearby. This is actually seen in the orientation of the centre and right C-boxes in *C. elegans dyf-5* and in the orthologs of several species. Unfortunately, if this was the case, one would predict both the centre and right C-boxes to show an equally strong effect, which was not observed.

### 5.2.3. Identification of ciliary genes

By understanding the motifs responsible for cilia gene expression, new cilia genes can be identified by their motifs. This strategy has previously been successful using the X-box (Efimenko et al. 2005; Chen et al. 2006). By searching for C-boxes or X-box and C-boxes together additional ciliary genes could be uncovered. For example, X-boxes and C-boxes together appear to identify pan-ciliary expressed genes (Burghoorn et al. 2012). Therefore, searching for this signature could allow more pan ciliary genes to be identified. Additionally, expression pattern of uncharacterised genes could be predicted by looking for this signature. Alternatively, it is possible some ciliary genes possess a C-box but no X-box. By looking for C-boxes additional ciliary genes could be uncovered.

# References

Aftab S, Semenec L, Chu JS-C, Chen N. 2008. Identification and characterization of novel human tissue-specific RFX transcription factors. *BMC Evol Biol* **8**: 226.

Ainsworth C. 2007. Cilia: tails of the unexpected. *Nature* **448**: 638–41.

Aldahmesh MA, Li Y, Alhashem A, Anazi S, Alkuraya H, Hashem M, Awaji AA, Sogaty S, Alkharashi A, Alzahrani S, et al. 2014. IFT27, encoding a small GTPase component of IFT particles, is mutated in a consanguineous family with Bardet-Biedl syndrome. *Hum Mol Genet* **23**: 3307–15.

Ansley SJ, Badano JL, Blacque OE, Hill J, Hoskins BE, Leitch CC, Kim JC, Ross AJ, Eichers ER, Teslovich TM, et al. 2003. Basal body dysfunction is a likely cause of pleiotropic Bardet-Biedl syndrome. *Nature* **425**: 628–33.

Ao W, Gaudet J, Kent WJ, Muttumu S, Mango SE. 2004. Environmentally induced foregut remodeling by PHA-4/FoxA and DAF-12/NHR. *Science* **305**: 1743–6.

Avasthi P, Marshall WF. 2012. Stages of ciliogenesis and regulation of ciliary length. *Differentiation* **83**: S30–42.

Bacaj T, Lu Y, Shaham S. 2008. The conserved proteins CHE-12 and DYF-11 are required for sensory cilium function in Caenorhabditis elegans. *Genetics* **178**: 989–1002.

Badano JL, Mitsuma N, Beales PL, Katsanis N. 2006. The ciliopathies: an emerging class of human genetic disorders. *Annu Rev Genomics Hum Genet* **7**: 125–48.

Badis G, Berger MF, Philippakis AA, Talukder S, Gehrke AR, Jaeger SA, Chan ET, Metzler G, Vedenko A, Chen X, et al. 2009. Diversity and complexity in DNA recognition by transcription factors. *Science* **324**: 1720–3.

Bae Y-K, Barr MM. 2008. Sensory roles of neuronal cilia: cilia development, morphogenesis, and function in C. elegans. *Front Biosci* **13**: 5959–74.

Baldari CT, Rosenbaum J. 2010. Intraflagellar transport: it's not just for cilia anymore. *Curr Opin Cell Biol* **22**: 75–80.

Barolo S. 2012. Shadow enhancers: frequently asked questions about distributed cis-regulatory information and enhancer redundancy. *Bioessays* **34**: 135–41.

Barrière A, Gordon KL, Ruvinsky I. 2012. Coevolution within and between regulatory loci can preserve promoter function despite evolutionary rate acceleration. *PLoS Genet* **8**: e1002961.

Barrière A, Yang S-P, Pekarek E, Thomas CG, Haag ES, Ruvinsky I. 2009. Detecting heterozygosity in shotgun genome assemblies: Lessons from obligately outcrossing nematodes. *Genome Res* **19**: 470–80.

Bektesh SL, Hirsh DI. 1988. C.elegans mRNAs acquire a spliced leader through a trans -splicing mechanism. *Nucleic Acids Res* **16**: 5692–5692.

Bell LR, Stone S, Yochem J, Shaw JE, Herman RK. 2006. The molecular identities of the Caenorhabditis elegans intraflagellar transport genes dyf-6, daf-10 and osm-1. *Genetics* **173**: 1275–86.

Berman SA, Wilson NF, Haas NA, Lefebvre PA. 2003. A novel MAP kinase regulates flagellar length in Chlamydomonas. *Curr Biol* **13**: 1145–9.

Blackwood EM, Kadonaga JT. 1998. Going the distance: a current view of enhancer action. *Science* **281**: 60–3.

Blacque OE, Perens EA, Boroevich KA, Inglis PN, Li C, Warner A, Khattra J, Holt RA, Ou G, Mah AK, et al. 2005. Functional genomics of the cilium, a sensory organelle. *Curr Biol* **15**: 935–41.

Blacque OE, Reardon MJ, Li C, McCarthy J, Mahjoub MR, Ansley SJ, Badano JL, Mah AK, Beales PL, Davidson WS, et al. 2004. Loss of C. elegans BBS-7 and BBS-8 protein function results in cilia defects and compromised intraflagellar transport. *Genes Dev* **18**: 1630–42.

Brahmachari SK, Sarkar PS, Raghavan S, Narayan M, Maiti AK. 1997. Polypurine/polypyrimidine sequences as cis-acting transcriptional regulators. *Gene* **190**: 17–26.

Bredrup C, Saunier S, Oud MM, Fiskerstrand T, Hoischen A, Brackman D, Leh SM,

Midtbø M, Filhol E, Bole-Feysot C, et al. 2011. Ciliopathies with skeletal anomalies and renal insufficiency due to mutations in the IFT-A gene WDR19. *Am J Hum Genet* **89**: 634–43.

Brenner S. 1974. The genetics of Caenorhabditis elegans. *Genetics* **77**: 71–94.

Brody SL, Yan XH, Wuerffel MK, Song SK, Shapiro SD. 2000. Ciliogenesis and left-right axis defects in forkhead factor HFH-4-null mice. *Am J Respir Cell Mol Biol* **23**: 45–51.

Brown JM, Witman GB. 2014. Cilia and Diseases. *Bioscience* **64**: 1126–1137.

Burghoorn J, Dekkers MPJ, Rademakers S, de Jong T, Willemsen R, Jansen G. 2007. Mutation of the MAP kinase DYF-5 affects docking and undocking of kinesin-2 motors and reduces their speed in the cilia of Caenorhabditis elegans. *Proc Natl Acad Sci U S A* **104**: 7157–62.

Burghoorn J, Piasecki BP, Crona F, Phirke P, Jeppsson KE, Swoboda P. 2012. The in vivo dissection of direct RFX-target gene promoters in C. elegans reveals a novel cis-regulatory element, the C-box. *Dev Biol* **368**: 415–26.

Buske FA, Mattick JS, Bailey TL. 2011. Potential in vivo roles of nucleic acid triple-helices. *RNA Biol* **8**: 427–39.

Cameron RA, Chow SH, Berney K, Chiu T-Y, Yuan Q-A, Krämer A, Helguero A, Ransick A, Yun M, Davidson EH. 2005. An evolutionary constraint: strongly disfavored class of change in DNA sequence during divergence of cis-regulatory modules. *Proc Natl Acad Sci U S A* **102**: 11769–74.

Chen J, Knowles HJ, Hebert JL, Hackett BP. 1998. Mutation of the mouse hepatocyte nuclear factor/forkhead homologue 4 gene results in an absence of cilia and random left-right asymmetry. *J Clin Invest* **102**: 1077–82.

Chen N, Mah A, Blacque OE, Chu J, Phgora K, Bakhoum MW, Newbury CRH, Khattra J, Chan S, Go A, et al. 2006. Identification of ciliary and ciliopathy genes in Caenorhabditis elegans through comparative genomics. *Genome Biol* **7**: R126.

Chen RA-J, Down TA, Stempor P, Chen QB, Egelhofer TA, Hillier LW, Jeffers TE, Ahringer J. 2013. The landscape of RNA polymerase II transcription initiation in C. elegans reveals promoter and enhancer architectures. *Genome Res* **23**: 1339–47.

Choksi SP, Lauter G, Swoboda P, Roy S. 2014. Switching on cilia: transcriptional networks regulating ciliogenesis. *Development* **141**: 1427–41.

Chu JSC, Baillie DL, Chen N. 2010. Convergent evolution of RFX transcription factors and ciliary genes predated the origin of metazoans. *BMC Evol Biol* **10**: 130.

Chu JSC, Tarailo-Graovac M, Zhang D, Wang J, Uyar B, Tu D, Trinh J, Baillie DL, Chen N. 2012. Fine tuning of RFX/DAF-19-regulated target gene expression through binding to multiple sites in Caenorhabditis elegans. *Nucleic Acids Res* **40**: 53–64.

Collet J, Spike CA, Lundquist EA, Shaw JE, Herman RK. 1998. Analysis of osm-6, a gene that affects sensory cilium structure and sensory neuron function in Caenorhabditis elegans. *Genetics* **148**: 187–200.

Dai X, Rothman-Denes LB. 1999. DNA structure and transcription. *Curr Opin Microbiol* **2**: 126–30.

Davidson EH. 2009. Network design principles from the sea urchin embryo. *Curr Opin Genet Dev* **19**: 535–40.

Davidson EH, Erwin DH. 2006. Gene regulatory networks and the evolution of animal body plans. *Science* **311**: 796–800.

Davidson EH, Levine MS. 2008. Properties of developmental gene regulatory networks. *Proc Natl Acad Sci U S A* **105**: 20063–6.

Davis EE, Katsanis N. 2012. The ciliopathies: a transitional model into systems biology of human genetic disease. *Curr Opin Genet Dev* **22**: 290–303.

Dawe AL, Caldwell KA, Harris PM, Morris NR, Caldwell GA. 2001. Evolutionarily conserved nuclear migration genes required for early embryonic development in Caenorhabditis elegans. *Dev Genes Evol* **211**: 434–41.

Drummond IA. 2012. Cilia functions in development. *Curr Opin Cell Biol* **24**: 24–30.

Duca M, Vekhoff P, Oussedik K, Halby L, Arimondo PB. 2008. The triple helix: 50 years later, the outcome. *Nucleic Acids Res* **36**: 5123–38.

Dwyer ND, Troemel ER, Sengupta P, Bargmann CI. 1998. Odorant receptor localization to olfactory cilia is mediated by ODR-4, a novel membrane-associated protein. *Cell* **93**: 455–66.

Efimenko E, Blacque OE, Ou G, Haycraft CJ, Yoder BK, Scholey JM, Leroux MR, Swoboda P. 2006. Caenorhabditis elegans DYF-2, an orthologue of human WDR19, is a component of the intraflagellar transport machinery in sensory cilia. *Mol Biol Cell* **17**: 4801–11.

Efimenko E, Bubb K, Mak HY, Holzman T, Leroux MR, Ruvkun G, Thomas JH, Swoboda P. 2005. Analysis of xbx genes in C. elegans. *Development* **132**: 1923–34.

Emery P, Durand B, Mach B, Reith W. 1996. RFX proteins, a novel family of DNA binding proteins conserved in the eukaryotic kingdom. *Nucleic Acids Res* **24**: 803–7.

Etchberger JF, Hobert O. 2008. Vector-free DNA constructs improve transgene expression in C. elegans. *Nat Methods* **5**: 3.

Etchberger JF, Lorch A, Sleumer MC, Zapf R, Jones SJ, Marra MA, Holt RA, Moerman DG, Hobert O. 2007. The molecular signature and cis-regulatory architecture of a C. elegans gustatory neuron. *Genes Dev* **21**: 1653–74.

Finetti F, Paccani SR, Rosenbaum J, Baldari CT. 2011. Intraflagellar transport: a new player at the immune synapse. *Trends Immunol* **32**: 139–45.

Finn RD, Clements J, Eddy SR. 2011. HMMER web server: interactive sequence similarity searching. *Nucleic Acids Res* **39**: W29–37.

Follit JA, San Agustin JT, Jonassen JA, Huang T, Rivera-Perez JA, Tremblay KD, Pazour GJ. 2014. Arf4 is required for Mammalian development but dispensable for ciliary assembly. *PLoS Genet* **10**: e1004170.

Forsythe E, Beales PL. 2013. Bardet-Biedl syndrome. *Eur J Hum Genet* **21**: 8–13.

Freund CL, Gregory-Evans CY, Furukawa T, Papaioannou M, Looser J, Ploder L, Bellingham J, Ng D, Herbrick JA, Duncan A, et al. 1997. Cone-rod dystrophy due to mutations in a novel photoreceptor-specific homeobox gene (CRX) essential for maintenance of the photoreceptor. *Cell* **91**: 543–53.

Frøkjær-Jensen C. 2013. Exciting prospects for precise engineering of Caenorhabditis elegans genomes with CRISPR/Cas9. *Genetics* **195**: 635–42.

Frøkjær-Jensen C, Davis MW, Ailion M, Jorgensen EM. 2012. Improved Mos1-mediated

transgenesis in C. elegans. *Nat Methods* **9**: 117–8.

Frøkjaer-Jensen C, Davis MW, Hopkins CE, Newman BJ, Thummel JM, Olesen S-P, Grunnet M, Jorgensen EM. 2008. Single-copy insertion of transgenes in Caenorhabditis elegans. *Nat Genet* **40**: 1375–83.

Fujiwara M, Ishihara T, Katsura I. 1999. A novel WD40 protein, CHE-2, acts cell-autonomously in the formation of C. elegans sensory cilia. *Development* **126**: 4839–48.

Gaudet J, Mango SE. 2002. Regulation of organogenesis by the Caenorhabditis elegans FoxA protein PHA-4. *Science* **295**: 821–5.

Gaudet J, Muttumu S, Horner M, Mango SE. 2004. Whole-genome analysis of temporal gene expression during foregut development. *PLoS Biol* **2**: e352.

Gerstein MB, Lu ZJ, Van Nostrand EL, Cheng C, Arshinoff BI, Liu T, Yip KY, Robilotto R, Rechtsteiner A, Ikegami K, et al. 2010. Integrative analysis of the Caenorhabditis elegans genome by the modENCODE project. *Science* **330**: 1775–87.

Goetz SC, Anderson K V. 2010. The primary cilium: a signalling centre during vertebrate development. *Nat Rev Genet* **11**: 331–44.

Grishkevich V, Hashimshony T, Yanai I. 2011. Core promoter T-blocks correlate with gene expression levels in C. elegans. *Genome Res* **21**: 707–17.

Guelen L, Pagie L, Brasset E, Meuleman W, Faza MB, Talhout W, Eussen BH, de Klein A, Wessels L, de Laat W, et al. 2008. Domain organization of human chromosomes revealed by mapping of nuclear lamina interactions. *Nature* **453**: 948–51.

Guenther CA, Tasic B, Luo L, Bedell MA, Kingsley DM. 2014. A molecular basis for classic blond hair color in Europeans. *Nat Genet* **46**: 748–52.

Hao L, Scholey JM. 2009. Intraflagellar transport at a glance. *J Cell Sci* **122**: 889–92.

Hartmann H, Guthöhrlein EW, Siebert M, Luehr S, Söding J. 2013. P-value-based regulatory motif discovery using positional weight matrices. *Genome Res* **23**: 181–94.

Haycraft CJ, Schafer JC, Zhang Q, Taulman PD, Yoder BK. 2003. Identification of CHE-13, a novel intraflagellar transport protein required for cilia formation. *Exp Cell Res*

**284**: 251–63.

Haycraft CJ, Swoboda P, Taulman PD, Thomas JH, Yoder BK. 2001. The C. elegans homolog of the murine cystic kidney disease gene Tg737 functions in a ciliogenic pathway and is disrupted in osm-5 mutant worms. *Development* **128**: 1493–505.

Hellman LM, Fried MG. 2007. Electrophoretic mobility shift assay (EMSA) for detecting protein-nucleic acid interactions. *Nat Protoc* **2**: 1849–61.

Higgins DG, Thompson JD, Gibson TJ. 1996. Using CLUSTAL for multiple sequence alignments. *Methods Enzymol* **266**: 383–402.

Hildebrandt F, Benzing T, Katsanis N. 2011. Ciliopathies. *N Engl J Med* **364**: 1533–43.

Hildebrandt F, Zhou W. 2007. Nephronophthisis-associated ciliopathies. *J Am Soc Nephrol* **18**: 1855–71.

Hobert O. 2008. Gene regulation by transcription factors and microRNAs. *Science* **319**: 1785–6.

Hobert O. 2010. Gene regulation: enhancers stepping out of the shadow. *Curr Biol* **20**: R697–9.

Hobert O. 2002. PCR fusion-based approach to create reporter gene constructs for expression analysis in transgenic C. elegans. *Biotechniques* **32**: 728–30.

Huber C, Cormier-Daire V. 2012. Ciliary disorder of the skeleton. *Am J Med Genet C Semin Med Genet* **160C**: 165–74.

Inglis PN, Ou G, Leroux MR, Scholey JM. 2007. The sensory cilia of Caenorhabditis elegans. *WormBook* 1–22.

Janky R, van Helden J. 2008. Evaluation of phylogenetic footprint discovery for predicting bacterial cis-regulatory elements and revealing their evolution. *BMC Bioinformatics* **9**: 37.

Jeziorska DM, Jordan KW, Vance KW. 2009. A systems biology approach to understanding cis-regulatory module function. *Semin Cell Dev Biol* **20**: 856–62.

Kagoshima H, Kohara Y. 2015. Co-expression of the transcription factors CEH-14 and TTX-1 regulates AFD neuron-specific genes gcy-8 and gcy-18 in C. elegans. *Dev*

*Biol* **399**: 325–36.

Kaletta T, Hengartner MO. 2006. Finding function in novel targets: C. elegans as a model organism. *Nat Rev Drug Discov* **5**: 387–98.

Katara P, Grover A, Sharma V. 2012. Phylogenetic footprinting: a boost for microbial regulatory genomics. *Protoplasma* **249**: 901–7.

Kato M, Hata N, Banerjee N, Futcher B, Zhang MQ. 2004. Identifying combinatorial regulation of transcription factors and binding motifs. *Genome Biol* **5**: R56.

Kessel RG, Kardon RH. 1979. *Tissues and Organs: A Text-atlas of Scanning Electron Microscopy*. W. H. Freeman.

Kiontke K, Fitch DHA. 2005. The phylogenetic relationships of Caenorhabditis and other rhabditids. *WormBook* 1–11.

Kiontke KC, Félix M-A, Ailion M, Rockman M V, Braendle C, Pénigault J-B, Fitch DHA. 2011. A phylogeny and molecular barcodes for Caenorhabditis, with numerous new species from rotting fruits. *BMC Evol Biol* **11**: 339.

Knowles MR, Daniels LA, Davis SD, Zariwala MA, Leigh MW. 2013. Primary ciliary dyskinesia. Recent advances in diagnostics, genetics, and characterization of clinical disease. *Am J Respir Crit Care Med* **188**: 913–22.

Lanjuin A, Sengupta P. 2004. Specification of chemosensory neuron subtype identities in Caenorhabditis elegans. *Curr Opin Neurobiol* **14**: 22–30.

Lee B-K, Iyer VR. 2012. Genome-wide studies of CCCTC-binding factor (CTCF) and cohesin provide insight into chromatin structure and regulation. *J Biol Chem* **287**: 30906–13.

Lee JE, Gleeson JG. 2011. A systems-biology approach to understanding the ciliopathy disorders. *Genome Med* **3**: 59.

Levine M, Tjian R. 2003. Transcription regulation and animal diversity. *Nature* **424**: 147–51.

Li E, Davidson EH. 2009. Building developmental gene regulatory networks. *Birth Defects Res C Embryo Today* **87**: 123–30.

Li JB, Gerdes JM, Haycraft CJ, Fan Y, Teslovich TM, May-Simera H, Li H, Blacque OE, Li L, Leitch CC, et al. 2004. Comparative genomics identifies a flagellar and basal body proteome that includes the BBS5 human disease gene. *Cell* **117**: 541–52.

Liefooghe A, Touzet H, Varré J-S. 2006. Large Scale Matching for Position Weight Matrices. In *Combinatorial Pattern Matching*, Vol. 4009 of, pp. 401–412.

Mak HY, Nelson LS, Basson M, Johnson CD, Ruvkun G. 2006. Polygenic control of Caenorhabditis elegans fat storage. *Nat Genet* **38**: 363–8.

Mello CC, Kramer JM, Stinchcomb D, Ambros V. 1991. Efficient gene transfer in C.elegans: extrachromosomal maintenance and integration of transforming sequences. *EMBO J* **10**: 3959–70.

Mukhopadhyay S, Lu Y, Qin H, Lanjuin A, Shaham S, Sengupta P. 2007. Distinct IFT mechanisms contribute to the generation of ciliary structural diversity in C. elegans. *EMBO J* **26**: 2966–80.

Murayama T, Toh Y, Ohshima Y, Koga M. 2005. The dyf-3 gene encodes a novel protein required for sensory cilium formation in Caenorhabditis elegans. *J Mol Biol* **346**: 677–87.

Nakagawa S, Gisselbrecht SS, Rogers JM, Hartl DL, Bulyk ML. 2013. DNA-binding specificity changes in the evolution of forkhead transcription factors. *Proc Natl Acad Sci U S A* **110**: 12349–54.

Nam J, Dong P, Tarpine R, Istrail S, Davidson EH. 2010. Functional cis-regulatory genomics for systems biology. *Proc Natl Acad Sci U S A* **107**: 3930–5.

Narasimhan K, Lambert SA, Yang AWH, Riddell J, Mnaimneh S, Zheng H, Albu M, Najafabadi HS, Reece-Hoyes JS, Fuxman Bass JI, et al. 2015. Mapping and analysis of Caenorhabditis elegans transcription factor sequence specificities. *Elife* **4**.

Natoli G, Andrau J-C. 2012. Noncoding transcription at enhancers: general principles and functional models. *Annu Rev Genet* **46**: 1–19.

Neph S, Vierstra J, Stergachis AB, Reynolds AP, Haugen E, Vernot B, Thurman RE, John S, Sandstrom R, Johnson AK, et al. 2012. An expansive human regulatory lexicon encoded in transcription factor footprints. *Nature* **489**: 83–90.

Newton FG, zur Lage PI, Karak S, Moore DJ, Göpfert MC, Jarman AP. 2012. Forkhead transcription factor Fd3F cooperates with Rfx to regulate a gene expression program for mechanosensory cilia specialization. *Dev Cell* **22**: 1221–33.

Nokes EB, Van Der Linden AM, Winslow C, Mukhopadhyay S, Ma K, Sengupta P. 2009. Cis-regulatory mechanisms of gene expression in an olfactory neuron type in Caenorhabditis elegans. *Dev Dyn* **238**: 3080–92.

O'Callaghan C, Sikand K, Rutman A. 1999. Respiratory and Brain Ependymal Ciliary Function. *Pediatr Res* **46**: 704–704.

Okkema PG, Krause M. 2005. Transcriptional regulation. *WormBook* 1–40.

Ou G, Koga M, Blacque OE, Murayama T, Ohshima Y, Schafer JC, Li C, Yoder BK, Leroux MR, Scholey JM. 2007. Sensory ciliogenesis in Caenorhabditis elegans: assignment of IFT components into distinct modules based on transport and phenotypic profiles. *Mol Biol Cell* **18**: 1554–69.

Ou G, Qin H, Rosenbaum JL, Scholey JM. 2005. The PKD protein qilin undergoes intraflagellar transport. *Curr Biol* **15**: R410–1.

Ozgül RK, Siemiatkowska AM, Yücel D, Myers CA, Collin RWJ, Zonneveld MN, Beryozkin A, Banin E, Hoyng CB, van den Born LI, et al. 2011. Exome sequencing and cis-regulatory mapping identify mutations in MAK, a gene encoding a regulator of ciliary length, as a cause of retinitis pigmentosa. *Am J Hum Genet* **89**: 253–64.

Praitis V, Maduro MF. 2011. Transgenesis in C. elegans. *Methods Cell Biol* **106**: 161–85.

Qin H, Rosenbaum JL, Barr MM. 2001. An autosomal recessive polycystic kidney disease gene homolog is involved in intraflagellar transport in C. elegans ciliated sensory neurons. *Curr Biol* **11**: 457–61.

Reidling JC, Said HM. 2003. In vitro and in vivo characterization of the minimal promoter region of the human thiamin transporter SLC19A2. *Am J Physiol Cell Physiol* **285**: C633–41.

Reidling JC, Subramanian VS, Dudeja PK, Said HM. 2002. Expression and promoter analysis of SLC19A2 in the human intestine. *Biochim Biophys Acta* **1561**: 180–7.

Reinke V, Krause M, Okkema P. 2013. Transcriptional regulation of gene expression in

C. elegans. *WormBook* 1–34.

Reith W, Herrero-Sanchez C, Kobr M, Silacci P, Berte C, Barras E, Fey S, Mach B. 1990. MHC class II regulatory factor RFX has a novel DNA-binding domain and a functionally independent dimerization domain. *Genes Dev* **4**: 1528–40.

Reith W, Mach B. 2001. The bare lymphocyte syndrome and the regulation of MHC expression. *Annu Rev Immunol* **19**: 331–73.

Rivolta C, Sharon D, DeAngelis MM, Dryja TP. 2002. Retinitis pigmentosa and allied diseases: numerous diseases, genes, and inheritance patterns. *Hum Mol Genet* **11**: 1219–27.

Robert VJP, Katic I, Bessereau J-L. 2009. Mos1 transposition as a tool to engineer the Caenorhabditis elegans genome by homologous recombination. *Methods* **49**: 263–9.

Rosenbaum JL, Witman GB. 2002. Intraflagellar transport. *Nat Rev Mol Cell Biol* **3**: 813–25.

Sawata M, Takeuchi H, Kubo T. 2004. Identification and analysis of the minimal promoter activity of a novel noncoding nuclear RNA gene, AncR-1, from the honeybee (Apis mellifera L.). *RNA* **10**: 1047–58.

Schafer JC, Haycraft CJ, Thomas JH, Yoder BK, Swoboda P. 2003. XBX-1 encodes a dynein light intermediate chain required for retrograde intraflagellar transport and cilia assembly in Caenorhabditis elegans. *Mol Biol Cell* **14**: 2057–70.

Seydoux G, Fire A. 1994. Soma-germline asymmetry in the distributions of embryonic RNAs in Caenorhabditis elegans. *Development* **120**: 2823–34.

She R, Chu JS-C, Uyar B, Wang J, Wang K, Chen N. 2011. genBlastG: using BLAST searches to build homologous gene models. *Bioinformatics* **27**: 2141–3.

Shiba D, Yokoyama T. 2012. The ciliary transitional zone and nephrocystins. *Differentiation* **83**: S91–6.

Signor D, Wedaman KP, Orozco JT, Dwyer ND, Bargmann CI, Rose LS, Scholey JM. 1999. Role of a class DHC1b dynein in retrograde transport of IFT motors and IFT raft particles along cilia, but not dendrites, in chemosensory neurons of living Caenorhabditis elegans. *J Cell Biol* **147**: 519–30.

Silverman MA, Leroux MR. 2009. Intraflagellar transport and the generation of dynamic, structurally and functionally diverse cilia. *Trends Cell Biol* **19**: 306–16.

Sleumer MC, Bilenky M, He A, Robertson G, Thiessen N, Jones SJM. 2009. Caenorhabditis elegans cisRED: a catalogue of conserved genomic elements. *Nucleic Acids Res* **37**: 1323–34.

Smith AD, Sumazin P, Zhang MQ. 2005. Identifying tissue-selective transcription factor binding sites in vertebrate promoters. *Proc Natl Acad Sci U S A* **102**: 1560–5.

Starich TA, Herman RK, Kari CK, Yeh WH, Schackwitz WS, Schuyler MW, Collet J, Thomas JH, Riddle DL. 1995. Mutations affecting the chemosensory neurons of Caenorhabditis elegans. *Genetics* **139**: 171–88.

Steimle V, Durand B, Barras E, Zufferey M, Hadam MR, Mach B, Reith W. 1995. A novel DNA-binding regulatory factor is mutated in primary MHC class II deficiency (bare lymphocyte syndrome). *Genes Dev* **9**: 1021–32.

Stein LD, Bao Z, Blasiar D, Blumenthal T, Brent MR, Chen N, Chinwalla A, Clarke L, Clee C, Coghlan A, et al. 2003. The genome sequence of Caenorhabditis briggsae: a platform for comparative genomics. *PLoS Biol* **1**: E45.

Stubbs JL, Oishi I, Izpisúa Belmonte JC, Kintner C. 2008. The forkhead protein Foxj1 specifies node-like cilia in Xenopus and zebrafish embryos. *Nat Genet* **40**: 1454–60.

Swoboda P, Adler HT, Thomas JH. 2000. The RFX-type transcription factor DAF-19 regulates sensory neuron cilium formation in C. elegans. *Mol Cell* **5**: 411–21.

Teif VB. 2010. Predicting gene-regulation functions: lessons from temperate bacteriophages. *Biophys J* **98**: 1247–56.

Thurman RE, Rynes E, Humbert R, Vierstra J, Maurano MT, Haugen E, Sheffield NC, Stergachis AB, Wang H, Vernot B, et al. 2012. The accessible chromatin landscape of the human genome. *Nature* **489**: 75–82.

Tucker BA, Scheetz TE, Mullins RF, DeLuca AP, Hoffmann JM, Johnston RM, Jacobson SG, Sheffield VC, Stone EM. 2011. Exome sequencing and analysis of induced pluripotent stem cells identify the cilia-related gene male germ cell-associated kinase (MAK) as a cause of retinitis pigmentosa. *Proc Natl Acad Sci U S A* **108**: E569–76.

Vannini A, Cramer P. 2012. Conservation between the RNA polymerase I, II, and III transcription initiation machineries. *Mol Cell* **45**: 439–46.

Vidal-Gadea A, Ward K, Beron C, Ghorashian N, Gokce S, Russell J, Truong N, Parikh A, Gadea O, Ben-Yakar A, et al. 2015. Magnetosensitive neurons mediate geomagnetic orientation in Caenorhabditis elegans. *Elife* **4**.

Walhout AJM. 2011. Gene-centered regulatory network mapping. *Methods Cell Biol* **106**: 271–88.

Wang J, Zhuang J, Iyer S, Lin X, Whitfield TW, Greven MC, Pierce BG, Dong X, Kundaje A, Cheng Y, et al. 2012. Sequence features and chromatin structure around the genomic regions bound by 119 human transcription factors. *Genome Res* **22**: 1798–812.

Weirauch MT, Yang A, Albu M, Cote AG, Montenegro-Montero A, Drewe P, Najafabadi HS, Lambert SA, Mann I, Cook K, et al. 2014. Determination and inference of eukaryotic transcription factor sequence specificity. *Cell* **158**: 1431–43.

Wenick AS, Hobert O. 2004. Genomic cis-regulatory architecture and trans-acting regulators of a single interneuron-specific gene battery in C. elegans. *Dev Cell* **6**: 757–70.

Wheatley DN, Wang AM, Strugnell GE. 1996. Expression of primary cilia in mammalian cells. *Cell Biol Int* **20**: 73–81.

Willaredt MA, Gorgas K, Gardner HAR, Tucker KL. 2012. Multiple essential roles for primary cilia in heart development. *Cilia* **1**: 23.

Williams CL, Li C, Kida K, Inglis PN, Mohan S, Semenec L, Bialas NJ, Stupay RM, Chen N, Blacque OE, et al. 2011. MKS and NPHP modules cooperate to establish basal body/transition zone membrane associations and ciliary gate function during ciliogenesis. *J Cell Biol* **192**: 1023–41.

Williams CL, Winkelbauer ME, Schafer JC, Michaud EJ, Yoder BK. 2008. Functional redundancy of the B9 proteins and nephrocystins in Caenorhabditis elegans ciliogenesis. *Mol Biol Cell* **19**: 2154–68.

Winkelbauer ME, Schafer JC, Haycraft CJ, Swoboda P, Yoder BK. 2005. The C. elegans homologs of nephrocystin-1 and nephrocystin-4 are cilia transition zone proteins involved in chemosensory perception. *J Cell Sci* **118**: 5575–87.

Xu X, Kim SK. 2011. The early bird catches the worm: new technologies for the Caenorhabditis elegans toolkit. *Nat Rev Genet* **12**: 793–801.

Yu X, Ng CP, Habacher H, Roy S. 2008. Foxj1 transcription factors are master regulators of the motile ciliogenic program. *Nat Genet* **40**: 1445–53.

Zahedi K, Bissler JJ, Prada AE, Prada JA, Davis AE. 1999. The promoter of the C1 inhibitor gene contains a polypurine.polypyrimidine segment that enhances transcriptional activity. *J Immunol* **162**: 7249–55.

Zhao Z, Fang L, Chen N, Johnsen RC, Stein L, Baillie DL. 2005. Distinct regulatory elements mediate similar expression patterns in the excretory cell of Caenorhabditis elegans. *J Biol Chem* **280**: 38787–94.

# Appendix

## Validated *daf-19* target genes

### Table A1. List of validated *daf-19* target genes

| Gene Name | Gene Model | X-box Sequence | WormBase Description | Reference |
|---|---|---|---|---|
| | ZK328.7 | GTTACCATGGCAAT | tetratricopeptide repeat domain 21B | (Blacque et al. 2005) |
| bbs-1 | Y105E8A.5 | GTTCCCATAGCAAC | Bardet-Biedl syndrome 1 protein ortholog | (Ansley et al. 2003; Efimenko et al. 2005; Blacque et al. 2005) |
| bbs-2 | F20D12.3 | GTATCCATGGCAAC | Bardet-Biedl syndrome 2 protein ortholog | (Ansley et al. 2003; Efimenko et al. 2005; Blacque et al. 2005) |
| bbs-5 | R01H10.6 | GTCTCCATGGCAAC | Bardet-Biedl Syndrome 5 protein ortholog | (Li et al. 2004; Efimenko et al. 2005) |
| bbs-8 | T25F10.5 | GTACCCATGGCAAC | Bardet-Biedl Syndrome 8 protein ortholog | (Ansley et al. 2003; Efimenko et al. 2005; Blacque et al. 2005) |
| bbs-9 | C48B6.8 | GTTTCCATGACAAC | parathyroid hormone-responsive B1 gene | (Blacque et al. 2005) |
| che-11 | C27A7.4 | ATCTCCATGGCAAC | Intraflagellar transport 140 homolog | (Qin et al. 2001; Efimenko et al. 2005; Chen et al. 2006) |
| che-12 | B0024.8 | GTTGCCCAGACTAC | Uncharacterised | (Bacaj et al. 2008) |
| che-13 | F59C6.7 | GTTGCTATAGCAAC | intraflagellar transport (IFT) complex B | (Haycraft et al. 2003; Efimenko et al. 2005) |
| che-2 | F38G1.1 | GTTGTCATGGTGAC | G-protein beta-like WD-40 | (Fujiwara et al. 1999; Swoboda et al. 2000; Efimenko et al. 2005) |
| daf-19 | F33H1.1b | GTTTCCATGGAAAC | RFX transcription factor | (Blacque et al. 2005) |
| dyf-11 | C02H7.1 | GTCTCCATGACAAC | intraflagellar transport (IFT) 54 protein ortholog | (Blacque et al. 2005; Chen et al. 2006; Bacaj et al. 2008) |
| dyf-2 | ZK520.3 | GTTACCAAGGCAAC | WDR19 | (Chen et al. 2006; Efimenko et al. 2006) |
| dyf-3 | C04C3.5 | GTTTCTATGGGAAC | clusterin-associated protein 1 ortholog | (Ou et al. 2005; Murayama et al. 2005) |
| dyf-5 | M04C9.5 | GTTACCATAGAAAC | MAP kinase | (Chen et al. 2006; Burghoorn |
| | | | orthologous to human MAK/ICK | et al. 2007) |
| dylt-2 | D1009.5 | GTTGCCATGACAAC | dynein light chain subunit | (Efimenko et al. 2005; Blacque et al. 2005) |
| fkh-2 | T14G12.4 | Not Reported | forkhead box transcription factor | (Mukhopadhyay et al. 2007) |
| ift-20 | Y110A7A.20 | GTCTCTATAGCAAC | Intraflagellar transport protein 20 homolog | (Chen et al. 2006; Blacque et al. 2005) |
| mks-1 | R148.1 | GTCACCATAGGAAC | Meckel syndrome 1 protein ortholog | (Williams et al. 2008; Efimenko et al. 2005) |
| mksr-1 | K03E6.4 | GTTCCCTTGGCAAC | B9 domain containing | (Blacque et al. 2005; Williams et al. 2008) |
| mksr-2 | Y38F2AL.2 | GTTGCCGTGGCAAC | B9 domain containing | (Blacque et al. 2005; Williams et al. 2008) |
| nhr-44 | T19A5.4 | GTCTTCATGGCAAC | hepatocyte nuclear factor 4, alpha | (Efimenko et al. 2005) |
| nphp-1 | M28.7 | GTTGCCAGGGGCAAC | nephrocystin-1 ortholog | (Winkelbauer et al. 2005) |
| nphp-4 | R13H4.1 | ATTTCCATGACAAC | NPHP4 | (Winkelbauer et al. 2005) |
| nud-1 | F53A2.4 | GTATCCATGAAAAC | Uncharacterised | (Dawe et al. 2001; Efimenko et al. 2005) |
| odr-4 | Y102E9.1 | ATCGTCATCGTAAC | type II membrane protein | (Dwyer et al. 1998; Efimenko et al. 2005) |
| osm-1 | T27B1.1 | GCTACCATGGCAAC | intraflagellar transport (IFT) complex B | (Signor et al. 1999; Swoboda et al. 2000; Efimenko et al. 2005; Bell et al. 2006) |
| osm-12 | Y75B8A.12 | GTTGCCATAGTAAC | Bardet-Biedl Syndrome 7 protein ortholog | (Ansley et al. 2003; Efimenko et al. 2005; Blacque et al. 2004) |
| osm-5 | Y41G9A.1 | GTTACTATGGCAAC | intraflagellar transport (IFT) 88 protein ortholog | (Haycraft et al. 2001; Efimenko et al. 2005; Qin et al. 2001) |
| osm-6 | R31.3 | GTTACCATAGTAAC | intraflagellar transport (IFT) 52 protein ortholog | (Collet et al. 1998; Swoboda et al. 2000; Efimenko et al. 2005; Blacque et al. 2004) |
| peli-1 | F25B4.2 | GTCTCCAATGGCAAC GTCCTCACAAGTAAC | Pellino family of E3 ubiquitin ligases | (Chu et al. 2012) |
| tub-1 | F10B5.4 | ATCTCCATGACAAC | TUBBY homolog | (Efimenko et al. 2005; Mak et al. 2006) |
| xbx-1 | F02D8.3 | GTTTCCATGGTAAC | dynein light intermediate chain | (Swoboda et al. 2000; Schafer et al. 2003; Efimenko et al. 2005) |
| xbx-3 | M04D8.6 | GTTGTCTTGGCAAC | Uncharacterised | (Efimenko et al. 2005) |
| xbx-4 | C23H5.3 | GTTGCCATGACAAC | Uncharacterised | (Efimenko et al. 2005) |
| xbx-5 | T24A11.2 | GTCTCCATGACAAC | Uncharacterised | (Efimenko et al. 2005) |
| xbx-6 | F40F9.1 | GTTTCCATGGAAAC | Fas apoptotic inhibitory molecule 2 | (Efimenko et al. 2005; Chen et al. 2006) |